\documentclass[11pt,a4paper]{article}
\pdfoutput=1
\usepackage{jheppub}
\usepackage[utf8]{inputenc}
\usepackage{tabularx}
\usepackage{array}
\usepackage{graphics}
\graphicspath{{figures/}}
\usepackage{graphicx}
\usepackage{psfrag}
\usepackage{epsfig}
\usepackage{amsmath}
\usepackage{amssymb}
\usepackage[normalem]{ulem}
\usepackage{setspace}
\usepackage{rotating}
\usepackage{colortbl}
\usepackage{tabularx}
\usepackage{longtable}
\usepackage{slashed}
\usepackage{lineno}
\makeatletter
\usepackage{textcomp}
\usepackage[usenames,dvipsnames]{xcolor}
\usepackage{relsize}
\usepackage{physics}
\usepackage{tikz-feynman}
\usepackage{caption}
\usepackage{subcaption}

\setlongtables

\newcommand{\be}{\begin{equation}}
\newcommand{\ee}{\end{equation}}
\newcommand{\bea}{\begin{equation} \begin{aligned}}
\newcommand{\eea}{\end{aligned} \end{equation}}	
	
\hypersetup{colorlinks,citecolor= nicegreen,linkcolor= nicegreen}
\definecolor{nicered}{rgb}{0.7,0.1,0.1}
\definecolor{nicegreen}{rgb}{0.1,0.5,0.1}
\definecolor{mygreen}{rgb}{0,0.72,0}

\newcommand {\red}

\def \eV{{\mathrm{eV}}}
\def \GeV{{\mathrm{GeV}}}
\def \TeV{{\mathrm{TeV}}}

\newcommand{\diag}{\operatorname{diag}}

\title{Exploring the Neutrino Sector of the Minimal Left-Right Symmetric Model}

\author[a]{Gustavo F. S. Alves}
\author[b]{\!\!, Chee Sheng Fong}
\author[a]{\!\!, Luighi P.  S. Leal}
\author[a]{and Renata Zukanovich Funchal}

\affiliation[a]{Departamento de F\'{\i}sica Matem\'atica, Instituto de F\'{\i}sica\\
Universidade de S\~ao Paulo, 05315-970 S\~ao Paulo, Brazil}

\affiliation[b]{Centro de Ciências Naturais e Humanas\\
Universidade Federal do ABC, 09.210-170, Santo André, Brazil}

\emailAdd{gustavo.figueiredo.alves@usp.br}
\emailAdd{sheng.fong@ufabc.edu.br}
\emailAdd{luighi.leal@usp.br}
\emailAdd{zukanov@if.usp.br}

\abstract{
We explore the neutrino sector of the minimal left-right symmetric model, with the additional charge conjugation
discrete symmetry, in the tuned regime where type-I and type-II seesaw mechanisms are equally responsible for the light 
neutrino masses. 
We show that unless the charged lepton mixing matrix is the identity and the right handed neutrino mass matrix has no phases, we expect sizable lepton flavor violation and electron dipole moment in this region.
We use results from recent neutrino oscillation fits, bounds 
on neutrinoless double beta decay, $\mu \to e \gamma$, 
$\mu \to 3 e$, $\mu \to e$ conversion in nuclei, the muon anomalous magnetic moment, the electron electric dipole moment and cosmology to determine the 
viability of this region. We derive stringent limits on the heavy neutrino masses and mixing angles as well as on the vacuum expectation value $v_L$, which drives the type-II seesaw contribution, using the current data. We discuss the 
perspectives of probing the remaining parameter space 
by future experiments.
}

 \arxivnumber{}
 
\begin{document}
\maketitle

\section{Introduction}
\label{sec:intro}

The neutrinos we know are massive and light.
Data on  neutrino flavor oscillations by a variety of different 
solar, atmospheric, reactor and accelerator neutrino experiments 
need two different mass squared difference scales $\Delta m^2_{21} \approx 7.5 \times 10^{-5}$ eV$^2$ and $\vert\Delta m^2_{31,32}\vert \approx 2.5 \times 10^{-3}$ eV$^2$ to be explained~\cite{Esteban:2020cvm}.
These measurements give rise to two possible mass ordering: 
$m_1<m_2<m_3$ (normal) or $m_3<m_1<m_2$ (inverted). Although it may seem that there is a slight data preference for normal ordering (NO), inverted ordering (IO) cannot be discarded~\cite{Kelly:2020fkv}. 
Complementary information comes from the KATRIN (KArlsruhe TRItium Neutrino) experiment that recently updated their limit on the effective electon neutrino mass to $m_\beta <0.8$ eV at 90\% CL~\cite{KATRIN:2021uub}.
Moreover, we need three light neutrinos to explain  cosmological observations that constrain the effective number of relativistic degrees of freedom to $N_{\rm eff}=2.99\pm 0.17$ and the sum of neutrino masses $\sum m_\nu< 0.12$ eV at 95\% CL~\cite{Planck:2018vyg}.

In spite of the fact that the theoretical origin of neutrino masses is unknown, the current very successful standard paradigm indicates that neutrino mass eigenstates are a particular admixture of the Standard Model (SM) neutrino flavor eigenstates. So neutrino oscillations present today the only solid experimental evidences of Lepton Flavor Violating (LFV) processes in nature. 

An unprecedented experimental campaign to probe rare LFV processes  with charged leptons is about to start. Experiments using a $\mu$ beam 
such as MEG II~\cite{MEGII:2018kmf}, Mu3e~\cite{Blondel:2013ia,Berger:2014vba}, COMET~\cite{COMET:2018auw} and Mu2e~\cite{Mu2e:2014fns} are prone to collect datasets of about $10^{15}-10^{17}$ muons, increasing the sensitivity to LFV transitions 
such as $\mu \to e \gamma$, $\mu \to 3 e$ and $\mu N \to e N$ by more than 
an order of magnitude down from current bounds.  

 However, although the leptonic charge associated to each lepton generation $L_\alpha\,  (\alpha = e, \mu, \tau)$ is  not conserved, 
oscillations do not allow us to say anything about the conservation 
of   the total leptonic charge $L=L_e+L_\mu +L_\tau$.
For this we need to seach for Lepton Number Violating processes (LNV).
Neutrinos can be Dirac or Majorana fermions. In the first case, 
neutrino interactions (including mass terms) must conserve $L$, in the second, neutrino interactions (including mass terms) do not conserve 
$L$. So the neutrino nature and $L$-conservation are interconnected.

Majorana neutrinos  can manifest themselves in $\vert\Delta L\vert= 2$
transitions of which the neutrinoless double-beta decay ($0\nu \beta \beta$)
is the most promising possibility. That is why there is a robust experimental program devoted to search for this decay using a variety of isotopes  ($^{76}$ Ge,$^{82}$Se,$^{130}$Te and $^{136}$Xe), with detector masses  in the range 100 kg -- 10000 kg, using different techniques, aiming to  reach sensitivies on the effective Majorana neutrino mass $m_{\beta \beta} \lesssim$ few meV~\cite{Agostini:2017jim}.

In the SM $L$-conservation is due to an 
accidental global $U(1)_L$ symmetry, which we expect to be broken 
by quantum fluctuations. The smallness of neutrino masses 
may be  understood in a plausible way in extensions of the SM 
by the introduction of new fields that are integrated out at low energies 
generating the so-called Weinberg~\cite{Weinberg:1979sa} operator. 
This operator leads to a Majorana mass term and encapsulates the so-called seesaw mechanisms~\cite{Minkowski:1977sc,Gell-Mann:1979vob,Yanagida:1979as,Mohapatra:1979ia,Schechter:1980gr,Foot:1988aq}. 
This mechanism emerges  inherently in theories where right-handed neutrinos are a necessity such as left-right symmetric models.
Although these models were initially motivated by a desired to understand the origin of parity violation in weak interactions, they turned out to be 
a natural framework to explain the pattern of neutrino masses. 

In the Minimal Left-Right Symmetric Model (MLRSM)~\cite{Pati:1974yy,Mohapatra:1974gc,Senjanovic:1975rk,Mohapatra:1979ia,Mohapatra:1980yp}, after left-right
and electroweak symmetry breaking, the light neutrino mass term  ($m_\nu$) is generated from the type-I and type-II seesaw mechanisms
\begin{eqnarray}
m_{\nu} & = & m_{{\rm I}}+m_{{\rm II}},
\label{eq:nu_mass}
\end{eqnarray}
where schematically, $m_{{\rm I}}\sim v^{2}/v_{R}$ and $m_{{\rm II}}\sim v_{L}$
with $v=246$ GeV the effective Higgs vacuum expectation value (vev),
and $v_{L}\ll v_{R}$ are the vevs associated with left and right
triplet Higgs which spontaneously break  the left-right symmetry. Depending
on the ratio $v_{L}/v_{R}$, one can have three scenarios: (i) a type-I
dominated scenario when $m_{{\rm I}}\gg m_{{\rm II}}$, (ii) a type-II dominated
scenario when $m_{{\rm I}}\ll m_{{\rm II}}$, or (iii) a tuned scenario when $m_{{\rm I}}\simeq m_{{\rm II}}$.
The mixing between the SM neutrinos and (mostly
right-handed) heavy neutrinos (in the following we will denote this
the active-sterile mixing) is proportional to $\sqrt{m_{{\rm I}}/v_{R}}$
and hence in the tuned scenario, if $m_{{\rm I}}\simeq m_{{\rm II}}\gg m_{\nu}$,
one can have a sufficiently large active-sterile mixing that leads
to signatures in present and forthcoming experiments. In contrast to previous works that assumed 
either type I~\cite{Barry:2013xxa, BhupalDev:2014qbx, Bambhaniya:2015ipg,Goswami:2020loc} or type II~\cite{Tello:2010am,Barry:2013xxa,Bambhaniya:2015ipg,Li:2020flq} seesaw dominance, although we also 
discuss  type-I seesaw here, we are mainly interested in exploring the tuned scenario where both types of seesaw are comparable\footnote{Requiring the Majorana neutrino mass of eq. \eqref{eq:nu_mass} with $v_L \neq 0$ and $v_R$ not far above the TeV scale, such that we have potentially experimental signatures, imply tuning in the Higgs potential parameters~\cite{Deshpande:1990ip} which we accept.}. This could have interesting experimental signatures due to active-sterile mixing and, as we shall show, allows us to study the role played by  
the different sectors of the MLRSM in various new 
physics processes implicating the leptonic sector.

This work is organized as follows. In section \ref{sec:model} we introduce the leptonic sector of the MLRSM, which is the focus of our work. We examine 
the physical parameters of this sector in the general case and for a discrete left-right symmetry realised by 
charge conjugation or parity. We describe how to 
reconstruct the neutrino mass in each of these cases. 
We will then assume the charge conjugation symmetry.
In section~\ref{sec:pheno} we describe the 
two regimes of our study (type I and tuned), the relevant parameters of the model used  in the 
phenomenological analysis and present a standard and three other benchmark cases. We discuss the model contributions to several observables in the neutrino sector, $0\nu \beta \beta$, LFV processes, the anomalous magnetic moment of the muon ($a_\mu$) and the electric dipole 
moment ($d_\ell$) and their dependence on the lightest heavy neutrinos mass ($m_4$) in view of current  and future experimental bounds. We also comment on the impact of the model in the new CDF II result on the W boson mass.
In section \ref{sec:results} we combine all available  
limits on the studied observables to determine the 
current allowed region for the heavy neutrino sector parameters. We also consider the effect of changing  the standard benchmark value of the most significant parameters on this allowed region.
We comment on the range of the heavy neutrino sector parameters that can be explored by future experiments and discuss some correlations among observables.
Finally, in section \ref{sec:conclusions} 
we draw our conclusions. In appendix~\ref{app:model} 
we describe for completeness the other sectors of the MLRSM, in appendix~\ref{app:formfac} we give approximate expressions for the form factors needed  
for the LFV, $a_\mu$ and $d_\ell$ calculations and in appendix~\ref{app:Plots} we show some supplementary 
plots to illustrate the small  dependence of our results on parameters otherwise fixed in our study.

\section{On the Leptonic Sector of the Minimal Left-Right Symmetric Model}
\label{sec:model}

The MLRSM is a gauge theory that, while keeping the fermionic 
content of the SM basically untouched (except for the neutrino sector as right-handed neutrinos are included), 
doubles its gauge sector by extending the gauge group to  
$SU(2)_{L}\times SU(2)_{R}\times U(1)_{B-L}$, possibly supplemented by  
 a discrete symmetry between the left and right sectors.
Here we will discuss the pattern of the mass matrices in the leptonic sector of the MLRSM. 
A description of the other sectors of the model relevant to our work can be found in Appendix~\ref{app:model}.

The  mass matrices in the leptonic sector of the MLRSM 
depends on the symmetries imposed at high energies. To understand that, let us examine the Yukawa interactions in the leptonic sector given by 
\begin{eqnarray}
{\cal L}_\ell & = & -\overline{L_{L}}\left(h_{l}\mathbf{\Phi}+\tilde{h}_{l}\tilde{\mathbf\Phi}\right)L_{R}-\overline{L_{L}^{c}}i\sigma_2\mathbf{\Delta}_{L}h_{M}L_{L}-\overline{L_{R}^{c}}i\sigma_2\mathbf{\Delta}_R\tilde{h}_{M}L_{R}+{\rm h.c.},
\label{eq:ell}
\end{eqnarray}
where $L_{L}\sim\left(2,1,-1\right)$, $L_{R}\sim\left(1,2,-1\right)$
are the left- and right-handed leptons and the scalars are $\mathbf\Phi\sim\left(2,2,0\right)$ with $\tilde{\mathbf \Phi} \equiv i \sigma_2 \mathbf{\Phi^*} \sigma_2 $,
$\mathbf{\Delta}_{L}\sim\left(3,1,2\right)$ and $\mathbf{\Delta}_{R}\sim\left(1,3,2\right)$, in parentheses we show their quantum numbers with respect to the model gauge symmetry groups.
In the above, we have left the flavor indices which run from 1 to 3 of $L_L$ and $L_R$ implicit and in the flavor space, both $h_{l}$ and $\tilde{h}_{l}$ are general $3\times3$
complex matrices while $h_{M}$ and $\tilde{h}_{M}$ are general $3\times3$
complex symmetric matrices. Let us count the physical parameters for
the following cases: general, charge conjugation 
(${\cal C}$) and parity (${\cal P})$ symmetry.
\begin{enumerate}
\item {\it General case}

Without loss of generality, we can rotate both left-handed
and right-handed lepton doublets to make both $h_{M}$ and $\tilde{h}_{M}$ real and diagonal. Let us count the parameters in the Lagrangian,
we have $9+9+3+3=24$ moduli and $9+9=18$ phases. The 24 moduli would correspond to 12 ``low-energy'' observables (3 charged lepton masses, 3 light neutrino masses, 3 heavy neutrino masses and 3 leptonic mixing angles) plus 12 additional ``high-energy'' parameters. The 18 phases would correspond to 3 leptonic CP phases $(\delta,\alpha_{1},\alpha_{2})$ plus 15 ``high-energy''
additional phases.

\item {\it ${\cal C}$ symmetry} 

Under ${\cal C}$ the fields in eq.~(\ref{eq:ell})
transform as 
\be
L_L \leftrightarrow L_R^c,\;\;\;\mathbf{\Delta}_L \leftrightarrow \mathbf{\Delta}_R^\dagger,\;\;\;
\mathbf\Phi \leftrightarrow \mathbf\Phi^T\,,
\ee
so in this case, we have $h_{l}=h_{l}^{T}$, $\tilde{h}_{l}=\tilde{h}_{l}^{T}$
and $h_{M}=\tilde{h}_{M}^{*}$. We will rotate the lepton fields such
that $h_{M}=\tilde{h}_{M}^{*}$ is diagonal but we will keep the three
phases. The Lagrangian parameters are now $6+6+3=15$ moduli and $6+6=12$
phases.  So under this extra symmetry there are only 3 
additional ``high-energy'' parameters and 9 additional ``high-energy'' phases. 

\item {\it ${\cal P}$ symmetry}

Under ${\cal P}$ the fields in eq.~(\ref{eq:ell})
transform as 
\be
L_L \leftrightarrow L_R,\;\;\;\mathbf{\Delta}_L \leftrightarrow \mathbf{\Delta}_R,\;\;\;
\mathbf\Phi \leftrightarrow \mathbf\Phi^\dagger\,,
\ee
so in this case, we have $h_{l}=h_{l}^{\dagger}$, $\tilde{h}_{l}=\tilde{h}_{l}^{\dagger}$
and $h_{M}=\tilde{h}_{M}$. We can again make $h_{M}$ real and diagonal.
The Lagrangian parameters reduce to $6+6+3=15$ moduli and $3+3=6$ phases. The difference from  the  
${\cal C}$ case is that now we have only 3 additional ``high-energy'' phases instead of 9.

\end{enumerate}

Assuming \emph{real} vevs for the scalar fields, after the left-right and 
electroweak symmetry breaking the charged lepton mass matrix becomes 
\be
M_l  = \frac{1}{\sqrt{2}}\left( h_l\, \kappa_2 + \tilde{h}_l \,\kappa_1 \right)\, ,
\label{eq:lepmass}
\ee
where $\kappa_1$ and $\kappa_2$ are the vevs of $\mathbf\Phi$. Assuming $v_R^2 \gg \kappa_1^2 + \kappa_2^2 \gg v_L^2$ where $v_R$ and $v_L$ are, respectively, the vevs of $\mathbf{\Delta}_R$ and $\mathbf{\Delta}_L$, we need
\be
\kappa_+^2 \equiv \kappa_1^2 + \kappa_2^2 = v^2 = (246\,{\rm GeV})^2\, ,
\label{eq:weak_scale}
\ee
to reproduce the observed masses of the $Z$ and $W$ bosons. Two of the four vevs can be complex. In the view that there are already several new phases which are not constrained in the leptonic sector, the possibility of complex vevs will not be explored in this work.

Eq.~\eqref{eq:lepmass} can be put in a diagonal form by a biunitary transformation
\be 
M_l = U_L^{\dagger} \hat{M_l}U_R \, ,
\ee
where $\hat{M_l}= {\text{diag}} (m_e, m_\mu, m_\tau)$ and $U_L^{*}=U_R$ if $M_l$ is symmetric (for ${\cal C}$) or 
$U_L=U_R$ if $M_l$ is hermitian (for ${\cal P}$).

In the {\em charged lepton mass basis}, the $6\times 6$ neutral lepton mass matrix ${\cal M}_\nu$ defined by $\bar\nu {\cal M}_\nu \nu^c$ where $\nu = \left(\nu_{eL},\nu_{\mu L}, \nu_{\tau L}, \nu_{eR}^{\quad c}, \nu_{\mu R}^{\quad c}, \nu_{\tau R}^{\quad c}\right)^{T}$, is given by
\be 
{\cal M}_\nu = {\cal V}\, {\cal M}_0 \, {\cal V}^T\, ,
\label{eq:fullnumass}
\ee
where ${\cal V}$ is the unitary matrix defined by
\be 
{\cal V} \equiv  \left( \begin{array}{cc} U_L & 0 \\ 0 & U_R^*\end{array}\right) \, ,
\label{eq:vcal}
\ee
and ${\cal M}_0$ is the mass matrix defined by
\be
{\cal M}_0 \equiv  \left( \begin{array}{cc} M_L^{\dagger} & M_D \\ M_D^T & M_R\end{array}\right) \, ,
\label{eq:m0}
\ee
with diagonal block matrices
\bea
M_L &=& \sqrt{2} \, h_M \, v_L\, ,\\
M_R &=& \sqrt{2}\, \tilde{h}_M \, v_R\, ,\\
\eea
and the Dirac neutrino mass matrix
\be
M_D  = \frac{1}{\sqrt{2}}\left(h_l\, \kappa_1 + \tilde{h}_l \,\kappa_2\right)\, .
\label{eq:Diracmass}
\ee
For ${\cal C}$ symmetry, $M_L=v_L M_R^*/v_R$ and $M_D$ is symmetric while for ${\cal P}$ symmetry, $M_L=v_L M_R/v_R$ and $M_D$ is hermitian.

We can block diagonalize ${\cal M}_0$ to obtain the $3\times 3$ light neutrino mass matrix. The leading term in the seesaw approximation 
$\vert M_D\vert \ll \vert M_R\vert$ is 
\be
m_\nu =  m_{{\rm I}}+m_{{\rm II}} = - M_D M_R^{-1} M_D^T+ M_L^\dagger =
- \frac{1}{\sqrt{2}\, v_R}
M_D {\tilde{h}_M}^{-1} M_D^T+ \sqrt{2} \,h_M^\dagger \,v_L \, ,
\label{eq:mnu}
\ee
the first being the type-I and the second the type-II seesaw contributions, which in the {\em charged lepton mass basis} defines the the leptonic mixing matrix $U_{\rm PMNS}$ by
\be 
U_L \, m_\nu \, U_L^T \equiv U_{\rm PMNS} \, \hat{m}_\nu \, U^T_{\rm PMNS}\,, 
\label{eq:UPMNSdef}
\ee
with $\hat{m}_\nu= {\text{diag}} (m_1, m_2, m_3)$ the diagonal light neutrino mass matrix. So we can also write 
in this basis $m_\nu$ as 
\be
m_\nu = U_{L}^\dagger\, U_{\rm PMNS} \, \, \hat{m}_\nu \, U_{\rm PMNS}^T \, U_{L}^*\,.
\label{eq:mnu2}
\ee

The heavy neutrino mass matrix at the leading order is just $M_R = \sqrt{2}\tilde h_M v_R$. 
At the minimum of the scalar potential, the vevs $v_R$, $v_L$ and $\kappa_+$ satisfy the relation~\cite{Deshpande:1990ip}
\be
v_L v_R = \gamma \kappa_+^2,
\ee
where 
\be
\gamma = \frac{\beta_1 \kappa_1 \kappa_2 + \beta_2 \kappa_1^2 + \beta_3 \kappa_2^2}{(2\rho_1 - \rho_3) \kappa_+^2},
\ee
with $\beta_i$ and $\rho_i$ the relevant dimensionless couplings in the scalar potential. In the absence of tuning, $\gamma$ would be of the order of one. 
In this work, we are exploring sub-TeV scale right-handed neutrinos which implies $\gamma \ll 1$ or tuning of the parameters of the scalar potential~\cite{Deshpande:1990ip}.

\subsection{Parametrization and Neutrino Mass Reconstruction}

From eq.~\eqref{eq:mnu}, we can write the type-I contribution as
\be
m_{\rm I} =- M_D M_R^{-1} M_D^T = m_\nu - M_L^\dagger.
\label{eq:mI}
\ee
Multiplying by $M_R^{-1}$ from the right, we have
\be
M_D M_R^{-1} M_D^T M_R^{-1} =  M_L^\dagger M_R^{-1} - m_\nu M_R^{-1}.
\label{eq:MDMR}
\ee
Since the active-sterile mixing (that we will discuss more later) is proportional to $M_D M_R^{-1}$, it can be enhanced if $|M_L^\dagger| \gg |m_\nu|$ i.e. if we are in the tuned scenario where $m_{\rm I} \simeq m_{\rm II}$.
Next, we discuss the solutions for $M_D$ in the three cases of interest.

\begin{enumerate}
     \item {\it General case}
     
     There are sufficient degrees of freedom to construct a general complex Dirac matrix $M_D$ based on the Casas-Ibarra parametrization~\cite{Casas:2001sr} 
\be
M_D = B^\dagger \sqrt{\hat{m}_{\rm I}}\,  R\,  \sqrt{M_R}\, ,\label{eq:CI}
\ee
where $R$ is a complex orthogonal matrix and $B$ is a unitary matrix which diagonalizes $m_\nu - M_L^\dagger$
\be
m_{\rm I} = - B^\dagger\, \hat{m}_{\rm I}\, B^*=
m_\nu - M_L^\dagger\, ,
\label{eq:master}
\ee
with $\hat{m}_{\rm I}$ a diagonal matrix containing the neutrino mass contributions from type-I seesaw.

     \item {\it ${\cal C}$ symmetry}
     
      $M_D$ is a symmetric matrix with six moduli and six phases. As we will see below, it will be completely fixed by $v_L$, $v_R$, six moduli and 12 phases residing in: three moduli in $U_L$, three moduli in $\tilde{h}_M$, three phases in $U_{\rm PMNS}$, six phases in $U_L$ and three phases in $\tilde{h}_M=h_M^*={\rm diag}(h_1,h_2,h_3)\, {\rm diag}(e^{i \phi_1},e^{i\phi_2},e^{i\phi_3})$.
     Setting $M_D^T = M_D$ and $M_L/v_L = M_R^\dagger/v_R$ in eq. \eqref{eq:MDMR}, we have
     \be 
     Z^2 =  \frac{v_L}{v_R} \; \mathbf{I} - m_\nu M_R^{-1},
     \label{eq:Z_C_scenario}
     \ee
     where we have defined $Z \equiv M_D M_R^{-1}$. From the equation above, we can solve for $M_D$ explicitly \cite{Nemevsek:2012iq}
     \be 
     M_D =  \left(\frac{v_L}{v_R} \; \mathbf{I} - m_\nu M_R^{-1}\right)^{1/2} M_R.
     \label{eq:MD_C_scenario}
     \ee
     Notice that $M_D$ is completely fixed by the rest of the parameters.
     In the tuned scenario $v_L/v_R \gg |m_\nu M_R^{-1}|$, we can approximate 
     \be
     M_D \simeq  \left(\sqrt{\frac{v_L}{v_R}} \; \mathbf{I} - \frac{1}{2} \sqrt{\frac{v_R}{v_L}} m_\nu M_R^{-1}\right) M_R 
     = \sqrt{\frac{v_L}{v_R}} M_R - \frac{1}{2} \sqrt{\frac{v_R}{v_L}} m_\nu,
     \label{eq:C-tuned}
     \ee
     where the second term is always subdominant.

    \item {\it ${\cal P}$ symmetry}
    
    $M_D$ is a hermitian matrix with six moduli and three phases. This scenario is more restrictive than the case of $\cal C$ symmetry since there are three less phases in $M_D$ and in principle, it will depend on $v_L$, $v_R$, six moduli and six phases residing in: three moduli in $U_L$, three moduli in $\tilde{h}_M=h_M={\rm diag}(h_1,h_2,h_3)$, three phases in $U_{\rm PMNS}$ and three phases in $U_L$. Setting $M_D^T = M_D^*$ and $M_L/v_L = M_R/v_R$ in eq. \eqref{eq:MDMR}, we have
    \be 
    ZZ^* =  \frac{v_L}{v_R} \; \mathbf{I} - m_\nu M_R^{-1},
    \label{eq:Z_P_scenario}
    \ee
    where we have worked in the basis where $M_R$ is real and diagonal. 
    A procedure for solving $M_D$ was laid out in ref.~\cite{Senjanovic:2016vxw} and we will defer the study of this case to future work.
\end{enumerate}
The parameters of the model not directly constrained by what we have already measured (beside  the three phases in $U_{\rm PMNS}$) are collected in table \ref{tab:pcouting}.

\begin{table}
\centering
\begin{tabular}{ |p{2cm}|p{7.0cm}|p{4.0cm}|}
\hline
\bf Case & \bf Free Moduli &  \bf Free Phases\\
\hline
General  & $U_L(3) + U_R(3) + R(3) + \tilde{h}_M(3) + h_M(3)$& $U_L(6) + U_R(3) + R(3)$\\
$\cal C$  & $U_L(3)+ h_M(3)$& $U_L(6)+h_M(3)$\\
$\cal P$  & $U_L(3)+ h_M(3)$& $U_L(3)$\\
\hline
\end{tabular}
\caption{\label{tab:pcouting} Parameters of the lepton sector of the MLRSM not constrained directly by what we have measured. Although the three Majorana phases of the PMNS matrix have not yet been measured, we do not include them in the table.}
\end{table}
From eqs.~\eqref{eq:lepmass} and \eqref{eq:Diracmass}, we can solve for 
\be
h_{l} =\frac{1}{\sqrt{2}}\frac{M_{D}\kappa_{1}-M_{l}\kappa_{2}}{\kappa_{1}^{2}-\kappa_{2}^{2}},\;\;\;\;\;
\tilde{h}_{l} = \frac{1}{\sqrt{2}}\frac{M_{l}\kappa_{1}-M_{D}\kappa_{2}}{\kappa_{1}^{2}-\kappa_{2}^{2}}.
\label{eq:hl_hlt}
\ee

Taking into account eq.~\eqref{eq:weak_scale}, there is still a freedom to choose $\kappa_1/\kappa_2$,  we will assume $\epsilon = 2\kappa_{1}\kappa_{2}/\kappa_{+}^{2}\ll 1$. The procedure described above to solve for $M_D$ allows us to construct $h_l$ and $\tilde h_l$ to recover the known observables in the lepton sector (charged lepton masses, neutrino mass differences and the $U_{\rm PMNS}$) at the \emph{leading order}. To strike a balance between the completely general case and the $\cal P$ symmetry case, we will focus on the case with ${\cal C}$ symmetry where the number of new parameters are reasonably restricted and work in the basis where $\tilde h_M = h_M^*$ is diagonal with three phases and leave the study of other scenarios (including CP-violating vevs) for future work.

We reconstruct $M_D$ and the neutrino mass matrix by the following steps:
\begin{enumerate}
    \item chose the ordering of the light neutrino masses (NO or IO);
    \item fix the $U_{\rm PMNS}$ matrix at the best fit values for the 
    corresponding selected ordering according to \cite{Esteban:2020cvm} as described in table~\ref{tab:params};
    \item chose $m_0$, the lightest neutrino mass and compute $m_2$ and 
    $m_3$ (or $m_1$) according to the best fit values of $\Delta m^2_{21}\equiv m^2_2-m^2_1$
    and $\Delta m^2_{31}\equiv  m^2_{3}-m^2_1$ ($\Delta m^2_{32}\equiv  m^2_{3}-m^2_2$) for  NO (IO);
    \item fix $v_R$, $v_L$;
    \item fix $U_L$ and $h_M$ according to the case with $\cal C$ symmetry; 
    \item determine $M_D$ from eq.~\eqref{eq:MD_C_scenario};
    \item construct and diagonalize the $6\times6$ neutrino mass matrix ${\cal M}_0$ in eq.~\eqref{eq:m0}.
\end{enumerate}

In possession of the unitary matrix  ${\cal U}$ that diagonalizes the neutrino sector
\be
{\cal U}^\dagger  {\cal M}_\nu\, {\cal U}^* =\hat M_\nu ={\rm diag}(m_1,m_2,m_3,m_4,m_5,m_6),
\label{eq:diagonalformnumass}
\ee
we have the mixtures and can finally compute all the relevant observables and subject them to the appropriate bounds. At the leading order, we have\footnote{The form of the mixing matrix below is general, independently of ${\cal C}$ or ${\cal P}$ discrete symmetry.}
\be
{\cal U} \equiv \left(\begin{array}{c} {\cal U}_L \\ {\cal U}^*_R \end{array}\right)\equiv \left(\begin{array}{cc} {\cal U}_{LL} & {\cal U}_{LR} \\ {\cal U}^*_{RL} & {\cal U}^*_{RR} \end{array}\right) =\left(\begin{array}{cc}
(\mathbf{I} - \delta U_L) U_{{\rm PMNS}}
& U_{L} Z \phi_R^\dagger\\
-U_{R}^{*}Z^{\dagger}U_{L}^{\dagger}U_{{\rm PMNS}} 
& U_{R}^{*}(\mathbf{I} - \delta U_R)\phi_R^{\dagger}
\label{eq:66_mixing}
\end{array}\right),
\ee
where
\begin{eqnarray}
\delta U_L &\equiv&
\frac{1}{2}U_{L}ZZ^{\dagger}U_{L}^{\dagger}, \\
\delta U_R &\equiv&
\frac{1}{2}Z^{\dagger} Z.
\end{eqnarray}
$\mathcal{U}_L$ and $\mathcal{U}_R$ are $3\times 6$ matrices while $\mathcal{U}_{LL}$, $\mathcal{U}_{LR}$, $\mathcal{U}_{RL}$ and $\mathcal{U}_{RR}$ are $3\times 3$ matrices.
$\phi_R$ is a unitary matrix which diagonalizes $M_R$ as $\phi_R M_R \phi_R^T = \hat M_R$ with $\hat M_R$ diagonal, real and positive. 
If we start with $M_R$ being a generic diagonal matrix with phases $\diag(e^
{i\phi_1},e^{i\phi_2},e^{i\phi_3})$ 
then $\phi_R = {\rm diag}(e^{-i\phi_1/2},e^{-i\phi_2/2},e^{-i\phi_3/2})$ and if $M_R$ is proportional to $\mathbf{I}$, then $\phi_R$ is an orthogonal complex matrix fixed by the higher order seesaw terms. 
If $m_4,m_5,m_6$ are degenerate, their ordering is immaterial while if they are not degenerate, our convention is $m_4 < m_5 < m_6$. The relation between neutrinos in the flavor and the mass basis can be written as $\hat\nu = {\cal U}^\dagger\nu$ where $\hat\nu = \left(\nu_1,\nu_2,\nu_3, N_4, N_5, N_6\right)$ are the neutrinos in the mass basis.

From eqs.~\eqref{eq:diagonalformnumass} and \eqref{eq:fullnumass}, we can write $M_L$, $M_R$ and $M_D$ in terms of $\hat{M}_\nu$ as
\begin{align}
    M_L & = U_L^T \mathcal{U}_L^* \hat{M}_\nu \mathcal{U}_L^\dagger U_L,\\
    M_R &= U_R^T \mathcal{U}_R^* \hat{M}_\nu \mathcal{U}_R^\dagger U_R,\\
    M_D &= U_L^\dagger \mathcal{U}_L \hat{M}_\nu \mathcal{U}_R^\dagger U_R.
    \label{eq:MLMRMD}
\end{align}
In eq.~\eqref{eq:hl_hlt}, we see that the couplings depend on both $M_l$ and $M_D$. In the basis where $M_l$ is diagonal with diagonal entries corresponding to the charged lepton masses, the rest of the contributions will come from $M_D$ which according to the eq.~\eqref{eq:MLMRMD}, can be written in terms of the mixing elements and the neutrino masses.

\section{On the contributions of the MLRSM to Leptonic Observables}
\label{sec:pheno}

In this section we will explore the different contributions of the MLRSM to observables involving the mixing in the leptonic sector. At the end of the section we will also comment on the effects on the model of the new limit on the $W$ boson mass by CDF II~\cite{CDF:2022hxs}.

To quantify the relative contributions of $m_{\rm I}$ with respect to $m_{\rm II}$ to the neutrino mass matrix let us define the $r$ ratio 
\be 
r \equiv \frac{{\rm max} \, \vert m_{\rm I}\vert}{{\rm max} \, \vert m_{\rm II}\vert} =\frac{{\rm max} \, \vert M_D {{\tilde h}_M}^{-1} M_D^T\vert}{2 v_R v_L{\rm max} \, \vert h_M\vert} \, ,
\label{eq:r}
\ee
where max$|B|$ denotes the maximum of the absolute value of the entries of matrix $B$.
This ratio can help identify three possible regimes:
\begin{itemize}
    \item[(i)] type-I seesaw dominance -  This is the case for $v_L \ll \frac{1}{\sqrt{2}}\vert m_\nu h_M^{-1}\vert$ when $\vert m_{\rm I}\vert \gg \vert m_{\rm II}\vert$ ($r \gg 1$);
    \item[(ii)] type-II seesaw dominance - This is the case when $v_L \simeq \frac{1}{\sqrt{2}}|m_\nu h_M^{-1}|$ which gives $\vert m_{\rm I}\vert \ll \vert m_{\rm II}\vert$ ($r \ll 1$);
    \item[(iii)] tuned -  This occurs for $v_L \gg  \frac{1}{\sqrt{2}}\vert m_\nu h_M^{-1}\vert$ when both contributions are important $\vert m_{\rm I}\vert \simeq \vert m_{\rm II}\vert$ ($r \sim 1$).
\end{itemize}
We will focus on two of them from now on, 
(i) the type-I seesaw dominance and (iii) the tuned regime such that the active-sterile mixing is large in the sense $|\mathcal{U}_{LR}|^2 \gtrsim |m_\nu M_R^{-1}|$. These two cases have interesting 
implications for the neutrino sector, LNV and LFV observables. They also impact the contributions of the model to the anomalous magnetic moment of the muon and the electron electric dipole moment. 
Before discussing these effects let us describe the relevant parameters of the model 
for our study and the standard values we have fixed them to.

Unless otherwise stated we have fixed $v_R=44$ TeV, which correspond
to $m_{W'}= 20$ TeV and $m_{Z'}=33$ TeV assuming the same left and right gauge couplings $g_L = g_R = g$.\footnote{These values are above 
the experimental limits on the masses of these bosons, {\em i.e.},~ $m_{Z'} \geq 1162$ GeV and $m_{W'}\geq 715$ GeV from eletroweak tests~\cite{delAguila:2010mx} or 
$m^{\rm exp}_{Z'} \geq 3.2$ TeV~\cite{Lindner:2016lpp} and $m^{\rm exp}_{W'}\geq 5.4$ TeV~\cite{CMS:2021dzb}  from direct searches at the LHC. 
Also a renormalization group 
evolution analysis imposes $m_{W'}>$ 6 TeV~\cite{Maiezza:2016ybz}.
Note that phenomenological studies of the MLRSM which focus on the hadronic sector have obtained even stronger bounds on $m_W' $ assuming ${\cal P}$ symmetry, i.e., $m_{W'} \geq 13$ TeV~\cite{Bertolini:2019out} and   $m_{W'} \geq 17$ TeV~\cite{Dekens:2021bro}.}
We have also fixed $m_{A^0}=m_{H}= 15$~TeV, at their minimum value allowed by their tree-level contribution to $K^0 -\overline K^0$ mixing~\cite{Zhang:2007da}, which also imposes $m_{H^+}= 15$ TeV,  
the doubly charged scalars masses $m_{\delta_L^{++}}=780$ GeV and $m_{\delta_R^{++}}=660$ GeV at the minimum value allowed by the LHC~\cite{ATLAS:2017xqs}, $m_{\delta_L^+}=780$ GeV and the mixing between 
the charged  vector bosons $ \xi \equiv  \kappa_1 \kappa_2/v_R^2 = 3.0 \times 10^{-8}$, this is still about one order of magnitude larger than the lower bound from electroweak radiative  corrections~\cite{Nemevsek:2012iq,Branco:1978bz}.

\begin{table}[htb]
    \begin{center}
      \begin{tabular}{|p{3cm}|p{3cm}|p{3cm}|}
            \hline \multicolumn{3}{|c|}{\textbf{Light neutrino sector}}\\
            \hline   & \hfil NO &\hfil IO \\
            \hline  \hfil $\sin^2{\theta_{12}}$ &\hfil 0.304&\hfil  0.304\\
            \hline  \hfil $\sin^2{\theta_{13}}$ &\hfil 0.02220&\hfil 0.02238\\
            \hline  \hfil $\sin^2{\theta_{23}}$ &\hfil 0.573&\hfil  0.578\\
            \hline  \hfil $\delta(^\circ)$ &\hfil 194&\hfil  287\\
                       \hline  \hfil $\alpha_{1,2}$ &\hfil 0&\hfil  0\\
            \hline  \hfil $\Delta m^2_{21} \, (\eV^2)$ &\hfil $7.42 \cross 10^{-5}$&\hfil  $7.42 \cross 10^{-5}$\\
            \hline  \hfil $\Delta m^2_{3l}\, (\eV^2)$ &\hfil $2.515 \cross 10^{-3}$&\hfil  $ -2.498 \cross 10^{-3}$\\
             \hline  \hfil $m_0 \, (\eV)$  &\multicolumn{2}{|c|}{0.01}\\
            \hline  \multicolumn{3}{|c|}{\textbf{Heavy Neutrino Sector}}\\
            \hline  \hfil $U_L$ &\multicolumn{2}{|c|}{$\mathbf{I}$}\\
            \hline \hfil $v_R \,  (\TeV)$ &\multicolumn{2}{|c|}{$44$}\\
            \hline \hfil ordering &\multicolumn{2}{|c|}{$h_1/h_3 = h_2/h_3 =1$}\\
            \hline \hfil phases &\multicolumn{2}{|c|}{$\phi_1 = 0,\ \phi_2 = 0,\ \phi_3 =0$}\\
            \hline  \multicolumn{3}{|c|}{\textbf{Gauge Sector}}\\
            \hline \hfil $\xi \, (\equiv  \kappa_1 \kappa_2/v_R^2)$&  \multicolumn{2}{|c|}{$3.0 \cross 10^{-8}$}\\
            \hline \hfil $m_{W'}\, (\TeV)$ &\multicolumn{2}{|c|}{$20$}\\
            \hline \hfil $m_{Z'}\, (\TeV)$ &\multicolumn{2}{|c|}{$33$}\\
            \hline  \multicolumn{3}{|c|}{\textbf{Scalar Sector}}\\
            \hline \hfil $m_{H}\, (\GeV)$ &\multicolumn{2}{|c|}{$15 \cross 10^3$}\\
            \hline \hfil $m_{A^0}\, (\GeV)$ &\multicolumn{2}{|c|}{$15 \cross 10^3$}\\
            \hline \hfil $m_{\delta_L^{+}}\, (\GeV)$ &\multicolumn{2}{|c|}{$780$}\\
            \hline \hfil $m_{H^{+}}\, (\GeV)$ &\multicolumn{2}{|c|}{$15 \cross 10^3$}\\
            \hline \hfil $m_{\delta^{++}_{L}}\, (\GeV)$ &\multicolumn{2}{|c|}{$780$}\\
            \hline \hfil $m_{\delta^{++}_{R}}\, (\GeV)$ & \multicolumn{2}{|c|}{$660$}\\
            \hline 
        \end{tabular}
    \end{center}
\caption{Standard values of the parameters used 
 in this work.  In the light neutrino sector 
 we show the best fit values for NO ($l = 1$) and for IO ($l = 2$)
 according to \cite{Esteban:2020cvm}. We will refer to the set of parameters above, assuming NO, as the \emph{standard case}. }
\label{tab:params}
\end{table}

We have also assumed ${\cal C}$-symmetry, $U_L=\mathbf{I}$, $U_{\rm PMNS}$ with all oscillation parameters 
fixed at their best fit values corresponding to NO 
according to ref.~\cite{Esteban:2020cvm}, with the lightest light neutrino mass $m_0=0.01$ eV.
The only source of CP violation is from the Dirac CP phase $\delta$ which we have fixed to the best fit value~\cite{Esteban:2020cvm} while setting the Majorana phases $\alpha_1 = \alpha_2 = 0$. Additionally, we work in the basis where $h_M = \tilde{h}_M^*$ are diagonal and real with no hierarchy between the heavy neutrinos i.e., $h_1/h_3=h_2/h_3=1$ and  $\phi_1=\phi_2=\phi_3=0$.
This choice of values for these parameters in NO, summarized in table~\ref{tab:params}, will be referred to as the {\em standard case}.

In what follows, in order to 
understand the overall behavior of LFV 
observables in the MLRSM, we will discuss the four 
different choices of parameters listed in table \ref{tab:benchmarkcases}. They differ 
with respect to lepton mixing matrix ($U_L=\mathbf{I}$ 
for  (a) and $U_L=U_{\rm PMNS}$ for the other cases), the heavy neutrino mass ordering (degenerate in mass for  (a) and (c) and slightly hierarchical for (b) and (d)) and the value of the left-right mixing parameter ($\xi$ in (d) is two orders of magnitude larger than in the other scenarios).

Although we have presented some approximate expressions 
in appendix~\ref{app:formfac} to help the reader understand certain behaviors, all our calculations 
were implemented  in \textsc{Mathematica}~\cite{Mathematica}
and the relevant form factors were computed without any
approximations with the help of the packages \textsc{FeynCalc}~\cite{Shtabovenko_2016, Shtabovenko_2020, MERTIG1991345} and 
\textsc{Package-X}~\cite{PATEL2015276}.

\begin{table}[!htb]
    \begin{center}
      \begin{tabular}{|p{1cm}|p{12cm}|}
            \hline Case & \hfil Benchmark Values\\
            \hline \hfil (a) &\hfil \textit{standard case}\\
            \hline \hfil (b) &\hfil $U_L = U_{\rm{PMNS}}$, $h_1/h_3 = 0.8, h_2/h_3 = 0.9 $\\
            \hline \hfil (c) &\hfil $U_L = U_{\rm{PMNS}}$\\
            \hline \hfil (d) &\hfil $U_L = U_{\rm{PMNS}}$, $h_1/h_3 = 0.8, h_2/h_3 = 0.9 $, $\xi = 3.0 \cross 10^{-6}$\\
            \hline 
        \end{tabular}
    \end{center}
\caption{Benchmark values for the cases (a)--(d) 
used in this work. All parameters not listed in each case, were fixed to their corresponding value given 
in table~\ref{tab:params}.}
\label{tab:benchmarkcases}
\end{table}

\subsection{Neutrino sector}
The neutrino sector can be divided into two parts: the light neutrino sector 
and the heavy neutrino sector.

In the light neutrino sector we have as parameters
three neutrinos masses $m_0 \equiv m_1<m_2<m_3$ ($m_0\equiv m_3<m_1<m_2$) in the NO (IO) and the $U_{\rm PMNS}$ matrix. Except for the value of 
$m_0$, the Majorana phases $\alpha_1$ and $\alpha_2$, which are currently unknown, the remaining of this sector (including the Dirac phase $\delta$ which has not yet been measured directly) is determined by data~\cite{Esteban:2020cvm}.
We expect, however, that the reactor neutrino experiment 
JUNO~\cite{JUNO:2022hxd}, besides determining $\Delta m^2_{21}$,
$\sin^2\theta_{12}$ and $\Delta m^2_{\rm ee}$~\cite{Nunokawa:2005nx}
to better than 0.5\%~\cite{JUNO:2022mxj}, will unravel the mass ordering, 
either alone or combined with other experiments by  2030~\cite{Forero:2021lax}. The forthcoming neutrino oscillation experiments DUNE~\cite{DUNE:2020jqi} and Hyper-Kamiokande~\cite{Kudenko:2020snj} are also expected to determine the Dirac phase $\delta$ in a few years after starting taking data.
In this sector, oscillation experiments~\cite{Antusch:2006vwa,Fernandez-Martinez:2007iaa,Goswami:2008mi,Antusch:2009pm,Escrihuela:2015wra,Parke:2015goa,Dutta:2016vcc,Fong:2016yyh,Ge:2016xya,Blennow:2016jkn,Fong:2017gke,Martinez-Soler:2018lcy,Martinez-Soler:2019noy,Ellis:2020hus} can capture the \emph{nonunitarity} of the $U_{\rm PMNS}$ by $\delta U_L = \frac{1}{2}\mathcal{U}_{LR} \mathcal{U}_{LR}^\dagger$, this is in general more challenging to be constrained by present and even future facilities.

In the heavy neutrino sector we have also three neutrinos $N_i\, (i=4,5,6)$ and their production depend on their masses $m_i\, (i=4,5,6)$, on the active-sterile mixing matrix $\mathcal{U}_{LR}$ and in the feasibility of producing the RH sector of the theory that couples directly to the heavy neutrino states. The former channel is suppressed due to the heaviness of the new states and the smallness of gauge boson mixing. 
Here there are experimental limits from $\beta$-decay, $\pi\to e(\mu)\nu$, $K \to e(\mu)\nu$, lepton universality tests and the $Z^0$ invisible width on the mixing 
\be
\vert U_{\alpha i}\vert^2 
\equiv \vert ({\cal U}_{L})_{\alpha i}\vert^2
=\vert ({\cal U}_{LR})_{\alpha (i-3)}\vert^2 ,
\ee
with $\alpha=e,\mu,\tau$ and $i=4,5,6$,
as a function of $m_i$, in the range $100\, {\rm eV} \leq m_i \leq 1 \, {\rm TeV} $~\cite{deGouvea:2015euy}.
The most stringent limits come from searches for decays of heavy neutral leptons that would be produced in beam dump experiments~\cite{Bernardi:1985ny,Bernardi:1987ek} and 
colliders~\cite{DELPHI:1996qcc,Belle:2013ytx,LHCb:2014osd,ATLAS:2018dcj,CMS:2018iaf,CMS:2018jxx,ATLAS:2019kpx, CMS:2022fut}. Recently the ArgoNeuT experiment was able to establish the most severe bound to date on $\vert {U}_{\tau i}\vert^2$ for $m_i$ between 280 MeV and 970 MeV looking for $N \to \nu \mu^+ \mu^-$~\cite{ArgoNeuT:2021clc}.
Future facilities such as FASER \cite{Ariga_2019} and MATHUSLA \cite{Curtin:2018mvb} will be competitive with planned accelerator experiments such as DUNE~\cite{Ballett:2019bgd}. In addition, if the promised conditions are met the next phase of the LHC will probe very small mixing in the higher mass range providing a complementary study to other searches \cite{Drewes:2019fou}. For a recent review on the physics of heavy neutral leptons see \cite{Abdullahi:2022jlv}.

Furthermore, cosmological data implications have to be considered for the light and heavy neutrino sectors.
In the light sector, cosmology current constrains $\sum_{i=1}^{3} m_i <0.12$~eV~\cite{Planck:2018vyg}, but the next generation of cosmological probes will probably be able to measure the sum of neutrino masses and even corroborate to the mass ordering, helping to determine $m_0$~\cite{CMB-S4:2016ple}. There are also limits on the heavy sector.
In the early Universe, since $N_{i}$ experience $SU(2)_{R}$ gauge
interaction, they could be in thermal equilibrium unless the reheating temperature is much smaller than the $W'$ mass in which case their abundance would be suppressed. Nonetheless, through $N_{i}$ mixing with the
active neutrinos $\mathcal{U}_{LR}$, they can still be produced efficiently through $SU(2)_{L}$
gauge interactions. If $N_{i}$ has a lifetime greater than $(10^{-2}-10)$~s, their decays to the SM particles will affect the Big Bang Nucleosynthesis
(BBN) by modifying the $p\leftrightarrow n$ conversion processes \cite{Boyarsky:2020dzc,Sabti:2020yrt}. The constraint from the effective
number of relativistic degrees of freedom $N_{{\rm eff}}$ from the
Cosmic Microwave Background (CMB) is very similar to the aforementioned
BBN constraints but interestingly, it can lead to an increase or decrease
in $N_{{\rm eff}}$ \cite{Boyarsky:2021yoh}. For longer lifetime $\tau\gtrsim10$~s, the modification to the cosmic expansion rate can affect the Supernovae type 
Ia luminosity distance, CMB shift parameter and the Baryon Acoustic
Oscillation (BAO) scale \cite{Vincent:2014rja}. 
The relevant bounds from refs.~\cite{Sabti:2020yrt,Vincent:2014rja} will be taken into account in our analysis.
It is important to note that
these limits have not taken into account the contribution from
the heavy $W'$, which could be relevant if its mass is not far above a few TeV. Future CMB measurements may impove the 
uncertainty on $N_{\rm eff}$ by a factor 10~\cite{CMB-S4:2016ple}, potentially significantly tightening  some of these bounds.

For the quasi-degenerate spectrum of $N_i$, ref.~\cite{Drewes:2021nqr} shows that leptogenesis is viable for $m_i \gtrsim 50\, \textrm{MeV}\, (i=4,5,6)$ for vanishing initial $N_i$ abundances and $m_i \gtrsim 2\, \textrm{GeV}\, (i=4,5,6)$ for thermal initial $N_i$ abundances, in essentially all the parameter space not excluded by the constraints considered in our work. We should caution that in this calculation, only type-I seesaw model is considered and new interactions pertaining to the MLRSM ($SU(2)_R$ gauge interactions and interactions with the new scalars) could potentially modify the viable parameter space of leptogenesis. We will leave this to future exploration.

\subsection{Lepton number violation}
\label{subsec:LNV}
The MLRSM naturally has various contributions to $0\nu \beta\beta$
decay of atomic nuclei $(Z,A) \to (Z+2,A)+ e^- + e^-$, which is a LNV process. New contributions proportional to heavy neutrino masses will involve $\mathcal{U}_{LR}$ and $\mathcal{U}_{RR}$. We can see from eq.~\eqref{eq:MDMR} [see also eqs.~\eqref{eq:Z_C_scenario} and \eqref{eq:Z_P_scenario}] that $\mathcal{U}_{LR}$ for the type-I seesaw dominant regime $(r \gg 1)$, $|\mathcal{U}_{LR}|^2 \simeq |m_\nu M_R^{-1}|$ is in general suppressed while in the tuned regime $(r \sim 1)$, $|\mathcal{U}_{LR}|^2 \simeq v_L/v_R$ can be large. There are also new contributions which do not depend on $\mathcal{U}_{LR}$ but depends on the mixing parameter
$\xi$ for those involving the heavy $SU(2)_R$ $W'$ boson, $h_M$ for those involving the left-handed double-charged scalar $\delta_L^{++}$ and $\tilde h_M$ for those involving the right-handed double-charged scalar from $\delta_R^{++}$.

The expression for the inverse half-life can be written as
\be 
\frac{1}{T^{0\nu\beta\beta}_{1/2}} = G_{0\nu}\frac{\vert M^{0\nu}\vert^2}{m^2_e}\, \vert m_{\beta \beta}\vert^2\, ,
\label{eq:halflife}
\ee
where $G_{0\nu}$ is the phase space factor\footnote{We use the results of the improved calculation given in \cite{Kotila:2012zza} only corrected  to the nuclear radius of ref. \cite{Pantis:1996py}.}  for the emitted electrons which depends on the isotope $Z$, $m_e$ is the electron mass,  $M^{0\nu}$ is the corresponding nuclear matrix element (NME) associated with a light neutrino exchange, which  can be divided in three parts the Fermi (F), Gamow-Teller (GT) and Tensor (T) part as
\be{\label{eq:mnu3}}
M^{0\nu}=-\frac{M^{0\nu}_{\rm F}}{g_A^2}+ M^{0\nu}_{\rm GT}+M^{0\nu}_{\rm T}\, ,
\ee
and $\vert m_{\beta \beta}\vert^2$ is, neglecting the contribution of the $C_1$ coefficient which is minute according to ref.~\cite{Pantis:1996py}, given by~\cite{Vergados:1985pq}
\begin{eqnarray}
\vert m_{\beta \beta}\vert^2 &\simeq& \vert m_{LL} \vert^2
+\vert m_{RR}\vert^2 + C_4 \vert m_{RL}^\lambda\vert^2 + C_5  \vert m_{RL}^\xi\vert^2 \nonumber \\ 
&+& 
C_2 \vert m_{RL}^\lambda\vert \vert m_{LL}\vert \cos \psi_1+ C_3 \vert m_{RL}^\xi\vert \vert m_{LL}\vert \cos \psi_2 + 
C_6 \vert m_{RL}^\lambda\vert \vert m_{RL}^\xi\vert \cos(\psi_1-\psi_2) 
\nonumber \\
&+& {\rm Re}[(C_2 \, m_{RL}^\lambda+C_3 \, m_{RL}^\xi) \, m_{RR}] \; ,
\label{eq:0nubb}
\end{eqnarray}
where
\begin{eqnarray}
m_{LL}&=&\eta^\nu_{LL}+ \eta^N_{LL}+\eta^{\delta_L}_{LL}\; ,\\
m_{RR}&=&\eta^N_{RR}+\eta^{\delta_R}_{RR}\; ,
\end{eqnarray}
are the light neutrino ($\nu$), heavy neutrino ($N$) and double-charged scalar ($\delta_{L,R}$) contributions with the same chirality electrons 
in the final state, mediated by either two $W_L$ (LL)  or two $W_R$ (RR), 
while 
\begin{eqnarray}
m_{RL}^{\lambda}&=& \lambda \, \eta_{RL}\; ,\\
m_{RL}^{\xi}&=&\xi \,\eta_{RL}\; ,
\end{eqnarray}
are the contributions to opposite chirality  electrons  involving one $W_L$ and one $W_R$, where $\lambda \equiv (m_W/m_W^\prime)^2$ and $\xi$ is mixing parameter previously defined (see Table \ref{tab:params}). In eq.~(\ref{eq:0nubb}) we also have interference terms between these contributions. The angles $\psi_1$ and $\psi_2$ 
are, respectively, the relative phases between $m_{RL}^\lambda$ and $m_{LL}$ and between $m_{RL}^\xi$ and $m_{LL}$. The explicit definition 
of all the $\eta$ contributions are listed bellow:

\begin{itemize}
    \item with two $W$ exchange
    \begin{eqnarray}
    \eta^\nu_{LL} &= &\sum_{i=1}^3\, ({\cal U}_L)_{ei}^2 \; m_i \, , \\
    \eta^N_{LL} & =& - \sum_{i=4}^6\, ({\cal U}_L)_{ei}^{2} \; \frac{\langle p\rangle^2}{m_i} \approx \sum_{i=4}^6\, ({\cal U}_L)_{ei}^{2} \frac{m_i}{\langle p\rangle^2-m_i^2}\, \langle p\rangle^2\, ,\\
     \eta_{LL}^{\delta_L}& =&-\langle p\rangle^2\, \frac{(h_M)_{ee}}{m_{\delta_{L}}^2} \, v_L\, ,
         \label{eq:mbb1}
    \end{eqnarray}
  \item involving two $W'$ exchange
 \begin{eqnarray}  
\eta^N_{RR} & =& -(\lambda^2 + 2 \xi \lambda + \xi^2 ) \sum_{i=4}^6\, ({\cal U}^*_R)_{ei}^{2} \; \frac{\langle p\rangle^2}{m_i} \nonumber \\
&\approx& (\lambda^2 + 2 \xi \lambda + \xi^2 )\sum_{i=4}^6\, ({\cal U}^*_R)_{ei}^{2} \frac{m_i}{\langle p\rangle^2-m_i^2}\;\langle p\rangle^2\, ,\\
     \eta^{\delta_R}_{RR} & =& -\langle p\rangle^2 \frac{(\tilde{h}_M)_{ee}}{m_{\delta_{R}}^2}\, v_R \, \lambda^2 \, ,
     \label{eq:mbb2}
    \end{eqnarray}
      \item involving one $W$ and one $W'$
   \begin{eqnarray}
   \eta_{RL} &=& \sum_{i=1}^3({\cal U}^*_R)_{ei}({\cal U}_L)_{ei}- \sum_{i=4}^6({\cal U}^*_R)_{ei}({\cal U}_L)_{ei} \frac{\langle p\rangle^2}{m_i^2} \nonumber \\
   &\approx& \sum_{i=1}^3({\cal U}^*_R)_{ei}({\cal U}_L)_{ei}+ \sum_{i=4}^6({\cal U}^*_R)_{ei}({\cal U}_L)_{ei} \frac{ \langle p\rangle^2}{\langle p\rangle^2-m_i^2}\, ,
    \label{eq:mbb3}
   \end{eqnarray}
\end{itemize}
using the very good approximation recommended in ref.~\cite{Mitra:2011qr}  with 
the definition $\langle p \rangle^2 \equiv  - m_e m_p \, M^{0\nu}_N/M^{0\nu}$, $m_p$ 
is the proton mass, $M^{0\nu}_N$ and $M^{0\nu}$ are, 
respectively, the nuclear matrix elements associated with a heavy and a light neutrino exchange that will depend on the particular nuclear model used.

The coefficients $C_i$ that enter in the interference terms are constants which depend on nuclear models and phase space factors, 
they are defined as :
\begin{eqnarray}
C_i& = & \tilde{C_i} \frac{\vert M^{0\nu}_{\rm GT} \vert}{M^{0\nu}}\,  m_e \quad (i=2,3) \, \\
C_i &=& \tilde{C_i} \frac{\vert M^{0\nu}_{\rm GT}\vert^2} {\vert M^{0\nu}\vert^2} \, m_e^2\quad (i=4,5,6)\, .
\end{eqnarray}
We use the values  for $^{136}$Xe$\to ^{136}$Ba transition 
calculated in the the context of the quasiparticle random phase approximation (QRPA) formalism with n-p paring given in 
ref.~\cite{Pantis:1996py} that can be found in table~\ref{tab:nme}.

\begin{table}[htb]
\centering
\begin{tabular}{|c|c|c|c|c|c|c|c|c|}
\hline
$\tilde{C_2}$ &  $\tilde{C_3}$ & $\tilde{C_4}$ & $\tilde{C_5}$ & $\tilde{C_6}$  & $M^{0\nu}_{\rm GT}$ & $M^{0\nu}$& $M^{0\nu}_N$\\
\hline
  -0.66 & -264. & 2.11 & $9.71  \times 10^4$ & -4.53 &1.346 &1.257& 47.6\\
     \hline
\end{tabular}
\caption{\label{tab:nme} $\tilde{C_i}$ coefficients and NME $0\nu \beta\beta$ for $^{136}$Xe$\to ^{136}$Ba  transition in the QRPA framework taken from ref.~\cite{Pantis:1996py}.}
\end{table}

The KamLAND-Zen experiment provides the most stringent bound on 
the $0\nu \beta\beta$-decay half-life $T^{0\nu \beta \beta}_{1/2} > 2.3 \times 10^{26}$ yr at 90\% CL~\cite{KamLAND-Zen:2022tow}. This 
corresponds to $m_{\beta \beta} < 126$ meV for the NME 
we are using here.

\begin{figure}[hbt]
    \centering
    \includegraphics[width=0.99\textwidth]{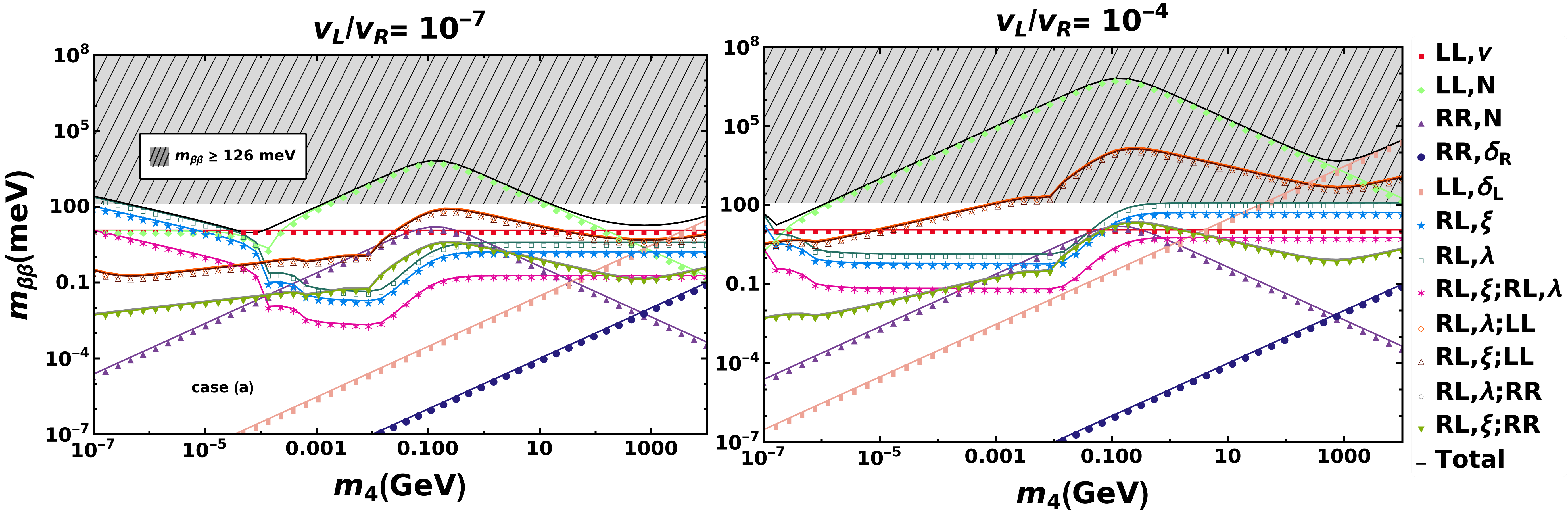}
    \caption{Contributions to $m_{\beta\beta}$ in the MLRSM within the \emph{standard case} or case (a) for $v_L/v_R= 10^{-7}$ (left panel) and $v_L/v_R=10^{-4}$ (right panel) as a function of $m_4$. The gray region is already excluded 
    by the current limit on $T^{0\nu \beta \beta}_{1/2} > 2.3 \times 10^{26}$ yr at 90\% CL~\cite{KamLAND-Zen:2022tow}.
    }
    \label{fig:NeutrinolessIndividual}
\end{figure}

In figure~\ref{fig:NeutrinolessIndividual} we show the 
behavior of the various contributions to $m_{\beta \beta}$
in the MLRSM for an intermediate value of $v_L$ ($v_L \sim 4$ MeV, left panel)
and a high value of  $v_L$ ($v_L \sim 4$ GeV, right panel) as a function of $m_4$ in the \emph{standard case} or case (a) and the KamLAND-Zen excluded region, for reference.

Let's discuss each contribution:
\begin{description}
\item (i) \:\: $\eta_{LL}^\nu$ ($LL,\nu$ - red line with filled squares) - this is the light neutrinos contribution which is independent of $v_L$ and $m_4$ and amounts to $\approx$ 10 meV = $m_1$ for the NO hypothesis;

\item (ii) $\eta_{LL}^N$ ($LL,N$ - light green line with filled diamonds) - this is the heavy neutrinos contribution for the $LL$ chirality. For $m_4 \ll \sqrt{-\langle p\rangle^2}$ and $r \ll 1$ (type-I dominance) 
this contribution is like the one for the light neutrinos, but when 
we pass to the region where $r \approx  1$ (tuned region), as the mixing becomes fixed, the contribution 
grows with $m_4$. On the other extreme, as $m_4 \gg \sqrt{-\langle p\rangle^2}$ (and $r \approx  1$) it decreases as $m_4^{-1}$;

\item (iii) $\eta_{RR}^{N}$ ($RR,N$ - violet line with filled triangles) - this is the heavy neutrinos contribution for the $RR$ chirality. For 
$m_4 \ll \sqrt{-\langle p\rangle^2}$ it 
grows with $m_4$ and as $m_4 \gg \sqrt{-\langle p\rangle^2}$ it  decreases as $m_4^{-1}$;

\item (iv) $\eta_{RR}^{\delta_R}$ ($RR,\delta_R$ - navy blue line with filled circles) - this is the $\delta_R^{++}$ contribution, which grows with $m_4$ but is suppressed by $(\lambda/m_{\delta_R})^2$; 

\item (v) $\eta_{LL}^{\delta_L}$ ($LL,\delta_L$ - light pink line with filled rectangles) - this is the $\delta_L^{++}$ contribution, which grows with $m_4$ and $v_L$; 

\item (vi) $m_{RL}^\xi$ ($RL,\xi$ - light blue line with five pointed stars) - this 
is the $RL$ chirality contribution which is suppressed by the mixing $\xi$.
For  $m_4 \ll \sqrt{-\langle p\rangle^2}$ and $r \ll 1$, it falls as $m_4^{-1/2}$. Here the behavior depends on $v_L$. 
In both cases shown in Fig.~\ref{fig:NeutrinolessIndividual}
one enters the tuned regime while $m_4 \ll \sqrt{-\langle p\rangle^2}$ but at different values of $m_4$, i.e. 
at $m_4 \sim 10^{-4} $ GeV ($m_4 \sim 10^{-7}$ GeV) for 
$v_L/v_R= 10^{-7}$ ($v_L/v_R= 10^{-4}$). After that the mixing becomes constant for a while. As one approaches $m_4 \sim \sqrt{-\langle p\rangle^2}$ this contribution grows again with $m_4$, becoming independent of $m_4$ when $m_4 > \sqrt{-\langle p\rangle^2}$. 
Note that $m_{RL}^\lambda$ ($RL,\lambda$ - green line with unfilled squares), the  $RL$ chirality contribution suppressed by the mass ratio $\lambda$,  
and  the mixed contribution ($RL,\xi$; $RL,\lambda$ - pink line with six pointed stars) have the same  behavior just suppressed by the appropriate constant factors;

\item (vii) $m_{RL}^\lambda m_{LL}$ ($RL, \lambda; LL$ - orange line with unfilled diamonds) 
and $m_{RL}^\xi m_{LL}$ ($RL, \xi; LL$ - brown line with unfilled triangles) - these are interference terms between $RL$ and $LL$ chiralities. There is some 
small dependence with $v_L$ and either $\eta_{RL}$ or $m_{LL}$ drive the behavior, depending on $m_4$.

\item (viii) $m_{RL}^\lambda m_{RR}$ ($RL, \lambda; RR$ - gray line with unfilled circles) 
and $m_{RL}^\xi m_{RR}$ ($RL, \xi; RR$ - olive green line with filled inverted triangles) - 
these are interference terms between $RL$ and $LL$ chiralities. Their 
behavior depends very little on $v_L$. 
\end{description}

 \begin{figure}[hbt]
    \centering
    \includegraphics[width=0.60\textwidth]{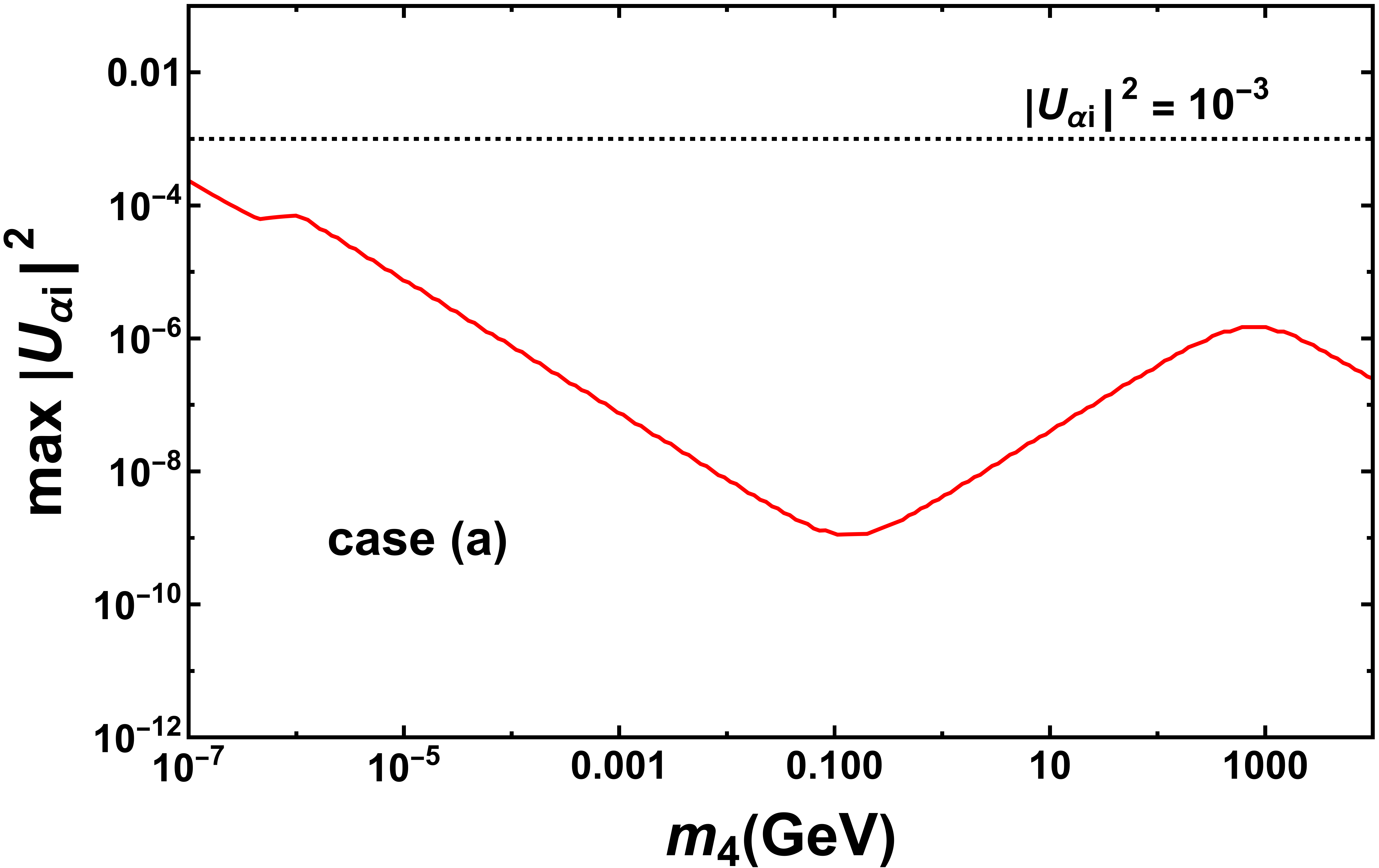}
    \caption{Maximum value of $\vert U_{\alpha i}\vert^2$ as a function of $m_4$ that can be reached respecting the current $0\nu\beta\beta$ decay limit for case (a). The dashed line 
    represents the minimum mixing value needed  
    to affect the determination of $m_W$ (see section~\ref{subsec:mW}). The behavior of this curve is explained in the text.}
    \label{fig:Nu0bb-Ue4}
\end{figure}

 In the region where type-I seesaw dominates, the main contribution to $m_{\beta \beta}$ comes from $m_{RL}^{\lambda}$. When we enter the tuned region, $m_{LL}^N$ will be the dominant contribution until  $\eta^\nu_{LL}$ or $\eta_{LL}^{\delta_L}$ can  pick up.

In figure~\ref{fig:Nu0bb-Ue4} we  show for  case (a)  the maximum value of $\vert U_{\alpha i}\vert^2$ as a function of  $m_4$ that can be reached respecting the current $0\nu \beta \beta$ decay bound.
For $m_4 \gtrsim 10^{-6}$ GeV the curve is smooth as the maximum mixing is given by the limit on $0\nu \beta \beta$ decay in the tuned region. For $m_4 \lesssim 6 \times 10^{-7}$ GeV the maximum  mixing is given 
by the limit on $0\nu \beta \beta$ decay in the type I seesaw region, where max$\vert U_{\alpha i}\vert^2$ increases with decreasing $m_4$. In the gap between this two regimes  max$\vert U_{\alpha i}\vert^2$ is simply a constraint by $v_L <$ 5 GeV which is imposed by the oblique parameters~\cite{Maiezza:2016ybz}.
We do not show cases (b), (c) or (d) here  because they exhibit similar behavior to (a).

\subsection{Lepton flavor violation}
\label{subsec:LFV}

In the MLRSM we also have different contributions to processes that violate \emph{charged} lepton flavor. As in the SM, the typical contributions to these processes from light neutrino masses are suppressed. The relevant new pieces will come from heavy neutrinos, single- and double-charged scalars which are captured by
\begin{eqnarray}
S_{LR} &\equiv& \mathcal{U}_{L} \hat M_\nu  \mathcal{U}_{R}^\dagger = U_L M_D U_R^\dagger,\\
S_{RR} &\equiv& {\cal U}_{R}^* \hat M_\nu {\cal U}_{R}^\dagger = U_R^* M_R U_R^\dagger,\\
S_{LL} &\equiv& {\cal U}_{L} \hat M_\nu {\cal U}_{L}^T = U_L M_L^\dagger U_L^T,
\end{eqnarray}
where we can see that $\phi_R$ does not explicitly appear. If the matrices above are diagonal, we will not have contributions to charged LFV. In the ${\cal C}$ symmetry case we have
\begin{eqnarray}
S_{LR} &=& U_L M_D U_L^T,\label{eq:LFVstudy1}
\\
S_{RR} &=& S_{LL} = U_L M_R U_L^T\, .
\label{eq:LFVstudy2}
\end{eqnarray}
In the tuned regime \eqref{eq:C-tuned} $M_D \propto M_R$ is diagonal (the non-diagonal part which is proportional to $m_\nu$ is subdominant), additionally,
if either $U_L = \mathbf{I}$ or $M_R \propto \mathbf{I}$ and $U_L$ is real, as in case (a), there is no contribution to LFV. 

In $\ell \to \ell' \gamma$ the pieces coming from the double-charged scalars ($\delta_L^{++}$ and $\delta_R^{++}$), and from the single-charged scalar $\delta_L^+$, have the leading terms (in the expansion of $m_{\ell(i)}/m_S \ll 1$, where $m_{\ell(i)}$ is the charged lepton (neutrino) mass and $m_S$ is the corresponding charged scalar mass) depending on the combination $S_{RR}S_{RR}^\dagger$ and will be diagonal if $|M_R| \propto \mathbf{I}$ (independently of $U_L$) and therefore, suppressed.

Note, however, that for $\ell \to 3\ell'$ and the nuclear transition, as the exchanged photon is not on-shell we have additional contributions that appear at leading order, even in  cases where $\ell \rightarrow \ell' \gamma$ is suppressed. This feature is particular noteworthy in the situation where $U_L \neq \mathbf{I}$ and $\vert M_R\vert \propto \mathbf{I}$ as in this case
the double- and single-charged  particles do not play a role in $\ell \rightarrow \ell' \gamma$ at leading order, but are not negligible to $\ell \to 3\ell'$ and $\mu \rightarrow e$ nuclei conversion. This is because the additional terms are not proportional to $(S_{RR}S_{RR}^\dagger)_{\mu e}$ but to $\sum_j (S_{RR})_{\mu j} (S_{RR}^\dagger)_{j e}\, f(m_j)$, where $f(m_j) \neq 1$ makes this combination non-diagonal. This has the important consequence of loss of correlation among the different LFV processes as will be seen in section~\ref{sec:results}.

Next we will present some details of our calculations 
for the charged LFV processes we studied. 
Because of the above features we will illustrate our discussion 
on LFV in the MLRSM with  cases (b), (c) and (d) of table \ref{tab:benchmarkcases}. 
We will comment on present constraints and future sensitivities of the relevant observables.

\subsubsection{$\ell \to \ell' \gamma$}
The partial width for these LFV radiative decays can be written as 
\be
\Gamma(\ell \to \ell' \gamma) = \frac{\alpha}{2\,m_\ell}(m_\ell-m_{\ell'})^2
(\vert F_2(0) \vert^2 +\vert G_2(0)\vert^2 )\, ,
\label{eq:partialwidth1}
\ee
where $m_\ell (m_{\ell'})$ is the initial (final) state lepton mass, 
$\alpha$ is the electromagnetic coupling constant and $F_2(0)$ and 
$G_2(0)$ are the two electromagnetic form factors that have to be calculated for the model at $q^2 \to 0$.

The MLRSM have several 1-loop contributions to these form factors. The particles that contribute to LFV are $H, A^0, \delta_L^+,H^+, \delta_{R,L}^{++}, W'$
and $W$, while $h, Z,Z'$ are 
flavor diagonal and do not play a role here. Neutral scalars $H_{L,R}$ and $A_L$ from scalar triplets do not couple to the charged leptons and hence do not contribute to  LFV processes. In appendix~\ref{app:formfac} we give  approximate 
expressions for the form factors for completion.

As already explained, because $U_L=\mathbf{I}$ and 
$M_R \propto \mathbf{I}$  in the {\em standard case} or case (a), 
only the SM-like vector boson $W$ can produce $\ell \to \ell'\gamma$ transition. This is at the most 
of ${\cal O}(10^{-23})$, way below the current experimental limit but much larger than the SM prediction, because of the presence of heavy neutrinos.

\begin{figure}[hbt]
\centering
\begin{subfigure}[b]{\textwidth}
   \includegraphics[width=1\linewidth]{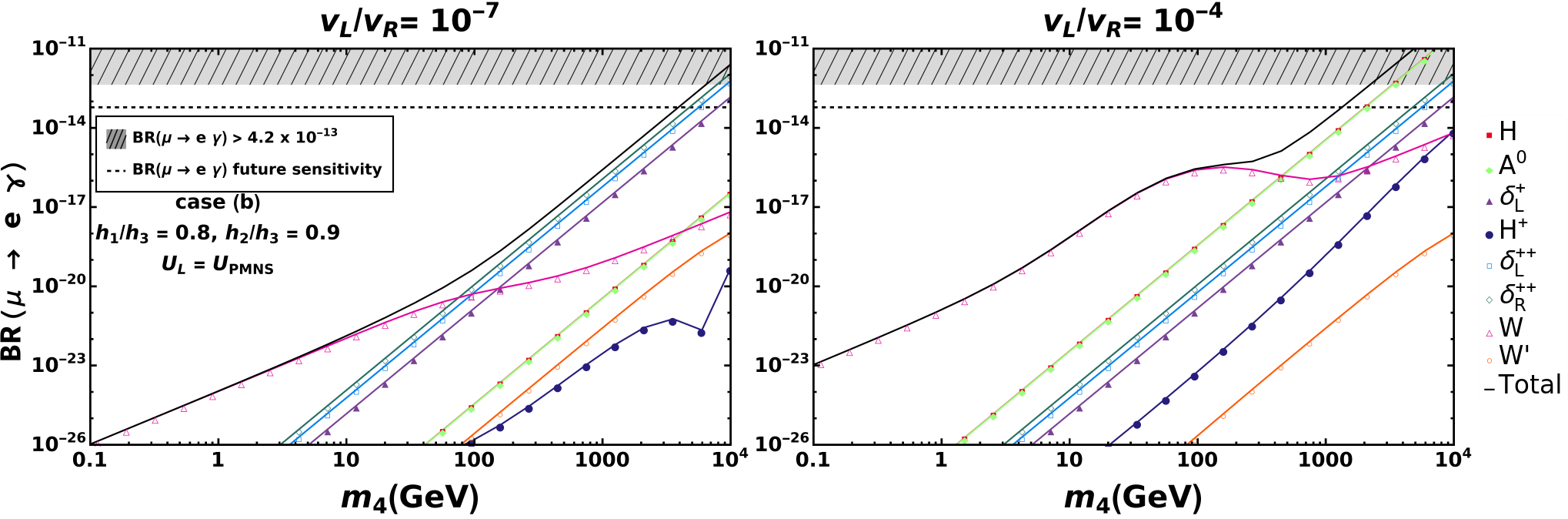}
\label{fig:casebmutoegamma} 
\end{subfigure}
\begin{subfigure}[b]{\textwidth}
   \includegraphics[width=1\linewidth]{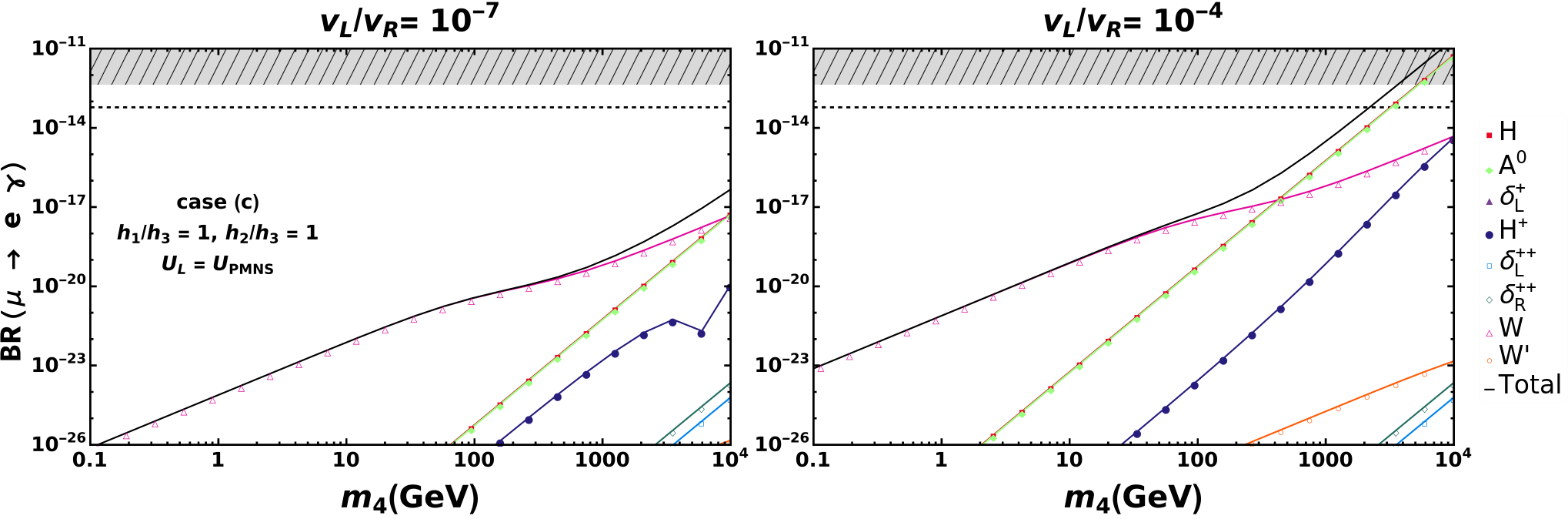}
   \label{fig:casecmutoegamma}
\end{subfigure}
\begin{subfigure}[b]{\textwidth}
   \includegraphics[width=1\linewidth]{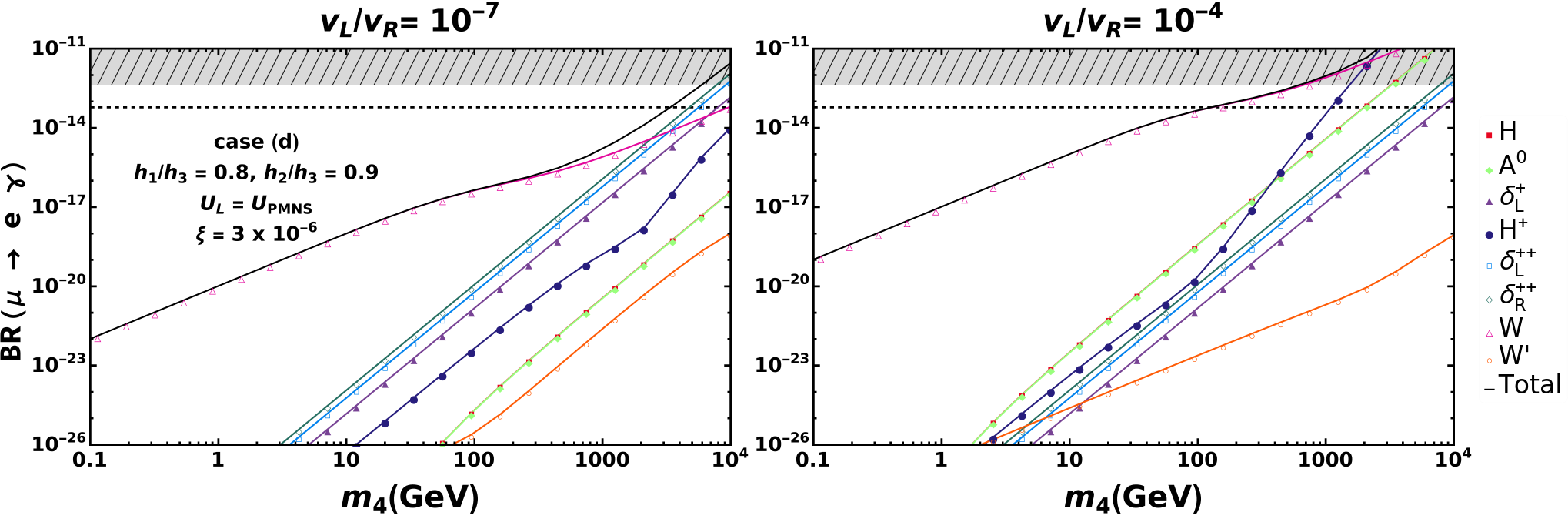}
   \label{fig:casec=dmutoegamma}
\end{subfigure}
\caption{Contributions to the BR$(\mu \to e \gamma)$ in the MLRSM for the parameters of  cases (b), (c) and (d). We show the behavior for an intermediate ratio $v_L/v_R= 10^{-7}$ (left panels) and for a larger one $v_L/v_R=10^{-4}$ (right panels) as a function of $m_4$. The gray region is already excluded at 90\% CL
by the current limit on BR($\mu \to e \gamma$) by the MEG experiment~\cite{MEG:2016leq},  
the dashed line shows the sensitivity of $6 \times 10^{-14}$, the goal of an upgraded version of MEG~\cite{Baldini:2013ke}.}
    \label{fig:BRmegamma}
\end{figure}

 To illustrate the general behavior, in figure~\ref{fig:BRmegamma} 
we show the various contributions of the model 
to BR$(\mu \to e \gamma)$ as a function of $m_4$ for the cases (b) (top panels), (c) (middle panels) 
and (d) (bottom panels) for $v_L \sim 4$ MeV (left panels) and $v_L \sim 4$ GeV (right panels).
We also show the current experimental bound 
BR$(\mu \to e \gamma)<4.3 \times 10^{-13}$~\cite{MEG:2016leq} 
and the expected future reach of the experiment $6\times 10^{-14}$~\cite{Baldini:2013ke}.

In cases (b) and (d) neutrinos are mildly 
hierarchical, in case (c) they are degenerate.
We see that as explained before in case (b) and (d) 
the charged scalars contributions $\delta_{L,R}^{++}$ and $\delta_L^+$ are not suppressed and will dominate in part of the parameter space. However, they do not 
depend on $v_L/v_R$ or $\xi$. For large enough $v_L/v_R$,  the neutral scalars (which grow with this ratio and are independent of $\xi$) can play a significant role depending on $\xi$. For a larger $\xi$
(case (d)) and large enough $v_L/v_R$
the $W$ contribution (which depends on both) will dominate, except when $H^+$ (that also depends on both) becomes significant at $m_4 \sim $ few TeV.
In case (c) since the heavy neutrinos are degenerate the charged scalar contributions $\delta_{L,R}^{++}$ and $\delta_L^+$ are always very suppressed. For large enough $v_L/v_R$ and $m_4$ the neutral scalar contributions can 
reach the experimental limit on BR$(\mu \to e \gamma)$.
We see that the current bound and the future experimental sensitivity allow to probe a narrow region of the parameter space, i.e., $m_4$ in the TeV range, unless $\xi$ is as large as in (d).

\subsubsection{Three body charged lepton decays}
\label{subsec:mu3e}
\begin{figure}
\centering
\begin{subfigure}{.5\textwidth}
  \centering
  \begin{tikzpicture}
    \begin{feynman}
      \vertex (a2) at (0, 0);
      \vertex (a1) at (-2,-.5) {$\mu^{-}$};
      \vertex (a3) at (2,-.5) {$e^{+}$};
      \vertex (b1) at (0.3, 1.5) ;
      \vertex (b2) at (2.2, 1.1) {$e^{-}$};
      \vertex (b3) at (2.2, 1.9) {$e^{-}$};
      ;
      \diagram* {
        (a1) -- [fermion] (a2) -- [fermion] (a3),
        (a1) -- [fermion] (a2) -- [fermion] (a3),
        (b2) -- [fermion] (b1) -- [fermion] (b3),
        (a2) -- [scalar,momentum'=\(q\),edge label=$\delta_{L,R}^{--}$] (b1),
      };
    \end{feynman}
  \end{tikzpicture}
  \caption{Tree-level $\mu \to 3e$ decay.}
  \label{fig:Mu3eAdiagram}
\end{subfigure}%
\begin{subfigure}{.5\textwidth}
  \centering
  \begin{tikzpicture}
    \begin{feynman}
      \vertex[blob] (a2) at (0, 0) {};
      \vertex (a1) at (-2,-.5) {$\mu^{-}$};
      \vertex (a3) at (2,-.5) {$e^{-}$};
      \vertex (b1) at (0.3, 1.5) ;
      \vertex (b2) at (2.2, 1.1) {$e^{-}$};
      \vertex (b3) at (2.2, 1.9) {$e^{+}$};
      ;
      \diagram* {
        (a1) -- [fermion] (a2) -- [fermion] (a3),
        (a1) -- [fermion] (a2) -- [fermion] (a3),
        (b2) -- [fermion] (b1) -- [fermion] (b3),
        (a2) -- [photon,momentum'=\(q\)] (b1),
      };
    \end{feynman}
  \end{tikzpicture}
  \caption{$\mu \to 3e$ through photon exchange.}
  \label{fig:Mu3eBdiagram}
\end{subfigure}
\caption{Different contributions to $\mu\rightarrow 3e$ decay.}
\label{fig:DiagramsMu3e}
\end{figure}
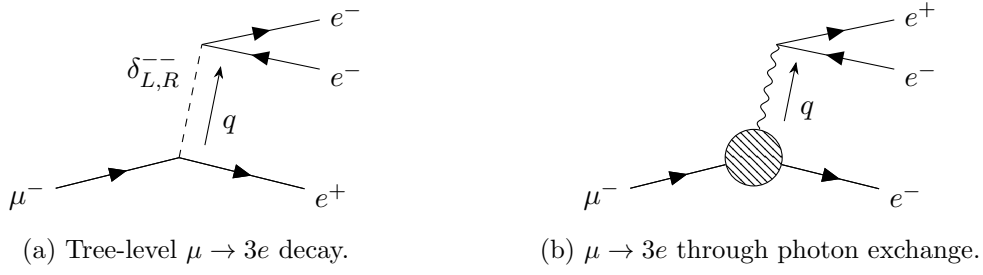

The MLRSM has three kinds of terms that can contribute 
to charged LFV three body decays such as $\mu \to 3 e$, $\tau \to 3 e$ or $\tau \to 3 \mu$: tree-level contributions, 1-loop contributions and interference
(see figure~\ref{fig:DiagramsMu3e}).

The tree-level contributions, which are not present in the {\em standard case}, take place through the exchange of the scalars $\delta_{L,R}^{++}$ and can be written as
\begin{eqnarray}
\Gamma(\ell \to 3 \ell')_{\delta_{R}^{++}}^{\rm{(tree)}} &=& \frac{m_{\ell}^5\left\vert\left(U_{R}^*\,\Tilde{h}_M\,U_{R}^\dagger\right)_{\ell\ell'}\left(U_{R}\,\Tilde{h}_M^*\,U_{R}^T\right)_{\ell'\ell'}\right\vert^2}{3 \,(32)^2\,\pi^3\, m_{\delta_{R}^{++}}^4} \nonumber\\
&=& \frac{m_{\ell}^5\left\vert\left(U_{R}^*\,M_R\,U_{R}^\dagger\right)_{\ell\ell'}\left(U_{R}\,M^*_R\,U_{R}^T\right)_{\ell'\ell'}\right\vert^2}{3 \,(64)^2\,\pi^3\, m_{\delta_{R}^{++}}^4 v_R^4},
\label{eq:partialwidth2}
\end{eqnarray}
with $\Gamma(\ell \to 3 \ell')_{\delta_{L}^{++}}^{\rm{(tree)}}$ obtained through the exchanges $U_R \rightarrow U_L$, $m_{\delta_{R}^{++}} \rightarrow m_{\delta_{L}^{++}}$ and $\Tilde{h}_M \rightarrow h_M$.
The 1-loop induced diagrams contain a vertex that has the same structure of the $l \rightarrow l' \gamma$ one, so the general decomposition given in eq.~(\ref{eq:GeneralAmplitude}) holds and we can write their part of the partial width as~\cite{Kuno:1999jp, Cornella:2019uxs}
\begin{eqnarray}
\Gamma(\ell \to 3\ell')_\gamma &=& \frac{\alpha^2 m_{\ell}^5}{48\,\pi}
\left\{3\left(\left\vert \dv{F_1(0)}{q^2}\right\vert^2+\left\vert\dv{G_1(0)}{q^2} \right\vert^2\right)\right. \nonumber\\
&-&\left.\frac{12}{m_\ell^2} \left( \text{Re} \left[ \dv{F_1(0)}{q^2}\,F_2(0)^*\right] - \text{Re} \left[ \dv{G_1(0)}{q^2}\,G_2(0)^*\right]\right)\right. \nonumber\\
&+&\left.\frac{8}{m_\ell^4}\left(\left\vert F_2(0) \right\vert^2+\left\vert G_2(0) \right\vert^2\right)\left(\ln{\frac{m_\ell^2}{m_{\ell'}^2}} - \frac{22}{8}\right)\right\}\, ,
\label{eq:partialwidth2_2}
\end{eqnarray}
where $F_{1,2}(0)$ and 
$G_{1,2}(0)$ are four electromagnetic form factors (see appendix \ref{app:formfac}) and we assumed $m_\ell \gg m_{\ell'}$. 
Finally the term $\Gamma(\ell \to 3\ell')_{\rm{interf.}}$ is the interference term between tree-level and loop level contributions. The full decay width is
\begin{equation*}
\Gamma(\ell \to 3\ell') = \Gamma(\ell \to 3\ell')_\gamma + \Gamma(\ell \to 3\ell')_{\delta_{R}^{++}}^{\rm{(tree)}} + \Gamma(\ell \to 3\ell')_{\delta_{L}^{++}}^{\rm{(tree)}} + \Gamma(\ell \to 3\ell')_{\rm{interf.}}\, .
\end{equation*}

\begin{figure}[hbt]
\centering
\begin{subfigure}[b]{\textwidth}
   \includegraphics[width=1\linewidth]{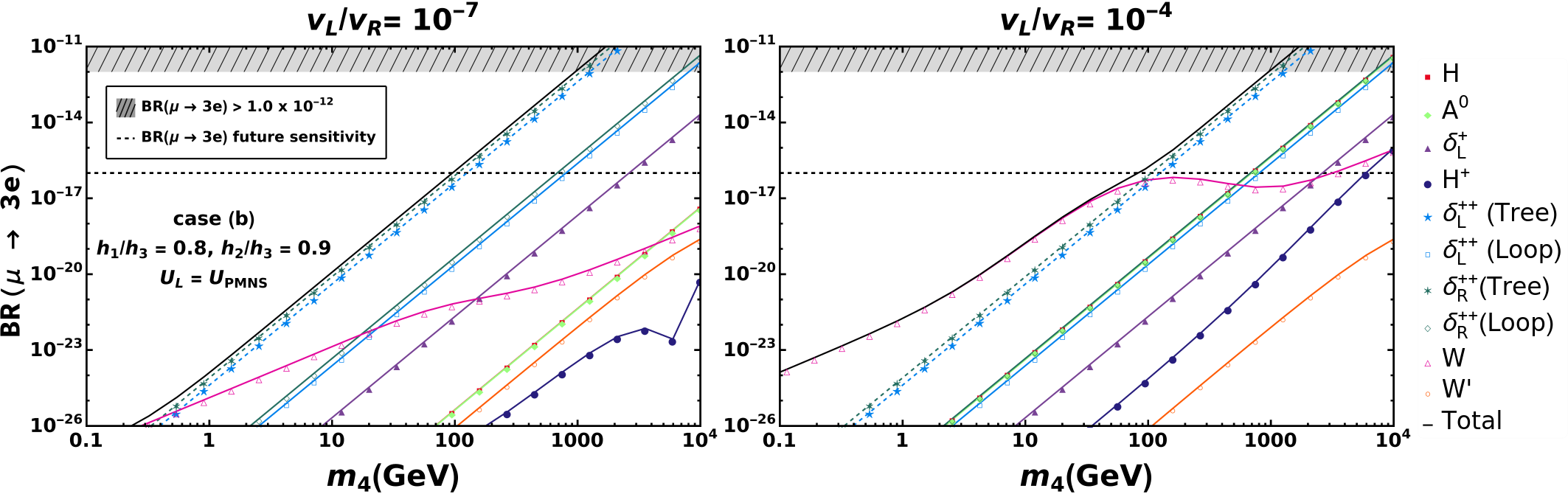}
   \label{fig:BRmuto3e_1} 
\end{subfigure}

\begin{subfigure}[b]{\textwidth}
   \includegraphics[width=1\linewidth]{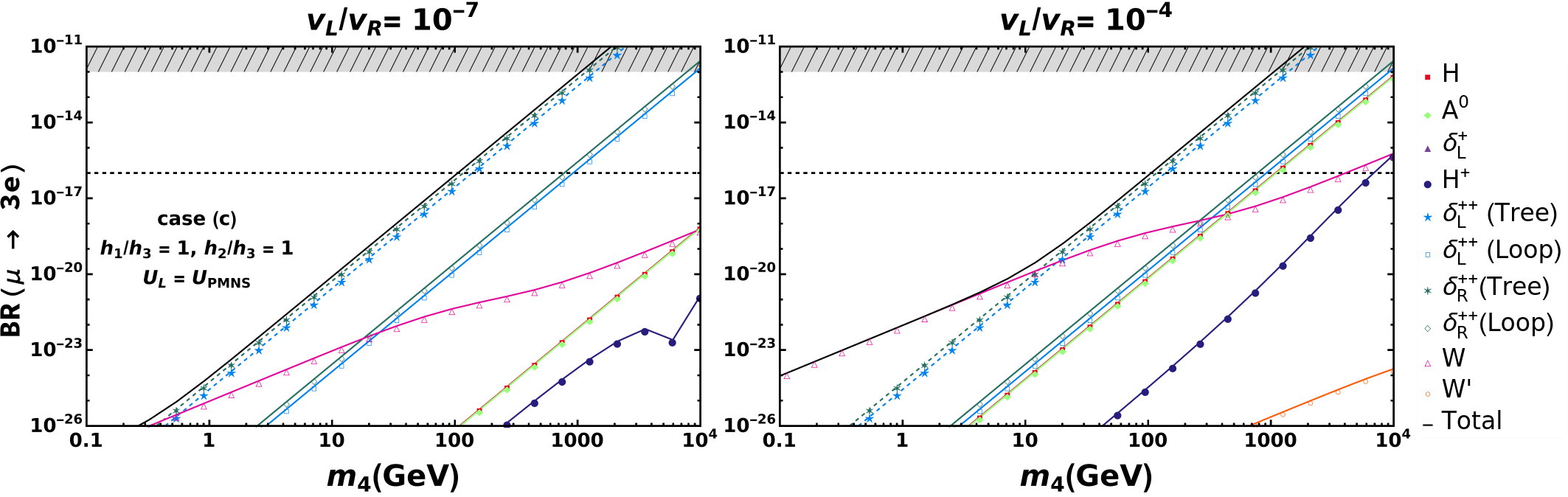}
   \label{fig:BRmuto3e_2}
\end{subfigure}

\begin{subfigure}[b]{\textwidth}
   \includegraphics[width=1\linewidth]{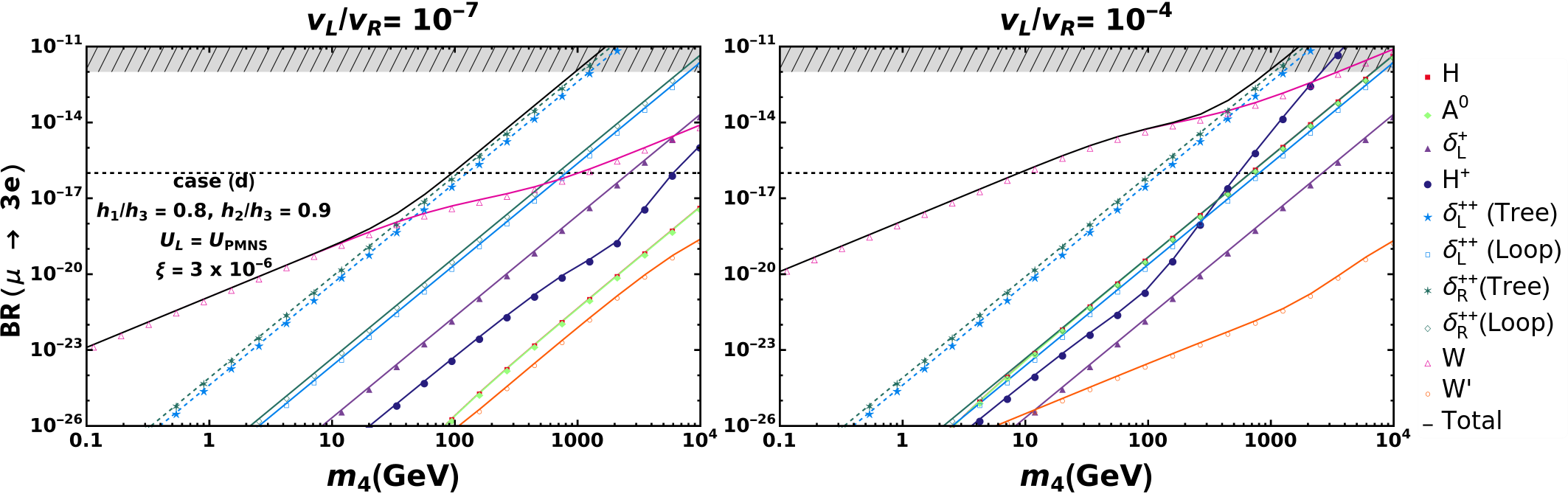}
   \label{fig:BRmuto3e_3}
\end{subfigure}

\caption{Contributions to BR$(\mu \to 3e)$ in the MLRSM for the parameters of cases (b), (c) and (d). We show the behavior for an intermediate ratio $v_L/v_R= 10^{-7}$ (left panels) and for a large one $v_L/v_R=10^{-4}$ (right panels) as a function of $m_4$. The gray region is already excluded at 90\%~CL by the current limit on BR($\mu \to 3e$) by the SINDRUM experiment~\cite{SINDRUM:1987nra}, the dashed line shows the sensitivity of the future Mu3e experiment~\cite{Mu3e:2020gyw}.}
    \label{fig:BRmu3e}
\end{figure}

In figure~{\ref{fig:BRmu3e}} we illustrate the behavior of each  MLRSM contribution to BR$(\mu \to 3 e)$  for the same benchmark cases (b) (top panels), (c) (middle panels) and (d) (bottom panels) of table \ref{tab:benchmarkcases} and the same two $v_L$ values as in figure~\ref{fig:BRmegamma}. We also show the current experimental limit on BR$(\mu \to 3 e)< 1.0 \times 10^{-12}$~\cite{SINDRUM:1987nra} as well 
as the expected future sensitivity BR$(\mu \to 3 e)= 1.0 \times 10^{-16}$~\cite{Mu3e:2020gyw}, for reference. 
We see that although the $W$ curve is similar  with respect 
to figure~\ref{fig:BRmegamma}, in contrast to BR$(\mu \to e \gamma)$, in all cases the double-charged scalar contributions $\delta_{L,R}^{++}$ always become dominant.
For larger $\xi$ (case (d)), it will surpass the $W$ 
contribution as $m_4$ increases.
This will happen, nevertheless, at lower values of $m_4$ than the corresponding cases (b) and (d) of  figure~{\ref{fig:BRmegamma}}.  Here there is no suppression of the double-charged scalar contributions when heavy neutrinos are degenerate  as compared to the case for $\mu \to e \gamma$ because of the presence of the tree level diagrams.
Again the current limit can exclude only a small part of the parameter space of the model, but this could increase substantially in the future.

\subsubsection{$\mu \to e$ conversion in nuclei}
\label{subsec:muenuc}
The MLRSM also has several 1-loop contributions to $\mu^- + A(N,Z) \to e^- + A(N,Z)$, where $A(N,Z)$ is a nucleus with $Z$ protons and $N$ 
neutrons. The experimental limits shown in table~\ref{tab:emuconv} are given on the ratio 
\be
B^{\rm A}_{\mu\to e} = \frac{\sigma(\mu^- A \to e^- A)}{\sigma(\mu^- A \to {\rm capture})}\, , \quad \quad A\rm ={Ti, Au, \; and \; Pb}\, .
\label{eq:ratio}
\ee

\begin{table}[htb]
\centering
\begin{tabular}{ |p{2cm}|p{3.8cm}|p{2.0cm}|}
\hline
\bf Process & \bf Experimental Limit & \bf Reference\\
\hline
$B^{\rm Ti}_{\mu\to e}$ & $< 4.3 \times 10^{-12}$& \cite{SINDRUMII:1993gxf}\\
$B^{\rm Au}_{\mu\to e}$ &$< 7.0 \times 10^{-13}$ & \cite{SINDRUMII:2006dvw}\\
$B^{\rm Pb}_{\mu\to e}$ &$< 4.6 \times 10^{-11}$ & \cite{SINDRUMII:1996fti}\\
\hline
\end{tabular}
\caption{\label{tab:emuconv} Limits 
from the SINDRUM-II experiment at PSI on the rate of $\mu \to e$ conversion 
on nuclei at 90\% CL.}
\end{table}

\begin{figure}[hbt]
\centering
\begin{subfigure}[b]{\textwidth}
   \includegraphics[width=1\linewidth]{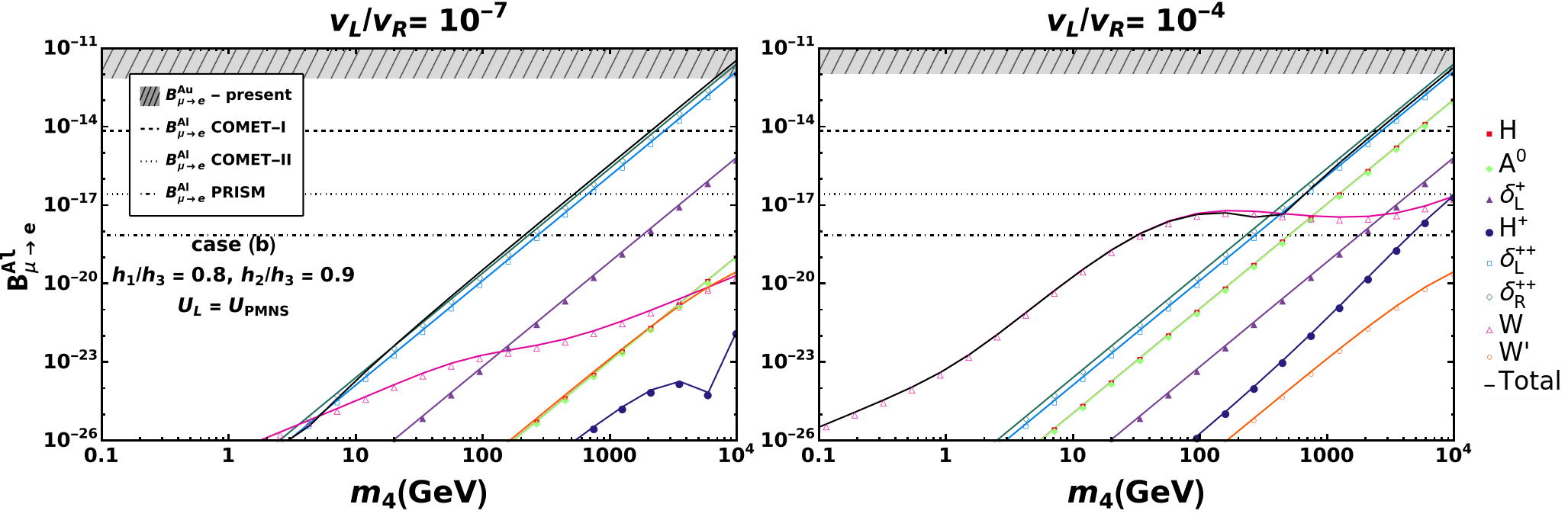}
   \label{fig:casebmuNtoeN_1} 
\end{subfigure}

\begin{subfigure}[b]{\textwidth}
   \includegraphics[width=1\linewidth]{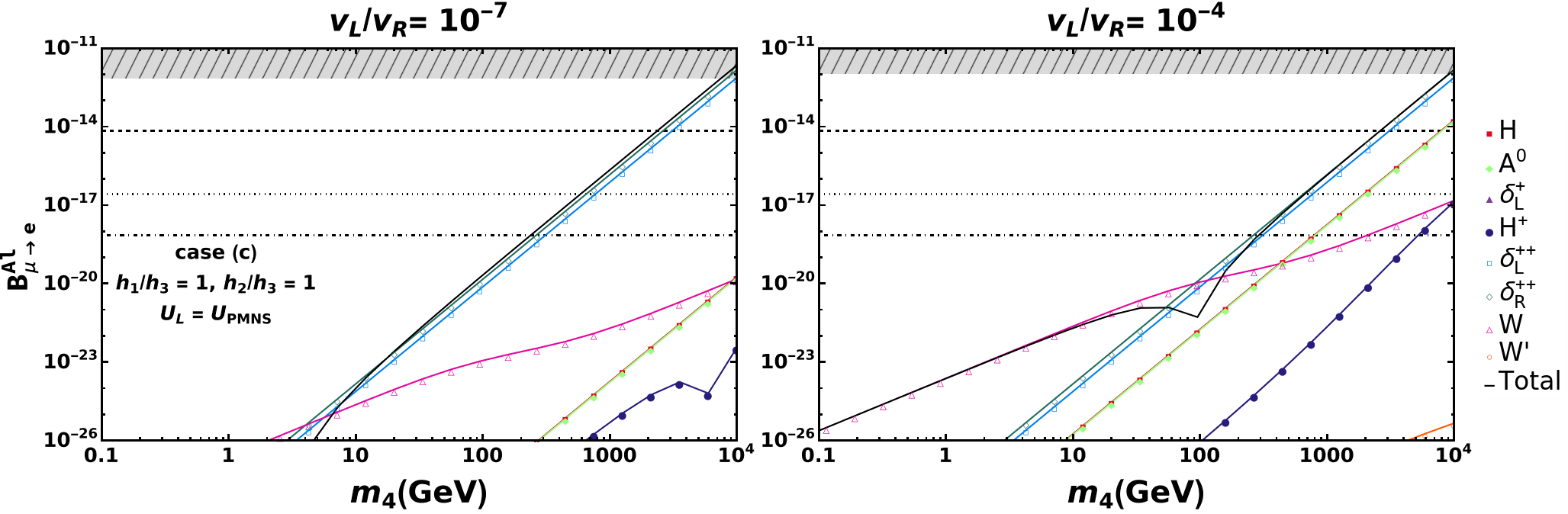}
   \label{fig:BRmuNtoeN_2}
\end{subfigure}

\begin{subfigure}[b]{\textwidth}
   \includegraphics[width=1\linewidth]{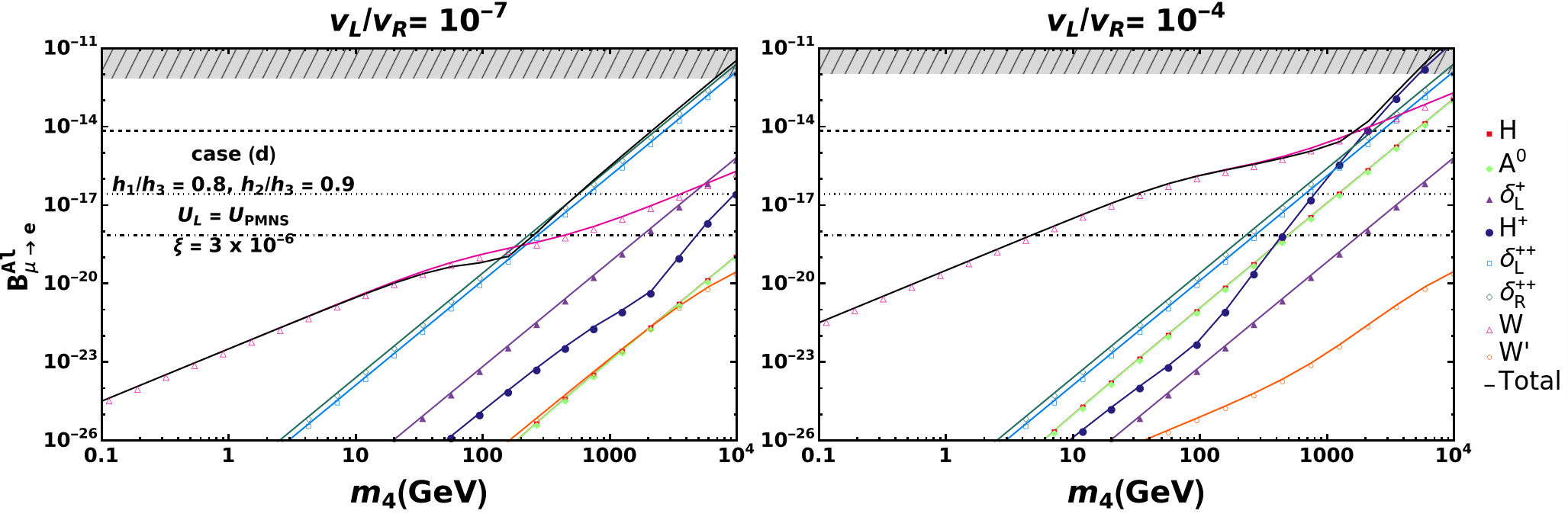}
   \label{fig:BRmuNtoeN_3}
\end{subfigure}

\caption{Contributions to $B^{\rm Al}_{\mu \to e}$ in the MLRSM for the parameters of cases (b), (c) and (d). We show the behavior for an intermediate ratio $v_L/v_R= 10^{-7}$ (left panels) and for a large one $v_L/v_R=10^{-4}$ (right panels) as a function of $m_4$. The gray region is already excluded at
90\%~CL by the SIMDRUM-II limit on $B^{\rm Au}_{\mu \to e}$~\cite{SINDRUMII:2006dvw}. We also show the future sensitivity of COMET-I (dashed line), COMET-II (dotted-line) and PRISM (dash-dotted line)~\cite{COMET:2018auw}.}
    \label{fig:BRmuNtoeN}
\end{figure}

\begin{figure}
  \centering
  \begin{tikzpicture}
    \begin{feynman}
      \vertex[blob] (a2) at (0, 0) {};
      \vertex (a1) at (-2,0) {$\mu$};
      \vertex (a3) at (2,0) {$e$};
      \vertex (b1) at (0.5, 2) ;
      \vertex (b2) at (-2, 2) ;
      \vertex (bl) at (-2.2, 2) ;
      \vertex (b3) at (2, 2) ;
      \vertex (c1) at (-2, 2.3) ;
      \vertex (c2) at (2, 2.3) ;
      \vertex (d1) at (-2, 2.6) ;
      \vertex (dl) at (-2.2, 2.6) ;
      \vertex (d2) at (2, 2.6) ;
      ;
      \diagram* {
        (a1) -- [fermion] (a2) -- [fermion] (a3),
        (b2) -- [fermion] (b1) -- [fermion] (b3),
        (c1) -- [fermion] (c2),
        (d1) -- [fermion] (d2),
        (a2) -- [photon,momentum'=\(q\)] (b1),
      };
      \draw [decoration={brace}, decorate] (bl.south west) -- (dl.north west)
          node [pos=0.5, left] {\(A(N,Z)\,\)};
     
    \end{feynman}
  \end{tikzpicture}
  \caption{$\mu \to e$ conversion in nuclei through photon exchange.}
  \label{fig:MuToeinNuclei}
\end{figure}
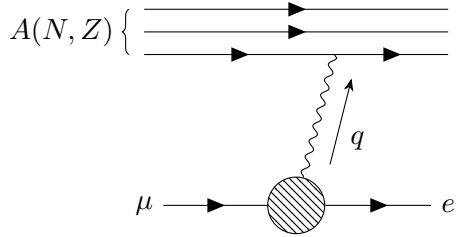

Here we are going to consider the 1-loop contributions through photon exchange with the nucleus (see figure~\ref{fig:MuToeinNuclei}). Again we have the same structure for the vertex as in 
eq.~(\ref{eq:GeneralAmplitude}), so the general contribution can be written as \cite{Kuno:1999jp}
\be
B_{\mu \rightarrow e}^\text{A} = \frac{8\,  V^{(p)\,2} \,e^4}{m_\mu^4\, \Gamma_{\text{cap.}}}
\left(\left\vert F_1 (-m_\mu^2) + F_2 (-m_\mu^2)\right\vert^2 + \left\vert G_1 (-m_\mu^2) + G_2 (-m_\mu^2)\right\vert^2\right),
\ee
where $\Gamma_{\rm cap.}$ is the capture rate of the nucleus considered, $F_1 (-m_\mu^2),\,F_2 (-m_\mu^2),$ $G_1 (-m_\mu^2)$, $G_2 (-m_\mu^2)$ are the form factors evaluated at $q^2 = -m_\mu^2$ and $V^{(p)}$ is a number that depends on the atomic number $Z$, the proton density, and the muon and electron wave functions. In the literature we found different methods of evaluating these contributions. This can be done either by using directly the form factors evaluated at $q^2 = -m_\mu^2$ or employing a first order expansion for the form factors $F_2 (-m_\mu^2)\approx F_2(0)$, $G_2 (-m_\mu^2)\approx G_2(0)$ and $F_1 (-m_\mu^2)\approx -m_\mu^2\, dF_1(0)/dq^2$, $G_1 (-m_\mu^2)\approx -m_\mu^2\, dG_1(0)/dq^2$. We have checked that both ways are consistent and the numerical difference is negligible.   To compute $B_{\mu \rightarrow e}^\text{A}$ here we used the values of $V^{(p)}$ and $\Gamma_{\rm cap.}$ given in table \ref{tab:VpandGammavalues}, taken from \cite{Kitano:2002mt}. 

There are two very aggressive experiments 
in the near future. COMET is an experiment at J-PARC in Japan which will operate with an Al target and is planed to have two phases: COMET-I which is supposed to achieve a sensitivity on $B^\text{Al}_{\mu\to e}$ of $7\times 10^{-15}$ and 
COMET-II which will bring the sensitivity further down to $2.6\times 10^{-17}$~\cite{COMET:2018auw}. A third phase, known as PRISM,  is under investigation 
aiming to a sensitivity of $7\times 10^{-19}$.
The other facility is Mu2e at Fermilab in the US. This experiment also aims to 
reach a sensitivity of $10^{-16}$  on $B^\text{Al}_{\mu\to e}$~\cite{Mu2e:2014fns}.

In figure~\ref{fig:BRmuNtoeN} we show the general behavior of the various MLRSM contributions to $B_{\mu \rightarrow e}^\text{Al}$ as a function of $m_4$ for the same cases of figures \ref{fig:BRmegamma} and \ref{fig:BRmu3e}.  We also show the region 
excluded by $B^\text{Au}_{\mu \to e}<7.0 \times 10^{-13}$
as well as the future sensitivities of COMET-I, COMET-II 
and PRISM. Here again in both cases $\delta_{L,R}^{++}$ 
surpass the $W$ curve at some point. This happens at about the same values of $m_4$ as that for BR$(\mu \to 3e)$. There is  again, as expected, no suppression of the double-charged scalars when neutrinos are degenerate as opposed to BR$(\mu \to e \gamma)$. 
We see that the current bound can exclude only a very modest part of the parameter space of the model, but this could increase very substantially in the future.

\begin{table}[htb]
\centering
\begin{tabular}{ |p{2cm}|p{2.0cm}|p{2.0cm}|}
\hline
\bf Nucleus & \bf $V^{(p)}\,(\text{{\tiny $\times$}} \,m_\mu^{5/2})$ & \bf $\Gamma_{\rm cap.}\, ({\rm eV})$\\
\hline
Au & 0.0974 & 8.60 $\times$ $10^{-9}$ \\
Al & 0.0161 & 4.64 $\times$ $10^{-9}$ \\
\hline
\end{tabular}
\caption{\label{tab:VpandGammavalues} Values of $V^{(p)}$ in units of $(m_\mu^{5/2})$  and $\Gamma_{\rm cap.}$ obtained from \cite{Kitano:2002mt}.}
\end{table}

\subsection{Anomalous magnetic moment of the muon ($a_\mu$)}
\label{subsec:amu}
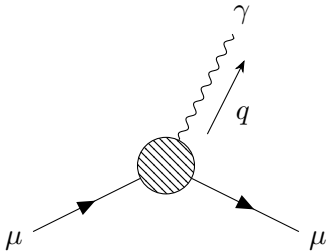
\begin{figure}[hbt]
  \centering
  \begin{tikzpicture}
    \begin{feynman}
      \vertex[blob] (a2) at (0, 0) {};
      \vertex (a1) at (-2,-1) {$\mu$};
      \vertex (a3) at (2,-1) {$\mu$};
      \vertex (b1) at (1, 2) {$\gamma$};
      ;
      \diagram* {
        (a1) -- [fermion] (a2) -- [fermion] (a3),
        (a2) -- [photon,momentum'=\(q\)] (b1),
      };
    \end{feynman}
  \end{tikzpicture}
  \caption{Generic diagram depicting the
  contributions to the anomalous magnetic moment of the muon.}
  \label{fig:MagneticMomentDiagram}
\end{figure}

The MLRSM contributions to the anomalous magnetic moment of the muon, $a_\mu \equiv (g-2)_\mu/2$,  can be evaluated by calculating  the form factor that enters in the vertex correction diagrams like the one depicted in  figure~\ref{fig:MagneticMomentDiagram} and is given by $a_\mu = F_2(0)$. In the SM $a_\mu^\text{SM} =4116591810(43) \times 10^{-11}$~\cite{Aoyama:2020ynm}, while the latest measurement 
by the Fermilab National Accelerator Laboratory (FNAL) Muon g-2 Experiment found $a_\mu^\text{FNAL}=116592040(54) \times 10^{-11}$~\cite{Muong-2:2021ojo}, which combined with 
previous experimental results lead to $\Delta a_\mu \equiv a_\mu^\text{exp}-a_\mu^\text{SM}= (251 \pm 59) \times 10^{-11}$.

\begin{figure}[htb]
    \centering
    \includegraphics[width=0.99\textwidth]{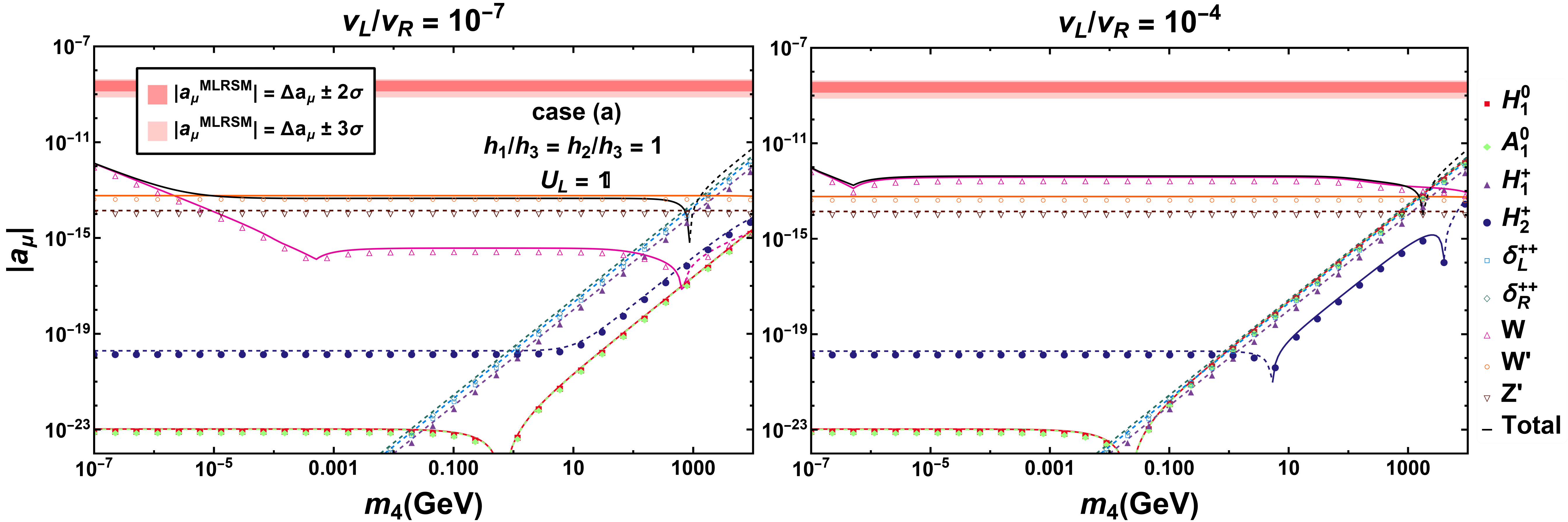}
    \caption{Contributions to  $\vert a_\mu\vert$ in the MLRSM in {\em standard case} (a) for 
    $v_L/v_R= 10^{-7}$ (left panel) and $v_L/v_R=10^{-4}$ as a function of $m_4$. The red (light red) region represents the value needed to explain $a_\mu^{\rm exp}$ within 2$\sigma$ (3$\sigma$)~\cite{Muong-2:2021ojo}. The solid (dashed) lines represent positive (negative) contributions.}
    \label{fig:MuonMagMoment}
\end{figure}

In figure~\ref{fig:MuonMagMoment} we show the various contributions to $\vert a_\mu \vert$ as a function of $m_4$ for case (a). The total contribution
which is overall positive, except for $m_4 \gtrsim 1 $ TeV, is dominated 
by the $W$ (for $m_4 \lesssim 10$ keV) or by the $W'$ (for $10$ keV $\lesssim m_4 \lesssim 1$ TeV) exchange. The $W'$ contribution is independent of $v_L$ and of 
$m_4$ but it is about three to four orders of magnitude smaller than what is needed to explain $a_\mu^{\rm exp}$. Except for the contributions from $W'$, $W$ and 
$H^+$, the other contributions are always negative, so they cannot help to decrease the discrepancy between theory and experiment. In particular, for $m_4 \gtrsim 1 $ TeV, the dominant contributions, from charged scalars ($\delta_{R,L}^{++},\delta_L^+$),
are all negative, but still too small to make any significant measurable effect.

\subsection{Electric dipole moment  ($d_\ell$)}
\label{subsec:edm}
Now we will examine the status of the electric dipole moment  $d_\ell, \ell =e, \mu$. 
In the case of the electron's electric dipole moment (eEDM) $d_e$, the 
ACME II collaboration used a beam of ThO molecules to set the best  limit to date at $\vert d_e\vert < 1.1 \times 10^{-29}$~e$\cdot$cm~\cite{ACME:2018yjb}.
The NL-eEDM collaboration~\cite{NL-eEDM:2018lno} claims  
a sensitivity of $5 \times 10^{-30}$~e$\cdot$cm  on 
$\vert d_e\vert $ is feasible
using an intense primary cold source of 
barium monofluoride molecules. There is an even more 
daring proposition, the authors of ref.~\cite{Fitch:2020jil}
affirm that using ultracold YbF molecules trapped in 
an optical lattice one can hope to bring down the sensitivity 
to the level of $(10^{-31}-10^{-32})$~e$\cdot$cm. 

For $d_\mu$ the most stringent limit  $\vert d_\mu \vert< 1.8 \times 10^{-19}$~e$\cdot$cm was set by 
the BNL Muon $g-2$ experiment~\cite{Muong-2:2008ebm}. 
The $g-2$  experiments at Fermilab and at J-PARC  aim to  achieve a similar sensitivity of $\vert d_\mu\vert \sim 10^{-21}$~e$\cdot$cm~\cite{Chislett:2016jau,Abe:2019thb}.
There are also plans to have a dedicated experiment using a frozen-spin technique at PSI to try to bring down the sensitivity to 
$\vert d_\mu \vert \sim 6 \times 10^{-23}$~e$\cdot$cm~\cite{Adelmann:2021udj,Sakurai:2022tbk}. 
J-PARC also proposed a dedicated experiment 
that claims to reach $\vert d_\mu\vert \sim 10^{-24}$~e$\cdot$cm  \cite{Farley:2003wt}.

Theoretically 
\begin{equation}
    \vert d_\ell\vert\equiv \frac{\vert G_2(0)\vert}{(2\, m_\ell)}\, \rm e \, ,
\label{eq:dmu}
\end{equation}
which can in  the MLSRM be enhanced by phases 
in $M_R$ or in $U_L$. For the approximate expressions for $d_\ell$ see appendix~\ref{app:formfac}. However, only $\vert d_e\vert$ can be significantly enhanced to reach values of the order of the current bound. The particles that impact this observable in the model are the vector bosons $W$ and $W'$, the charged scalar $H^+$ and the neutral scalars $H,A^0$.

\begin{figure}[hbt]
\centering
\begin{subfigure}[b]{\textwidth}
   \includegraphics[width=1\linewidth]{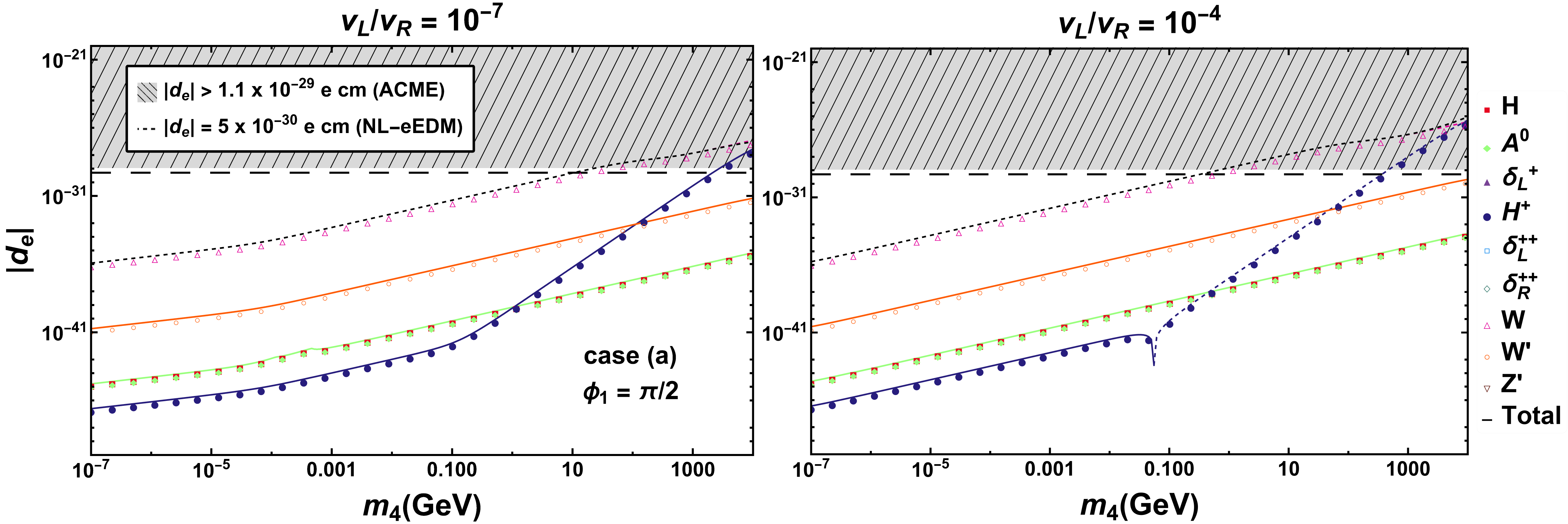}
\end{subfigure}
\begin{subfigure}[b]{\textwidth}
   \includegraphics[width=1\linewidth]{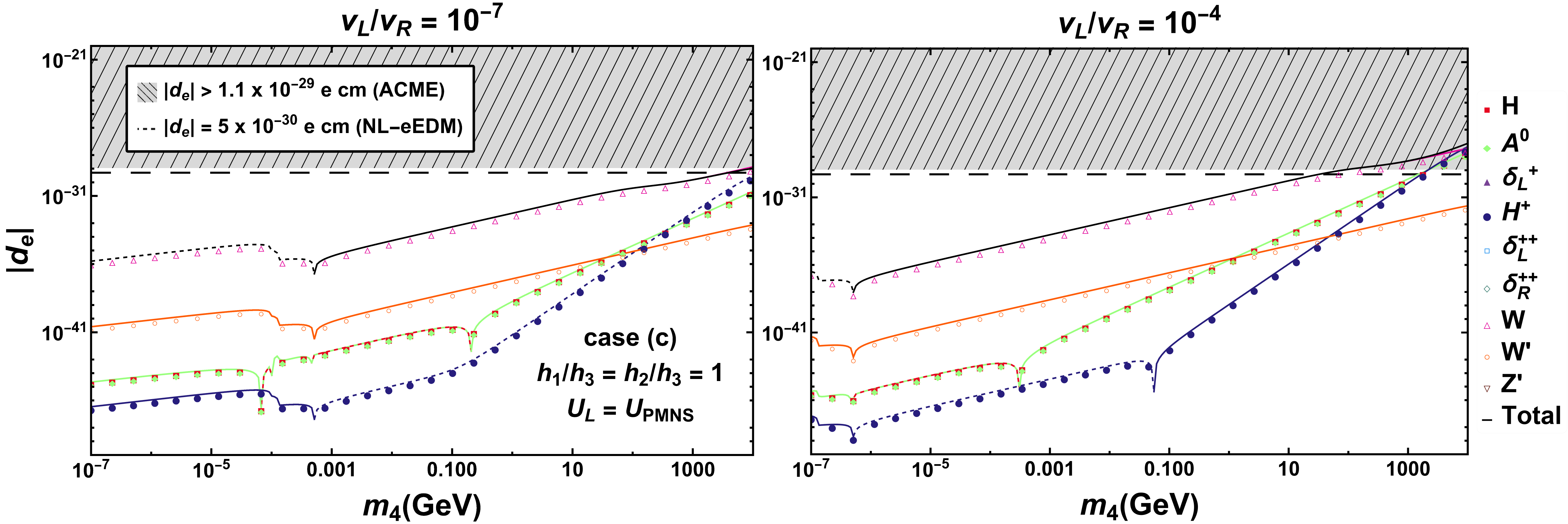}
\end{subfigure}
\begin{subfigure}[b]{\textwidth}
   \includegraphics[width=1\linewidth]{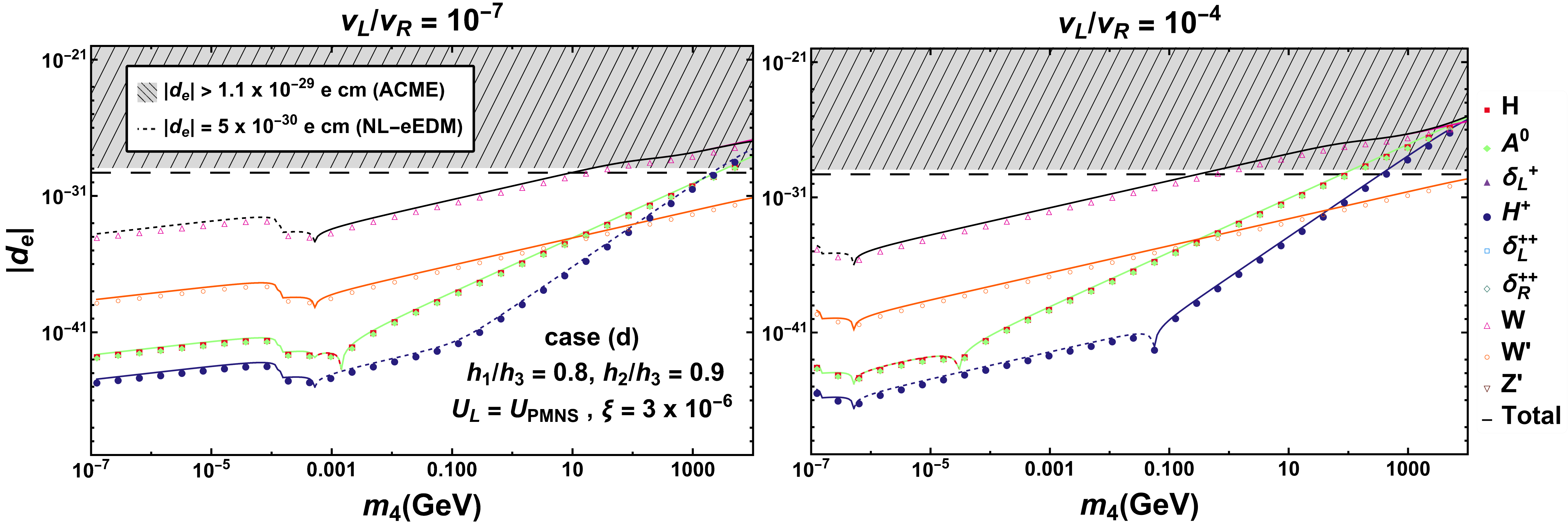}
\end{subfigure}
\caption{Contributions to $\vert d_e\vert$ in the MLRSM for the parameters of cases (a) but with $\phi_1=\pi/2$, (c) and (d). We show the behavior for an intermediate ratio $v_L/v_R= 10^{-7}$ (left panels) and for a large one $v_L/v_R=10^{-4}$ (right panels) as a function of $m_4$.
The solid (dashed) lines represent positive (negative) values of $d_e$.  
The gray region is already excluded at by the ACME II collaboration~\cite{ACME:2018yjb}, while the dashed line show the 
possible reach of the NL-eDM experiment using BaF molecules~\cite{NL-eEDM:2018lno}.}
    \label{fig:EDM}
\end{figure}

In figure~\ref{fig:EDM} we show the various contributions of the model to $\vert d_e\vert$  as a function of $m_4$ for $v_L/v_R= 10^{-7}$ (left panel) and 
$v_L/v_R= 10^{-4}$ (right panel) for the 
case (a) but with $\phi_1=\pi/2$ (top panels) and for case (c) (middle panels) and case (d)(bottom panels).
We observe that case (a) with $\phi_1=\pi/2$ and case (d) 
where $U_L = U_{\rm PMNS}$ and larger mixing $\xi$ produce similar constraints on $\vert d_e\vert$. The contribution of $H^+$ for small $m_4$ is dominated by the couplings that depend on $M_\ell$ (see eqs.~(\ref{eq:coupHpS}) and (\ref{eq:coupHpP})) and, depending on $\phi_R$, can change the sign when crossing the transition region from the type-I seesaw to the tuned regime due to the change of behavior of $M_D$. On the other hand, for big $m_4$, contributions of $H^+$ that grow with the heavy neutrino mass become important and dominate. Among those there are two different combinations of couplings that generate $d_e$'s with opposite signs and what dictates which one dominates is the value of $\vert U_{\alpha i} \vert^2 \sim v_L/v_R$, see for example case (c) in figure \ref{fig:EDM}. Regarding the neutral scalars $H$ and $A^0$, their contributions are the same but with  opposite signs so even when their individual contributions have sizable effect, they cancel each other when summed. In case (c) since $\xi$ is two orders of magnitude smaller than in case (d) the limit is weakened in the same proportion. We do not show case (b) as it is very similar to case (c).

\begin{figure}[hbt]
\centering
   \includegraphics[width=1\linewidth]{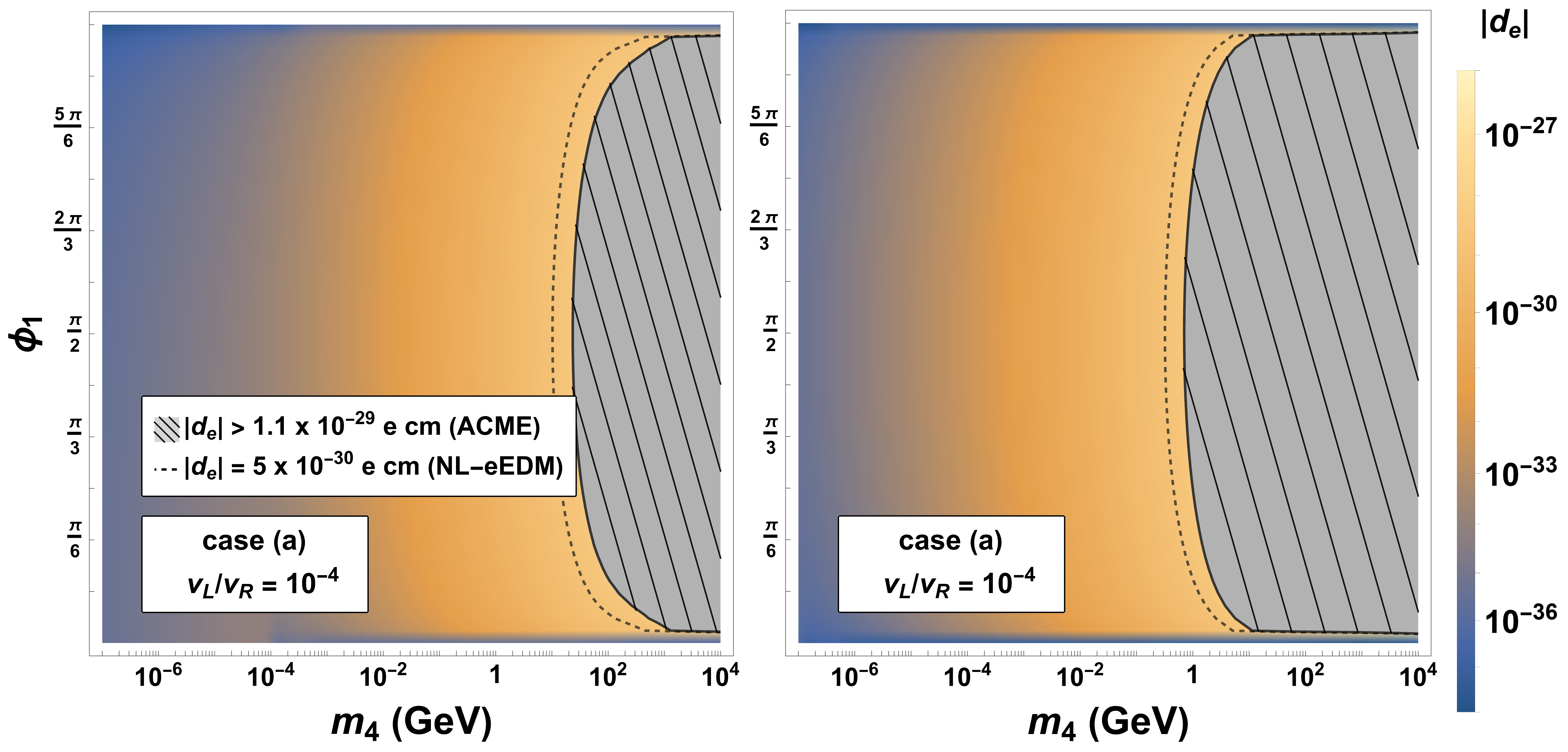}
\caption{Region in the plane $m_4 \times \phi_1$ excluded by the current limit on $\vert d_e\vert$~\cite{ACME:2018yjb} for the case (a) with varying $\phi_1$, we also show by a dashed line the expected sensitivity for a future experiment~\cite{NL-eEDM:2018lno}. The panel on the left (right) is for $v_L/v_R= 10^{-7}$ ($v_L/v_R= 10^{-4}$).}
    \label{fig:EDMcont}
\end{figure}

In figure~\ref{fig:EDMcont} we show the region in the
plane $m_4 \times \phi_1$ excluded by  
the current limit on $\vert d_e\vert$ with the remaining parameters fixed as in case (a). It is interesting to notice that it 
is not very dependent on the exact value of the phase $\phi_1$, as long as 
$\frac{\pi}{50} \lesssim \phi_1 \lesssim \frac{49\pi}{50}$, away from the CP-conserving values.

Besides the case where both $U_L=\mathbf{I}$ and $\phi_1\neq0$, the maximum contributions to $\vert d_\mu \vert $ are similar to those for $\vert d_e\vert$ and so do not reach a value that can be probed by neither current nor future experiments. When $U_L = \mathbf{I}$ and $\phi_1 \neq 0$ only the imaginary part of the electron couplings are affected, keeping the muon contributions untouched and therefore suppressed. In order to modify the imaginary part of the muon couplings keeping $U_L = \mathbf{I}$, the phase $\phi_2$ needs to be invoked,  though the maximum value $\vert d_\mu\vert $ achieved is still well below experimental reach.

\subsection{$W$ boson mass}
\label{subsec:mW}
Finally, we would like to address the influence 
the exact value of $m_W$ has on the model parameters used in our study.
The recent analysis of CDF II data resulted on a new value for the $W$ boson mass~\cite{CDF:2022hxs}
\begin{eqnarray}
m_{W}^\textrm{CDF} & = & \left(80.4335\pm0.0094\right)\,{\rm GeV},\label{eq:W_mass_CDFII}
\end{eqnarray}
which is in 7$\sigma$ tension with the SM prediction
$m_{W}^\textrm{SM} =\left(80.361\pm0.006\right)$ GeV from a global electroweak fit~\cite{ParticleDataGroup:2020ssz}.

The shift in the $W$ boson mass from the SM value $m_{W}^\textrm{SM}$ can be written in terms of $S$, $T$, $U$ parameters~\cite{Peskin:1990zt,Peskin:1991sw}
as~\cite{Maksymyk:1993zm}
\begin{eqnarray}
m_{W} & = & m_{W}^{\rm SM}\left[1-\frac{\alpha S}{2\left(c_{W}^{2}-s_{W}^{2}\right)}+\frac{c_{W}^{2}\alpha T}{c_{W}^{2}-s_{W}^{2}}+\frac{\alpha U}{4s_{W}^{2}}\right]^{1/2},\label{eq:Wmass_STU}
\end{eqnarray}
where $s_W$ is the sine of the Weinberg angle, i.e., $s_{W}^{2}=0.2351$ ($c_{W}^{2}=1-s_{W}^{2}$)
and the fine structure constant $\alpha=1/128.18$ is taken at the $Z$ boson mass scale $m_{Z}$. The $W$ boson mass in the SM electroweak fit is determined from the relation~\cite{Awramik:2003rn}
\begin{eqnarray}
m_{W}^\textrm{SM} & = & m_{Z}\left[\frac{1}{2}+\frac{1}{2}\sqrt{1-\frac{2\sqrt{2}\pi\alpha}{G_{F}m_{Z}^{2}}\left(1+\Delta r\right)}\right]^{1/2},
\end{eqnarray}
where $G_{F}$ is the Fermi constant and $\Delta r$ takes into account the radiative corrections. $G_{F}$ is determined from the muon lifetime $G_{F}=G_{\mu}$ which can be sensitive to the nonunitarity of the $U_{\rm PMNS}$ and can affect the determination of $m_{W}^\textrm{SM}$. Ref.~\cite{Blennow:2022yfm}
showed that a deviation from unitarity of the $U_{\rm PMNS}$ of the order of $10^{-3}$ is required to be consistent with the new result \eqref{eq:W_mass_CDFII} which translates to $\left|\delta U_{L}\right|\sim\left|{\cal U}_{LR}\right|^{2}\sim10^{-3}$
in our scenario. This possibility is ruled out by neutrinoless double beta observable (see \cite{deGouvea:2015euy} and also figure \ref{fig:Nu0bb-Ue4} of our earlier analysis)  and hence will not be considered further.

\begin{figure}
    \centering
    \includegraphics[width =0.8\textwidth]{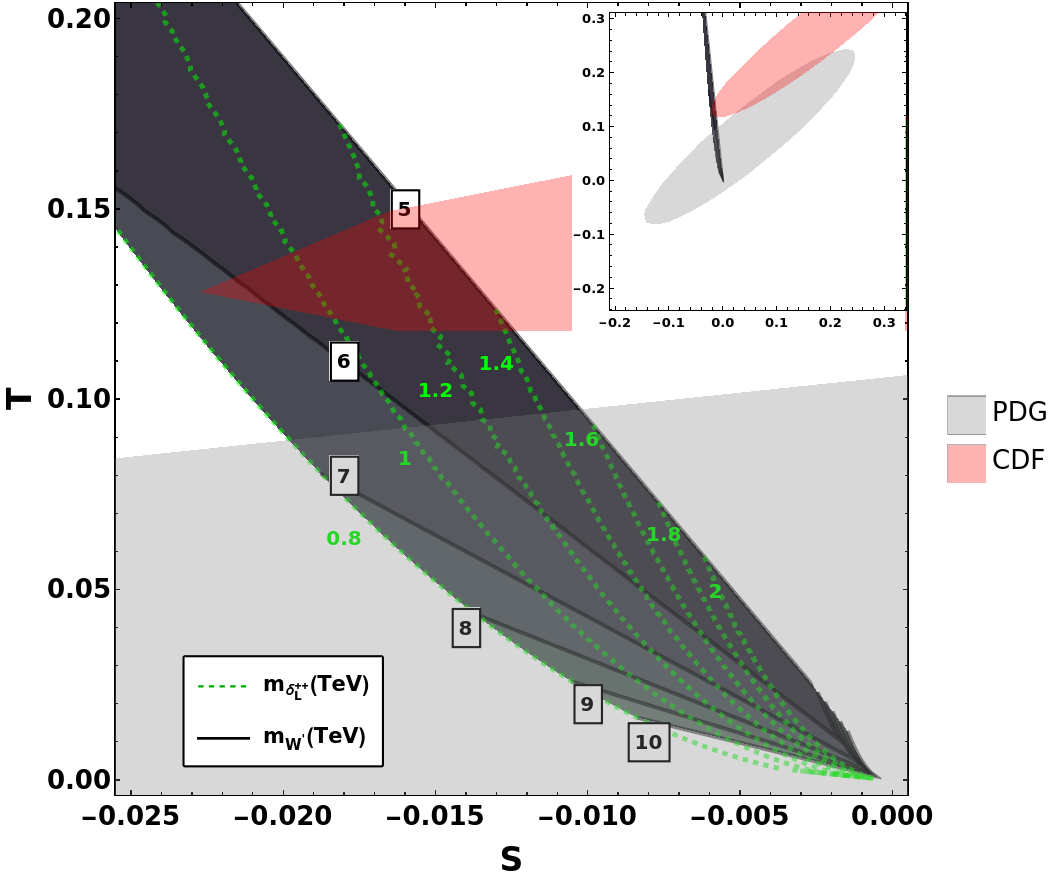}
    \caption{Contribution to S and T in the MLRSM. The blue and red ellipses are the $2 \sigma$ preferred regions from the electroweak fit for the PDG and the new CDF II $m_W$ measurent, respectively \cite{Asadi:2022xiy}. }
    \label{fig:STLRmodel}
\end{figure}

In the MLRSM, since $v_{L}\ll v_{R}$, the main contributions to $W$ boson mass will come from loop corrections from the left-handed scalar triplet $\mathbf \Delta_{L}$ expressed in term of $S$, $T$ and $U$ as in eq. (\ref{eq:Wmass_STU}). Using the expressions from ref.~\cite{Lavoura:1993nq},
we can determine $S$, $T$ and $U$ with the mass spectrum of the
field components of $\mathbf \Delta_{L}$
\begin{eqnarray}
m_{\delta_{L}^0}^{2} & = & m_{\delta_{L}^{++}}^{2}-\frac{\alpha_{3}}{2}\kappa_{+}^{2},\\
m_{\delta_{L}^{+}}^{2} & = & m_{\delta_{L}^{++}}^{2}-\frac{\alpha_{3}}{4}\kappa_{+}^{2},
\end{eqnarray}
where $\alpha_3$ is a dimensionless coupling in the scalar potential, $m_{\delta_L^{0}}^2 \equiv m^2_{H_L} = m^2_{A_L}$ and
we recall that $\epsilon = 2\kappa_{1}\kappa_{2}/\kappa_{+}^{2}\ll 1$. The contribution to $U$ is in general small and is at the level $\lesssim10^{-3}$ for our scenario and can be neglected. Furthermore, we find $S < 0$ while $T > 0$ and that $-S \ll T$. So, the contribution to $W$ mass is essentially determined by $T > 0$.  In the isospin conserving limit $\alpha_{3}\to0$, $T\to 0$ and hence one will need a large $\alpha_{3}$ to have a large $T$. This will make the heavy neutral Higges $H$, $A^0$ correspondingly heavy since their squared masses are $\alpha_3 v_R^2/2$. The $H$, $A^0$ and heavy $W'$ contributions to $B-\bar{B}$ mixing lead to a lower bound on $\alpha_{3}$ as a function of $m_{W'}$~\cite{Bertolini:2014sua,Maiezza:2016bzp}
\begin{eqnarray}
\alpha_{3} & \approx & \frac{30}{\left(m_{W'}/2.23\,{\rm TeV}\right)^{2}-1}.
\end{eqnarray}
Hence we can express the compatible $S$ and $T$ in terms of $m_{W'}$ and $m_{\delta_L^{++}}$. Various groups, for example \cite{Balkin:2022glu,Asadi:2022xiy},  provide the best-fit 
for $S$ and $T$  at 1$\sigma$ assuming $U=0$ taking into account the new CDF result (\ref{eq:W_mass_CDFII}). In figure~\ref{fig:STLRmodel}, we show the $S$ and $T$ parameter space of the MLRSM and the 1 $\sigma$ parameter space that can explain the $W$ mass is obtained from ref.~\cite{Asadi:2022xiy}.

We see that for the model to be compatible with $m_W^{\rm CDF}$
we need $m_{W'} \sim (5-6)$~TeV and $m_{\delta_L^{++}} \sim (0.9 -1.4)$ TeV. Both ranges of values are still allowed by experiments.

\section{Discussion of Results }
\label{sec:results}

Now we will discuss the allowed regions in the plane 
$v_L/v_R \times m_4$ after imposing the following experimental limits on the model: (a)  
$m_{\beta \beta}<126$ meV~\cite{KamLAND-Zen:2022tow};
(b) BR$(\mu \to e \gamma)<4.2 \times 10^{-13}$~\cite{MEG:2016leq}; (c) BR$(\mu \to 3 e)<1.0 \times 10^{-12}$~\cite{SINDRUM:1987nra} ; $B^{\rm Au}_{\mu \to e}< 7.0 \times 10^{-13}$~\cite{SINDRUMII:2006dvw} and 
$\vert d_e\vert< 1.1 \times 10^{-29}$ e$\cdot$cm~\cite{ACME:2018yjb}.
We have also forced the mixing elements 
$\vert U_{\alpha i}\vert$ to pass the laboratory 
constraints given in ref.~\cite{deGouvea:2015euy}.
We show on the plots the region of the parameter space that can be explored 
by the future accelerator experiments taken from ref.~\cite{Ballett:2019bgd}.
For completeness we also show  the 
BBN limits on $m_4 \times \vert U_{\alpha 4}\vert^2$ for 
$m_4$ in the range 3 MeV to 1 GeV taken from ref.~\cite{Sabti:2020yrt} (in teal blue) and 
on $m_4 \times \sum_{\alpha}\vert U_{\alpha 4}\vert^2$ for $m_4$ in the range $10^{-8}$ GeV to 1 GeV
from an analysis of heavy neutral leptons in consistence with measurements of the Hubble constant, Supernovae Ia luminosity distances, the CMB shift parameter, the BAO scale taken from ref.~\cite{Vincent:2014rja} (in blue grey) .

\begin{figure}[!htb]
\centering
\begin{subfigure}[b]{0.45 \textwidth}
   \includegraphics[width=\linewidth]{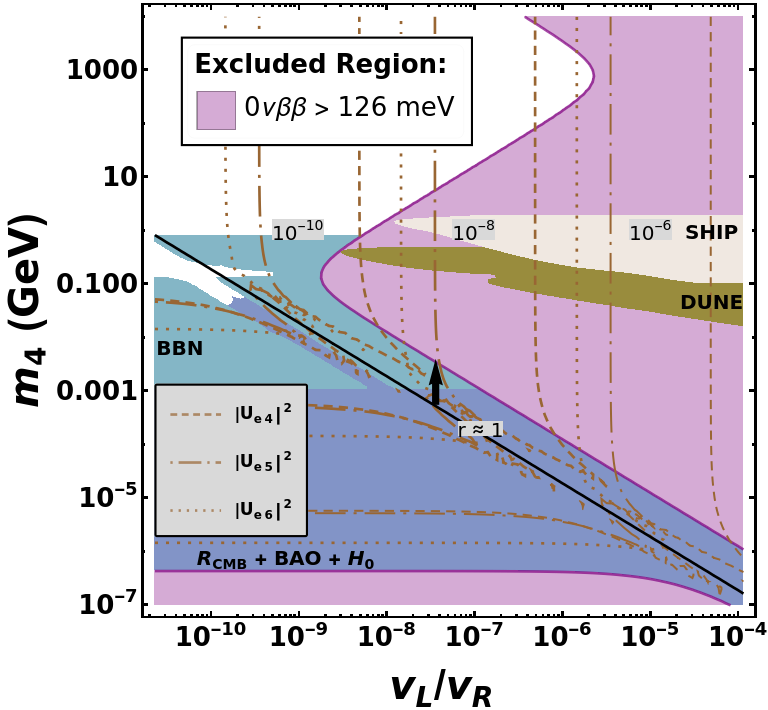}
\end{subfigure}
\begin{subfigure}[b]{0.45\textwidth}
   \includegraphics[width=\linewidth]{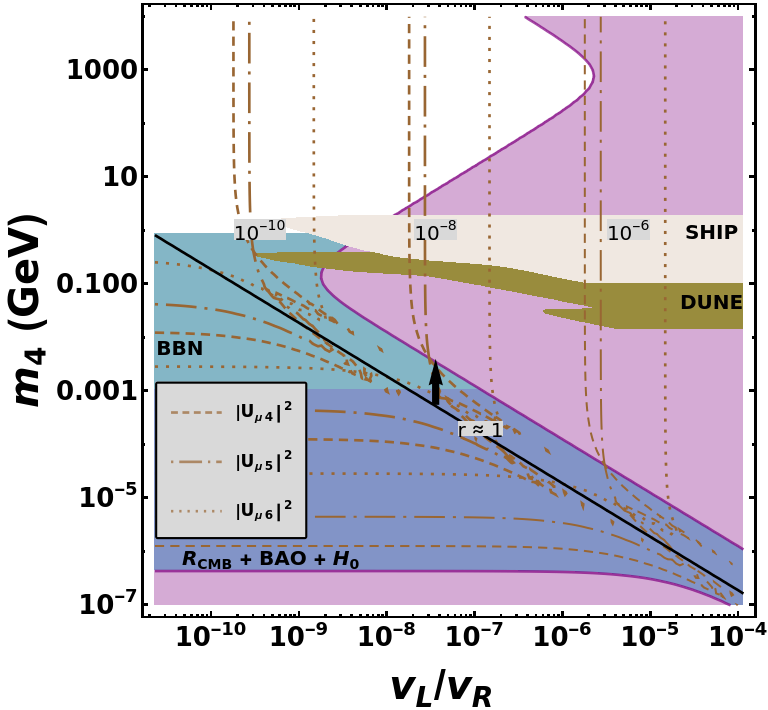}
\end{subfigure}
\begin{subfigure}[b]{0.45\textwidth}
   \includegraphics[width=\linewidth]{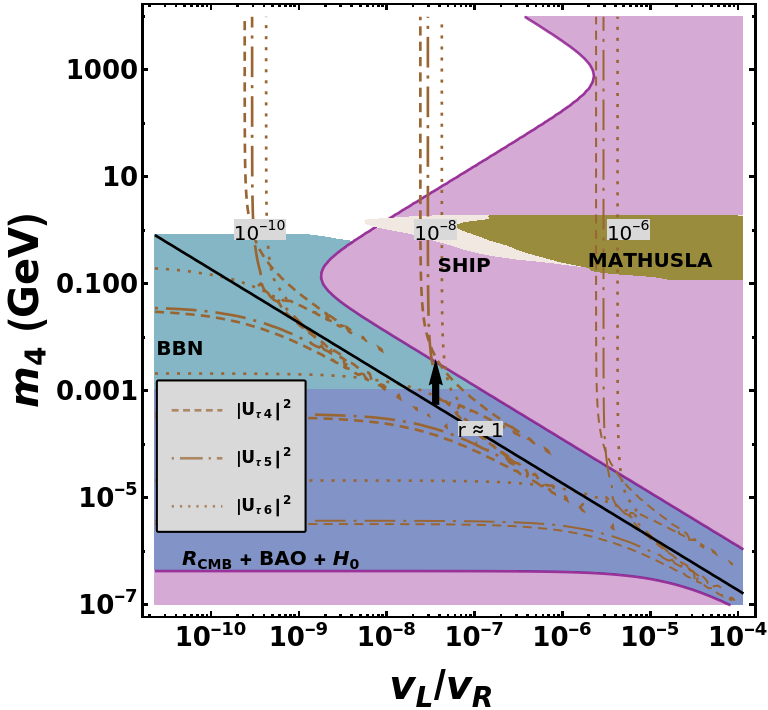}
\end{subfigure}
\caption{Excluded regions for case (a) by the current bound on $0\nu\beta\beta$ decay (pink)~\cite{KamLAND-Zen:2022tow}, BBN~\cite{Sabti:2020yrt}(teal blue) and other cosmological observables ($H_0$, BAO, Supernovae type Ia and the CMB shift parameter)~\cite{Vincent:2014rja} (blue grey). The black line separates the tuned region ($r \sim 1$), above the line, from the type-I seesaw region ($r < 1$), below the line. In the top left panel we show lines for $\vert U_{ei}\vert^2=10^{-10}, 10^{-8}$ and $10^{-6}$ as dashed ($i=4$), dashed-dotted ($i=5$) and dotted ($i=6$). The same is shown in the top right (bottom) panel for 
$\vert U_{\mu i}\vert^2$($\vert U_{\tau i}\vert^2$). The sensitivity regions for SHIP, DUNE and MATHUSLA were taken from ~\cite{Ballett:2019bgd}.}
    \label{fig:casea}
\end{figure}

We will start by looking at the benchmark cases.
In these cases we have set $v_R$ to the minimum value 
and $\xi$ to the maximum value for which $0\nu \beta \beta$ decay can restrict the plane $v_L/v_R \times m_4$ the least, in the tuned region. If we increase $v_R$ or decrease $\xi$ from their  standard values, we have verified that this 
region will not change. Nevertheless an increase of $v_R$ may remove the constraint in the type-I regime.
The black diagonal line in the plots  separates the tuned regime ($r \sim 1$), the region above this line, from the type-I seesaw  ($r < 1$),  the region below this line.

\begin{figure}[htb]
     \centering
         \includegraphics[width=0.9\textwidth]{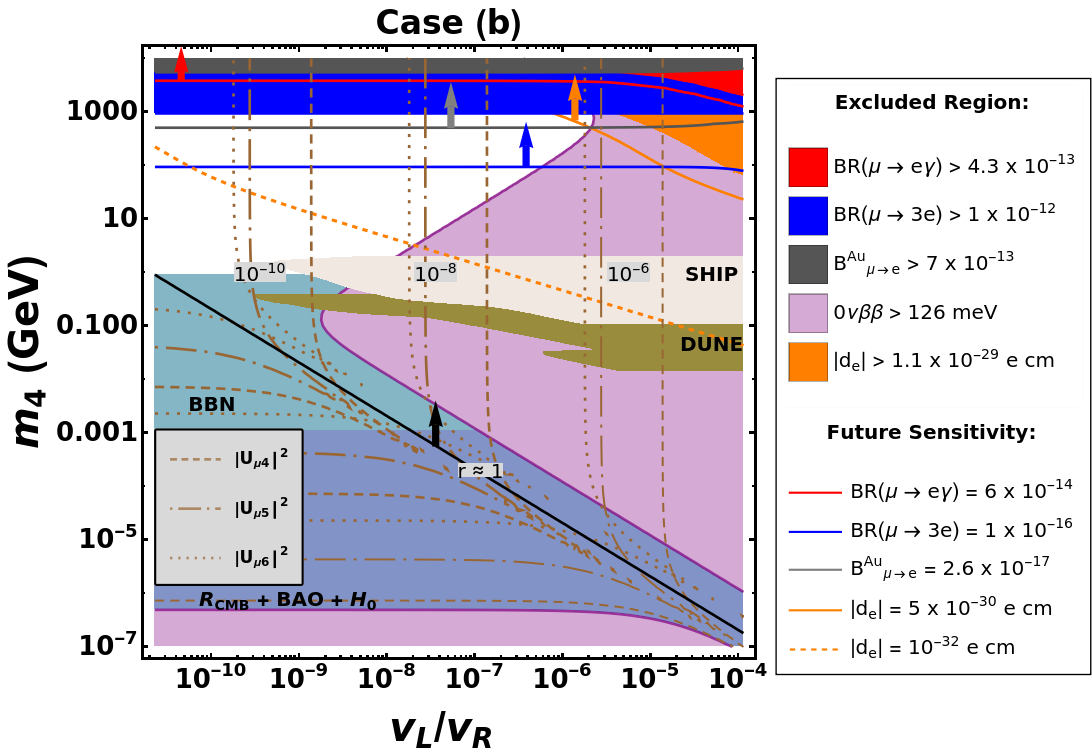}
        \caption{Excluded regions for case (b). We include the same bounds used in  figure~\ref{fig:casea} as well as the current 
        limits on $\mu \to e \gamma$~\cite{MEG:2016leq} (red),
        on $\mu \to 3 e$~\cite{SINDRUM:1987nra}  (blue), on 
        $B^{\rm Au}_{\mu \to e}$~\cite{SINDRUMII:2006dvw} (dark grey) and on $\vert d_e\vert$~\cite{ACME:2018yjb}. The expected sensitivity of future LFV and eEDM experiments are  shown by the corresponding color line with an arrow.
        }
        \label{fig:caseb}
\end{figure}

In figure~\ref{fig:casea} we 
see our results for case (a) where the 
$0\nu \beta \beta$ bound excludes a significant portion of the tuned region as well as a small portion of the type-I seesaw region (below $m_4 \lesssim 4 \times 10^{-7}$ GeV). As previously explained, there are no LFV or $\vert d_e \vert$ limits for this case. For degenerate neutrinos and 
$U_L=\mathbf{I}$, in the tuned region all matrix elements $\vert U_{\alpha i}\vert^2, i=4,5,6$ are  
of the order  $(v_L/v_R)\vert \phi_R\vert^2$ (see eqs.~\eqref{eq:66_mixing}) and \eqref{eq:C-tuned}).
In the top left panel we show lines of constant $\vert U_{e i}\vert^2=10^{-10}, 10^{-8}$ and $10^{-6}$, as dashed ($i=4$), dashed-dotted ($i=5$)
and dotted ($i=6$), for reference. 
In the top right (bottom) panel the same is shown for 
$\vert U_{\mu i}\vert^2$ ($\vert U_{\tau i}\vert^2$ ).
These lines are independent 
of $v_L/v_R$ in the type-I seesaw region and in the transition there is a cancellation which corresponds 
to one of the three light neutrino masses $m_1$, $m_2$ or
$m_3$~\footnote{This happens when seesaw type II accounts for one of the neutrino masses in such a way that the mixing is suppressed.}.
 We also show the expected sensitivity of the 
accelerator experiments most effective to probe 
each of those mixing matrix elements according to \cite{Ballett:2019bgd}.
The cosmological limits exclude the entire type-I seesaw region and a small part of the tuned region that was not excluded by $0\nu \beta \beta$. In the parameter space still allowed all the elements $\vert U_{\alpha i}\vert^2 \lesssim 10^{-5}$. 
\begin{figure}[hbt]
     \centering
         \includegraphics[width=0.9\textwidth]{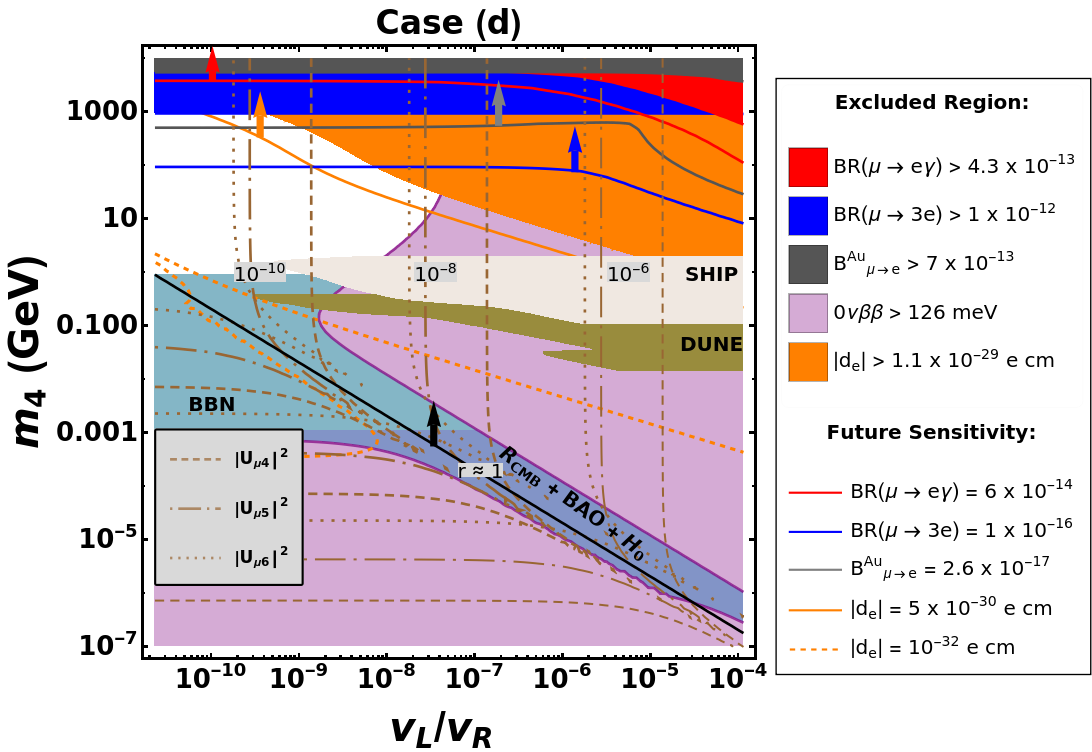}
        \caption{Same as figure~\ref{fig:caseb} but for case (d).}
        \label{fig:cased}
\end{figure}
In case (b), shown in figure~\ref{fig:caseb},  $U_L \neq \mathbf{I}$ and the heavy neutrinos are mildly hierarchical. We observe that the limits from 
$0 \nu \beta \beta$ are not affected by this choice 
due to the low value we took for the mixing $\xi$,
but there are now limits from LFV. These limits concern 
exactly the part of the tuned region that is free from cosmological constraints. The reach of some of the future charged LFV experiments is also shown in this figure.
They have the potential to  probe  a significant  
portion of the parameter space still allowed in the tuned region. The current bound on $\vert d_e \vert$ excludes 
regions that are already not allowed by $0\nu \beta \beta$ decay or LFV. 
Here we only show lines of constant 
$\vert U_{\mu i}\vert^2$ as for the top right panel of figure ~\ref{fig:casea}.
In the tuned region all matrix elements $\vert U_{\alpha i}\vert^2$ are   of the order  $v_L/v_R \vert U_L\vert^2$. In the region still allowed all the elements $\vert U_{\alpha i}\vert^2 \lesssim 10^{-6}$ ($v_L \lesssim 200$ MeV).
We do not show explicitly case (c) here, as it is very 
similar to (b), so the heavy neutrino mass hierarchy 
does not play a significant role. See figures 
\ref{fig:casecwithoutphases} in appendix~\ref{app:Plots}.

In figure~\ref{fig:cased} we can see that case (d) is qualitatively similar to (b), however, the larger value of $\xi$ 
results in a larger area excluded by $0\nu\beta\beta$ decay, most notably in the type-I seesaw region, but also in the tuned one.
Due to this the LFV bounds also increase for $v_L/v_R \gtrsim 10^{-5}$ (the $W$ term is enhanced by the larger mixing $\xi$ ), covering a part of the parameter space  already excluded by $0\nu \beta \beta$ decay. In the window still allowed by data we have all the elements $\vert U_{\alpha i}\vert^2 \lesssim 10^{-8}$ ($v_L \lesssim 2$ MeV). Likewise the eEDM 
limits becomes more significant in this case, already excluding part of the parameter space in the reach of 
future LFV experiments, in particular, $v_L/v_R \gtrsim 10^{-7}$
is discarded.
\begin{figure}[htb]
     \centering
         \includegraphics[width=0.9\textwidth]{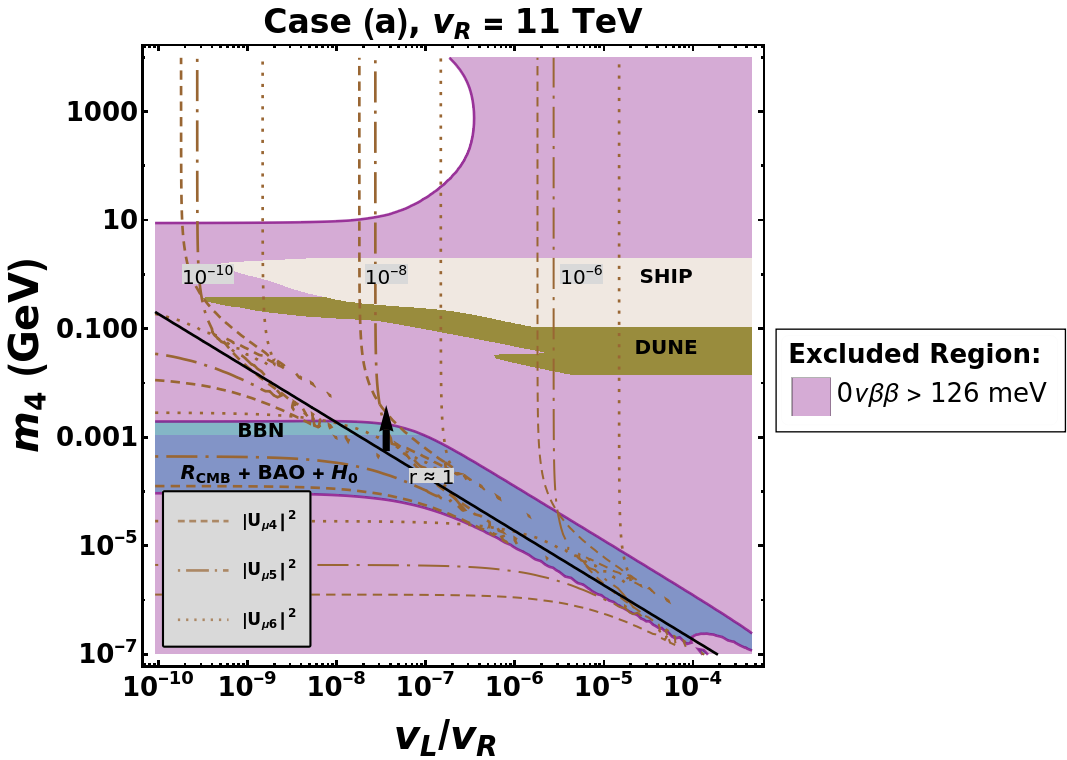}
        \caption{Same as figure~\ref{fig:casea} but for $v_R= 11$ TeV.}
        \label{fig:casea-VR}
\end{figure}
Let us discuss now the impact of changing the value of 
one of the fixed parameters of case (a) on our results:
\begin{enumerate}
    \item  decrease $v_R$: lower values 
    of $v_R$ will increase the part of the parameter space excluded by $0\nu\beta\beta$ decay (as $\lambda \propto v_R^{-2}$), specially in the     tuned region, but also in the type I seesaw regime. This is mainly due to the 
    contributions $RR,N$, that is independent of $v_L/v_R$, and $ RL, \lambda$ (see figure~\ref{fig:NeutrinolessIndividual}) which increases as $v_L/v_R \to 0$.
    In figure~\ref{fig:casea-VR} we illustrate this effect for $v_R=$ 11 TeV ($m_{W'}= 5$ TeV). The only region that still survives is  $m_4 \gtrsim 10$ GeV
    and $v_L/v_R \lesssim 3 \times 10^{-7}$.
    If $U_L \neq \mathbf{I}$, LFV bounds would restrict  
    the allowed region even further.
    \item change $U_L$: different $U_L$ can 
    have two effects: modify the $0\nu\beta\beta$ decay 
    exclusion regions, connecting them in the transition
    region, and increasing the bounds from LFV and eEDM processes excluding  lower values of $m_4$. In figure~\ref{fig:casea-UL} we show 
    an example for a random unitary $U_L$ where we see 
    how the LFV and eEDM bounds can be more restrictive depending on $U_L$.
    \item change $\phi_i$: when $U_L \neq \mathbf{I}$ the effect     of the extra phases in $M_R$ is somewhat degenerate with the 
    phases in $U_L$ (see figure \ref{fig:casecwithphases} in appendix \ref{app:Plots}), nevertheless, when $U_L=\mathbf{I}$, $\phi_1\neq 0$ can produce sizable $\vert d_e\vert$. In figure~\ref{fig:casea-phi}
    we illustrate this effect for the case (a) with $\phi_1=\pi/2$ while $\phi_2 = \phi_3 =0$. We see that 
    eEDM  and $0\nu \beta \beta$ rule out $v_L/v_R \gtrsim 10^{-7}$.
    \begin{figure}[hbt]
     \centering
         \includegraphics[width=0.9\textwidth]{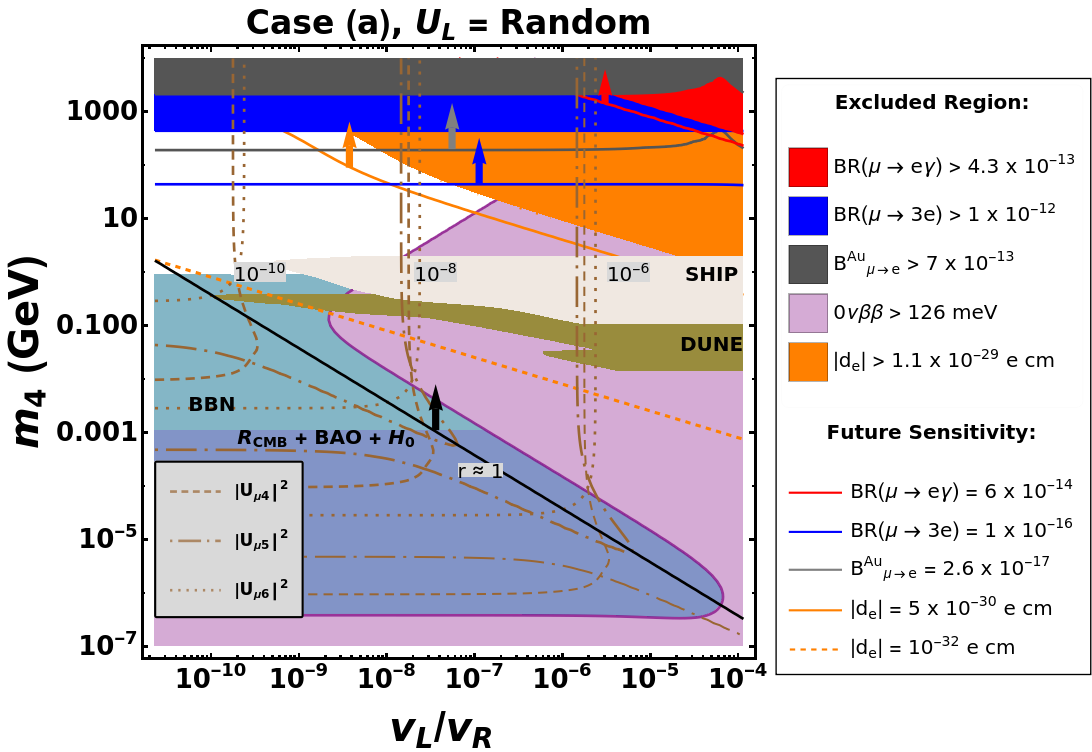}
        \caption{Same as figure~\ref{fig:casea} but for a random $U_L$.}
        \label{fig:casea-UL}
\end{figure}    
\begin{figure}[htb]
     \centering
         \includegraphics[width=0.9\textwidth]{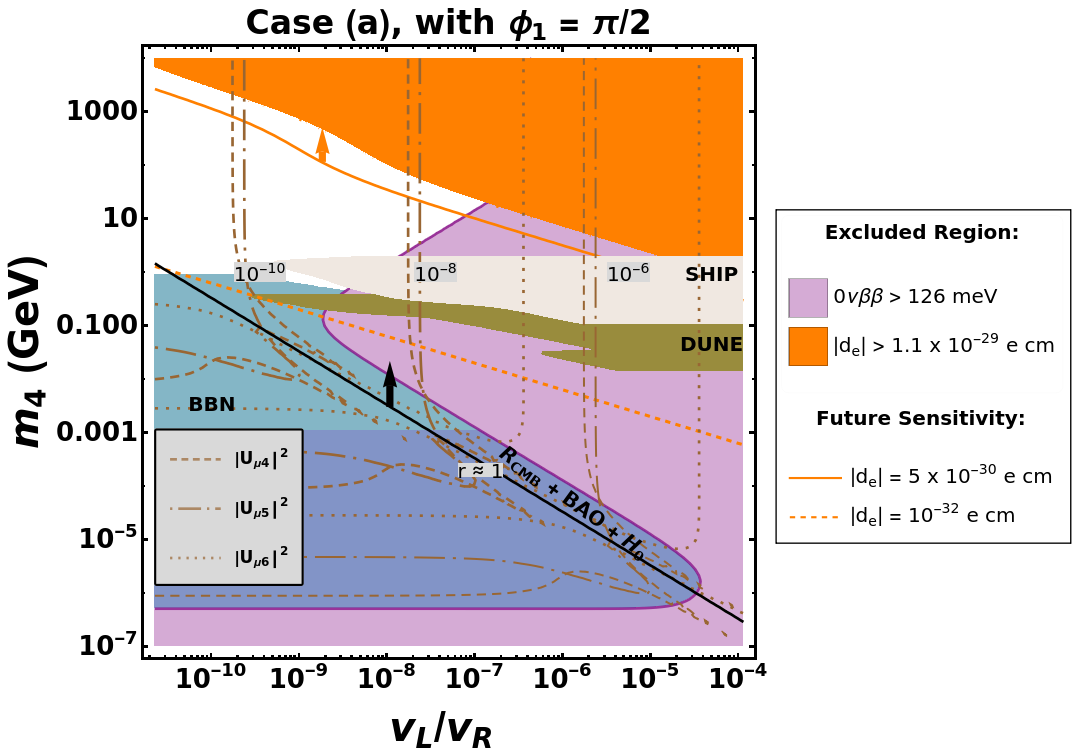}
        \caption{Same as figure~\ref{fig:casea} but with $\phi_1=\pi/2$.}
        \label{fig:casea-phi}
\end{figure}
    \item change the heavy neutrinos mass hierarchy: the
    results are basically unaffected by the assumed 
    heavy neutrinos mass hierarchy, 
    the only  feature worth mentioning is the fact that for non-degenerate heavy neutrinos and $U_L=\mathbf{I}$, in the tuned region the $U_{\alpha i}$ matrix is diagonal, i.e.,    we will only have $\vert U_{e4}\vert^2 =\vert U_{\mu 5}\vert^2=     \vert U_{\tau 6}\vert^2 =v_L/v_R$. See figure \ref{fig:finalplotdifHierarchies} in appendix \ref{app:Plots}.
    \item change the light neutrinos mass scale and mass ordering: this will only affect the mixing matrix 
    $U_{\alpha i}$ in the type-I seesaw region 
    and in the transition zone (for IO or quasi-degenerate we will see two or one cancellation lines). It is worth mentioning 
    that for NO as $ m_0 \to 0$  the $0\nu \beta \beta$ decay 
    exclusion region in the type I seesaw region disappears.
    See the figures in appendix~\ref{app:lightneutrinosplots}.
\end{enumerate}

\begin{figure}[htb]
    \centering
    \includegraphics[width=0.99\textwidth]{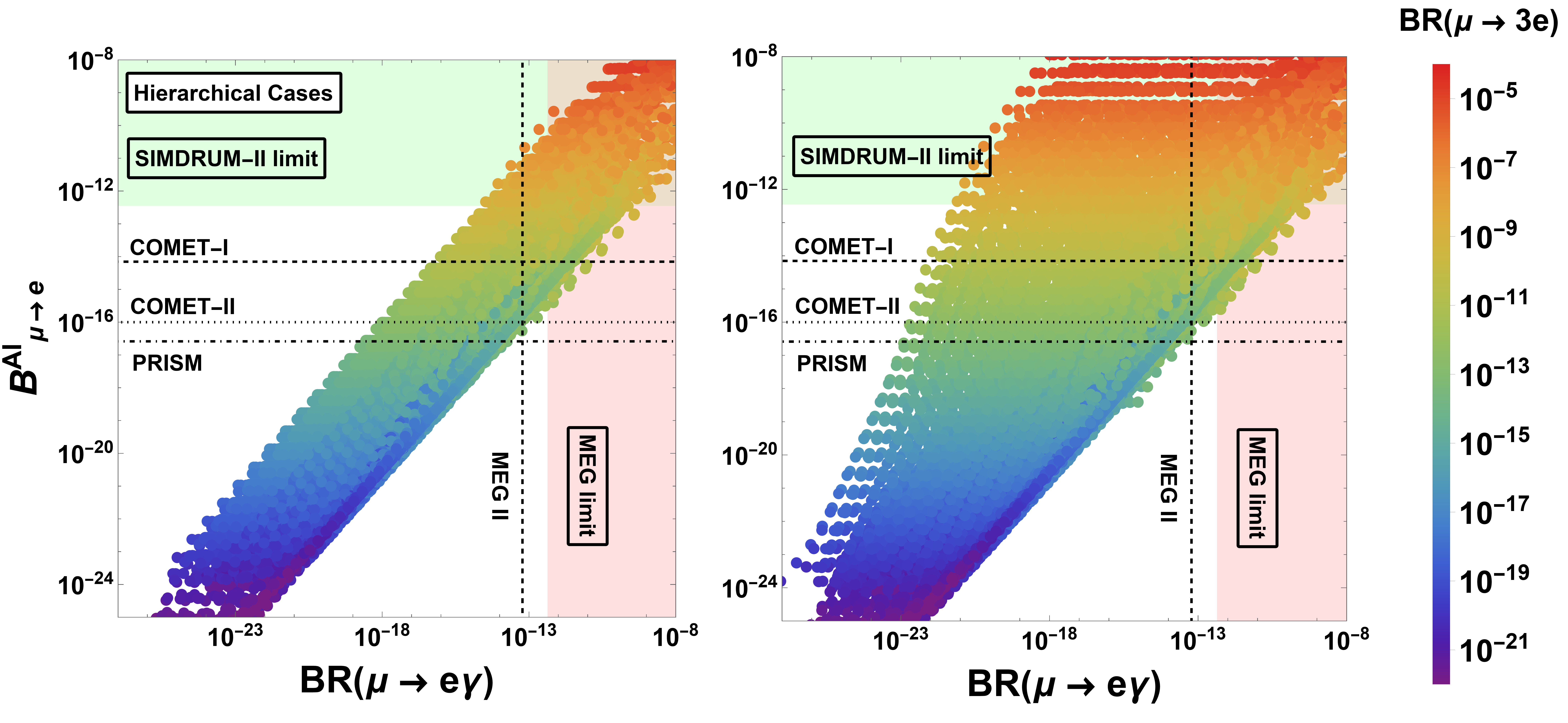}
    \caption{Correlation among the three LFV observables $\mu \to e \gamma$, $B^{\rm Al}_{\mu \to e}$ and $\mu \to 3e$ (color scale). 
    We also show the current limits and the sensitivities expected 
    for experiments in the near future. On the left panel 
    heavy neutrinos are hierarchical and on the right panel we have included also the cases when heavy neutrinos are degenerate.}
    \label{fig:LFVCorrelationPlots}
\end{figure}

In the auspicious occurrence of a positive observation of 
LFV in $\mu \to e \gamma$, $\mu \to 3 e$ or $B^{\rm Al}_{\mu \to e}$ in the near future we provide the correlations among these observables in the tuned region in figure~\ref{fig:LFVCorrelationPlots}.
We have constructed this plot by varying: 
$v_R = 11$ and  $44$ TeV, $\xi = (3 \times 10^{-8},3\times 10^{-7},  3\times 10^{-6})$,
$m_4 \in [1, 10^{4}]$ GeV, $v_L/v_R \in [10^{-10},10^{-4}]$, $(h_1/h_3$,$h_2/h_3)=\{(1,1),(0.9,0.95),(1/3,1/2),(1/5,1/3)\}$   
and $U_L=U_{\rm PMNS}$ and 4 other different random unitary matrices for $U_L$.
On the left panel we consider only the cases when heavy neutrinos are hierarchical with $(h_1/h_3$, $h_2/h_3)$ = $\{(0.9,0.95),$ $(1/3,1/2),(1/5,1/3)\}$ and on the right panel we consider all mentioned cases including degenerate heavy neutrinos when $(h_1/h_3$, $h_2/h_3) = (1,1)$.
The correlation between $\mu \to e \gamma$ and 
$B^{\rm Al}_{\mu \to e}$  decreases when 
heavy neutrinos are degenerate and the left-right mixing $\xi$ is smaller. In particular for $\xi= 3 \times 10^{-8}$ it is possible to have $B^{\rm Al}_{\mu \to e} \sim 10^{-15}$ while having  BR$(\mu \to e \gamma )\lesssim 10^{-23}$.

We see that we can have $\mu \to 3 e$ and $B^{\rm Al}_{\mu \to e}$ sizable, well within the reach of future experiments, while having $\mu \to e \gamma$ much bellow 
the future experimental sensitivity of MEG-II. This is even 
more so if heavy neutrinos are degenerate. The reason
being the fact that $\mu \to e \gamma$ depends only on the form factors $F_2$ and $G_2$ and these for the doubly scalar contributions are, at leading order, suppressed when $M_R$ is proportional to the identity (see, for instance,  case (c) in figure~\ref{fig:BRmegamma}). 
This is not true for the other processes.
But if MEG-II observes a signal, $\mu \to 3 e$ must be observed with BR$(\mu \to 3e) \lesssim 10^{-14}$. However, in this case COMET-I may not observe anything, while COMET-II must see something. On the other hand, if COMET-I (or COMET-II) observe a signal,  $\mu \to 3 e$ must be around the corner, then again,  in this case MEG-II may not observe anything.
Finally, if $\mu \to 3 e$ is observed it will tell us where to 
look for $\mu \to e \gamma$ and $B^{\rm Al}_{\mu \to e}$.

\section{Conclusions}
\label{sec:conclusions}

We have explored the neutrino sector of the MLRSM, in particular, in 
the tuned regime where type-I and type-II seesaw mechanisms supply similar contributions  to neutrino masses. 
We have discussed how to parametrize and reconstruct the 
neutrino mass matrix in the general case as well as under ${\cal C}$ and ${\cal P}$ discrete symmetries.

Imposing the model to be additionally 
invariant under  ${\cal C}$ discrete symmetry, we have reconstructed and diagonalized neutrino mass matrices,
subject to several assumptions on the heavy neutrino sector, which are  compatible with the requirements of the light neutrino sector:
the mass squared differences and mixing angles indicated by  
neutrino oscillation data, KATRIN limits and the cosmological 
bound on the sum of light neutrino masses.
The unknowns in the light neutrino sector ($m_0$, 
mass ordering, $\delta$, $\alpha_{1,2}$) do not alter our 
general conclusions.

Fixing the masses of the relevant scalars 
to the minimum value allowed  by data, i.e., $m_H=m_{A^0}=m_{H^+}=$ 
15 TeV, $m_{\delta^{+}_L}=m_{\delta^{++}_L}=780$ GeV and 
$m_{\delta^{++}_R}=660$ GeV, we have investigated 
for different choices of $U_L$, values of $v_R$, values 
of the left-right mixing $\xi$,
assumptions on the heavy neutrino mass hierarchy, and on the $\phi_i$ phases, the  constraints from
$0\nu \beta \beta$ decay, $\mu \to e \gamma$,
$\mu \to 3 e$, $B^{\rm Al}_{\mu \to e}$, $a_\mu$ and $\vert d_e \vert$.
For completeness, 
we have also included BBN and other cosmological limits in our study.

To facilitate the discussion, we have defined four benchmark cases (a)-(d), according to table~\ref{tab:benchmarkcases}.
As the $0\nu \beta \beta$ decay limit depends mostly on $v_R$ and $\xi$, we have fixed $v_R=44$ TeV and $\xi=3 \times 10^{-8}$ at the minimum and maximum values of these parameters, respectively, for which the  constraint from 
$0\nu \beta \beta$ decay exclude the least the parameter space of the model in the plane $v_L/v_R \times m_4$ in the tuned regime. 
Since the value of $U_L$, the heavy neutrino mass 
hierarchy and the $\phi_i$ phases are practically unimportant
here, we set $U_L=\mathbf{I}$, heavy neutrinos as degenerate and $\phi_i=0$ for the benchmark values of our standard case, case (a). In this situation there is no LFV or eEDM, so this particular scenario is the least restrictive for the neutrino sector of the model. Nevertheless, cosmological data and $0\nu \beta \beta$ decay exclude the entire type-I seesaw region and a substantial part of the tuned region for $v_L/v_R \gtrsim 10^{-11}$. For instance, 
if $m_4 =10$ TeV then $v_L/v_R \lesssim 2 \times 10^{-5}$, 
if $m_4 = 1$ GeV then $v_L/v_R \lesssim 2 \times 10^{-8}$.
If $\frac{\pi}{50}\lesssim \phi_1 \lesssim \frac{49\pi}{50}$, however, the current limit on  eEDM does not
allow for $v_L/v_R \gtrsim 10^{-7}$.

In general, we have shown that unless the charged lepton mixing matrix $U_L=\mathbf{I}$ and the phases $\phi_i$ are all zero, 
we expect to have sizable LFV 
and eEDM in the tuned region. This is exemplified by cases (b) 
and (c) where $U_L = U_{\rm PMNS}$, the first with mildly hierarchical, the second with degenerate neutrinos. We have observed that the exact mass hierarchy is not very significant 
for the final results. 
Here the non-observation of LFV processes further discard $m_4 \gtrsim 1$ TeV and imposes $v_L/v_R \lesssim 5 \times 10^{-6}$, with the perspective in the future to probe down to $m_4 \sim 100$ GeV independent of $v_L/v_R$. In these cases eEDM can only discard 
a smaller corner of the $v_L/v_R \times m_4$ parameter space, 
already ruled out by LFV and $0\nu \beta \beta$ decay. 
If one allows for higher values of $\xi$, as in case (d), 
eEDM becomes important and can exclude a part of the parameter 
space still allowed  by other types of data.
From the future proposed accelerator experiments only SHIP can probe a small region in 
$\vert U_{\mu i}\vert^2 \times m_4$ that is still allowed 
for $10^{-9}\lesssim v_L/v_R \lesssim 10^{-8}$ and $m_4 \sim $ few GeV.

Different random unitary $U_L$ matrices,  $v_R < 44$ TeV, $\xi > 3 \times 10^{-8}$,
$\phi_i \neq 0$, all tend to increase the regions 
affected by LFV and eEDM and as a consequence reduce 
the allowed tuned region. 
The contributions of MLRSM in the tuned region to $a_\mu$ 
are too small to be of any experimental consequence.
Future $\mu \to e \gamma$, $\mu \to 3e$, $B^{\rm Al}_{\mu \to e}$ and eEDM experiments will 
have the sensitivity to further probe the remaining 
parameter space in the tuned regime, in some scenarios, either 
making a discover or excluding almost the entire parameter space. 

As a side note, we have investigated the recent CDF II measurement of $m_W$ in view of the MLRSM and concluded that to be compatible with this measurement $m_{W'} \sim (5-6)$ TeV and $m_{\delta_L^{++}} \sim (0.9 -1.4)$ TeV. In this region the $0\nu \beta \beta$ decay bound  implies $m_4 \gtrsim 10$ GeV and $v_L \lesssim 4$ MeV.

Finally, we have shown that there  are interesting 
correlations among the three LFV observables $\mu \to e \gamma$,
$\mu \to 3e$ and $B^{\rm Al}_{\mu \to e}$ in the tuned regime
that can be used to help experiments to indirectly 
constrain other LFV observable in the MLRSM or even guide discovery.

\section*{Acknowledgement}
G.F.S.A. acknowledges financial support from Funda\c{c}\~ao de Amparo \`a Pesquisa do Estado de S\~ao Paulo (FAPESP) under contracts 2019/04837-9 and 2020/08096-0, 
RZF is partially supported by FAPESP and Conselho Nacional de Ci\^encia e Tecnologia (CNPq) and
L.P.S.L. is fully supported by Coordena\c{c}\~ao de Aperfei\c{c}oamento de Pessoal de N\'ivel Superior (CAPES).
C.S.F. acknowledges the support by grant 2019/11197-6 and 2022/00404-3 from FAPESP, and grant 301271/2019-4 from CNPq.

\appendix
\section{MLRSM Description}
\label{app:model}
The MLRSM considered here is based on the $SU(2)_L \times SU(2)_R \times U(1)_{B-L}$ gauge group, where the eletromagnetic  charge $Q_{\rm em}$ becomes
$$Q_{\rm em} = T_{3L} + T_{3R} + \frac{B-L}{2}\, ,$$
where $T_{3L,3R}$ are the third component of $SU(2)_{L,R}$ generators 
and $B$ and $L$ are, respectively, baryon and lepton number.

There are seven gauge fields $\mathbf W_{L,R}$ and $B$ associated with 
the gauge symmetry groups, the left-handed (right-handed) fermion fields transform as doublets of $SU(2)_L$ ($SU(2)_R$ ) and the scalar sector 
contains a bidoublet $\mathbf \Phi$, a $SU(2)_L$ triplet $\mathbf \Delta_{L}$ and a $SU(2)_R$ triplet $\mathbf \Delta_{R}$.  
The total Lagrangian of the model can be described as 
\be
{\cal L} = {\cal L}_{\rm c} + {\cal L}_{\rm G} + {\cal L}_{\rm Y} + {\cal L}_{\rm H}\, 
\label{eq:lagrangian}
\ee
where ${\cal L}_{\rm c}$ is the kinetic part that contains the 
gauge invariant interactions between leptons  and gauge bosons 
\be 
{\cal L}_{\rm c} \supset \sum_{j =e,\mu,\tau} (\bar L_{j L} i \slashed{D} L_{j L} +\bar L_{j R} i \slashed{D} L_{j R} ) \, ,
\label{eq:Leptonic}
\ee
where $D_\mu \equiv \displaystyle \partial_\mu - i g_L \, \frac{\pmb\sigma}{2} \cdot \mathbf{W}_{\mu L}- 
i g_R \, \frac{\pmb\sigma}{2} \cdot \mathbf{W}_{\mu R}- 
i \frac{g'}{2} B_\mu$ is the appropriate covariant derivative and $g_L$, $g_R$ and $g'$ are, respectively, 
the coupling constants of $SU(2)_L$, $SU(2)_R$ and $U(1)_{B-L}$. 
Here $\pmb\sigma \equiv (\sigma_1, \sigma_2, \sigma_3)$ are the Pauli matrices. Manifest left-right symmetry imply $g_R=g_L=g$.
${\cal L}_{\rm G}$ is the gauge boson Lagrangian
\be
{\cal L}_{\rm G} = -\frac{1}{4}  W^a_{\mu \nu L} W^{a\mu \nu}_L
-\frac{1}{4}  W^a_{\mu \nu R} W^{a\mu \nu}_R-\frac{1}{4}  B_{\mu \nu} B^{\mu \nu}\, ,
\label{eq:gauge}
\ee
where the field tensors $W_{\mu \nu L,R}$ and $B_{\mu \nu}$ are defined as 
$$W^a_{\mu \nu L,R} \equiv \partial_\mu W^a_{\nu L,R}- \partial_\nu W^a_{\mu L,R}+ \epsilon^{abc}\,  W^{b}_{\mu L,R}W^{c}_{\nu L,R}\, ,$$
with $\epsilon^{abc}$ the structure constants of $SU(2)$ and 
$$B_{\mu \nu}\equiv \partial_\mu B_\nu - \partial_\nu B_\mu.$$

The Yukawa interaction lagrangian ${\cal L}_Y$ is built by the most 
general possible couplings of the Higgs multiplets to bilinear fermion 
field products forming singlets under the symmetry group. This includes the lepton part 
\be
{\cal L}_{\rm Y} \supset 
-\overline{L}_{j L} \left( (h_l)_{j k} {\mathbf \Phi} + (\tilde{h}_l)_{j k} \tilde{\mathbf \Phi}\right) L_{k R}
- \overline{(L_{j R})^c} i\sigma_2{\mathbf \Delta}_R (\tilde{h}_M)_{j k} L_{k R} - \overline{(L_{j L})^c} i\sigma_2{\mathbf \Delta}_L (h_M)_{j k} L_{k L} + \rm h. c.  \, , 
\label{eq:yukawa}
\ee
where $\tilde{\mathbf \Phi} \equiv i \sigma_2 \mathbf{\Phi^*} \sigma_2 $ and 
$h_M$, $\tilde{h}_M$, $h_l$ and $\tilde{h}_l$ are $3\times 3$ matrices that 
mix lepton flavors $j,k = e, \mu,\tau$.

Finally, the Higgs lagrangian ${\cal L}_{\rm H}$ can be written as
\be 
{\cal L}_{\rm H} = {\rm Tr} \vert D_\mu \mathbf \Phi\vert^2 
+ {\rm Tr} \vert D_\mu \mathbf \Delta_L\vert^2 
+ {\rm Tr} \vert D_\mu \mathbf \Delta_R\vert^2 
-V_{\rm H}\, 
\label{eq:higgs}
\ee
where $V_{\rm H}$ is the scalar potential that can be found, for instance in Ref.~\cite{Zhang:2007da}. The kinetic terms of eq.~(\ref{eq:higgs}),  
after spontaneous symmetry breaking (SSB), give rise to the mass of the 
gauge bosons of the model. First $SU(2)_L \times SU(2)_R \times U(1)_{B-L}$
is broken down to $SU(2)_L \times U(1)_Y$ by the vev of $\mathbf \Delta_R$,  then electroweak symmetry breaking is induced by $\mathbf \Phi$.
The corresponding vacuum expectation values of the scalar fields are given by
\be
\langle {\mathbf \Phi} \rangle = \frac{1}{\sqrt{2}} 
\left( \begin{array}{cc} \kappa_1 & 0 \\
0 & \kappa_2 \end{array} \right) \, , \quad  \langle {\mathbf \Delta_{L,R}} \rangle = \frac{1}{\sqrt{2}} 
\left( \begin{array}{cc} 0 & 0 \\
v_{L,R} & 0 \end{array} \right) \, ,
\label{eq:vev}
\ee
where we are taking all the vevs to be positive real numbers here, this corresponds to no spontaneous CP violation.

After the charged and neutral gauge boson mass matrices diagonalization we obtain the  mass values at tree-level
\be
m^2_{W,W'} = \frac{g^2}{4} \left[\kappa^2_+ +v_R^2 +v_L^2\mp \sqrt{(v_R^2-v_L^2)^2 + 4 \kappa_1 \kappa_ 2}\, \right]\, ,
\label{eq:mw}
\ee
where $\kappa_+^2=\kappa_1^2 + \kappa_2^2$ and 
\be
m^2_{Z,Z'} = \frac{1}{4} \left\{ [g^2\kappa^2_+ +2 v_R^2(g^2+g'^{2})] \mp \sqrt{[g^2\kappa^2_+ + 2 v_R^2(g^2+g'^{2})]^2 -4 g^2(g^2+2g'^{2})\kappa^2_+ v^2_R} \, \right\} \, ,
\label{eq:mz}
\ee
and the photon remains massless.
In the limit $v_R \gg \kappa_+ \gg v_L$, 
$$m_W \simeq \frac{g \, \kappa_+}{2}\, , \quad m_{W'} \simeq \frac{g \,  v_R}{\sqrt{2}},$$
$$m_Z \simeq \frac{g \, \kappa_+}{2 \cos \theta_W}\, , \quad m_{Z'} \simeq \frac{g \,  v_R \cos \theta_W}{\sqrt{\cos 2\theta_W}},$$
where $g=e/ \sin \theta_W$ as in the SM and $\kappa_+ = v = 246$ GeV in order 
to reproduce the experimental values for the masses $m_W$ and $m_Z$.
In this limit we can also write
\be
\left( \begin{array}{c} W \\ W' \end{array} \right) \simeq 
\left( \begin{array}{cc} 1 &  \xi \\ -\xi & 1 \end{array} \right) \left( \begin{array}{c} W_L \\ W_R \end{array} \right) \, ,
\label{eq:wmixing}
\ee
with the mixing parameter $\displaystyle \xi \simeq  \frac{\kappa_1 \kappa_2}{v_R^2}=\lambda \,\sin 2 \beta $, where we have defined 
$\lambda \equiv \left(\frac{m_W}{m_{W'}}\right)^2$, $\sin \beta\equiv \frac{\kappa_2}{\sqrt{\kappa_1^2 +\kappa_2^2}}$ 
and $\cos \beta\equiv \frac{\kappa_1}{\sqrt{\kappa_1^2 +\kappa_2^2}}$ .

In the scalar sector, we have the CP even neutral scalars ($h$ is
the SM-like Higgs discovered at the LHC in 2012)
\begin{eqnarray}
 h & \simeq & \frac{\sqrt{2}}{\kappa_{+}}\textrm{Re}\left(\kappa_{1}\phi_{1}^{0}+\kappa_{2}\phi_{1}^{0*}\right),\\
H & \simeq & \frac{\sqrt{2}}{\kappa_{+}}\textrm{Re}\left(\kappa_{1}\phi_{2}^{0*}-\kappa_{2}\phi_{1}^{0}\right),\\
H_{L} & \simeq & \sqrt{2}\textrm{Re}\left(\delta_{L}^{0}\right),\\
 H_{R} & \simeq & \sqrt{2}\textrm{Re}\left(\delta_{R}^{0}\right),
\end{eqnarray}
with masses
\begin{eqnarray}
m_{h}^{2} & \simeq & 2\kappa_{+}^{2}\left[\lambda_{1}+2\lambda_{2}\epsilon+\left(2\lambda_{2}-\lambda_{3}\right)\epsilon^{2}\right],\\
m_{H}^{2} & \simeq & \frac{1}{2\sqrt{1-\epsilon^{2}}}\alpha_{3}v_{R}^{2},\\
m_{H_{L}}^{2} & \simeq & \frac{1}{2}\left(\rho_{3}-2\rho_{1}\right)v_{R}^{2},\\
m_{H_{R}}^{2} & \simeq & 2\rho_{1}v_{R}^{2},
\end{eqnarray}
where we have defined
\begin{eqnarray}
\epsilon & \equiv & \frac{2\kappa_{1}\kappa_{2}}{\kappa_{+}^{2}}.\label{eq:epsilon}
\end{eqnarray}

We also have the CP odd neutral scalars
\begin{eqnarray}
A^0 & \simeq & \frac{\sqrt{2}}{\kappa_{+}}\textrm{Re}\left(\kappa_{1}\phi_{2}^{0*}-\kappa_{2}\phi_{1}^{0}\right),\\
 A_{L} & \simeq & \sqrt{2}\textrm{Im}\left(\delta_{L}^{0}\right),
\end{eqnarray}
with masses
\begin{eqnarray}
m_{A^0}^{2} & \simeq & \frac{\kappa_{+}^{2}}{2\kappa_{-}^{2}}\alpha_{3}v_{R}^{2}-2\kappa_{+}^{2}\left(2\lambda_{2}-\lambda_{3}\right),\\
m_{A_{L}}^{2} & \simeq & \frac{1}{2}\left(\rho_{3}-2\rho_{1}\right)v_{R}^{2}.
\end{eqnarray}
where we have defined $\kappa_{-}^{2}\equiv\kappa_{2}^{2}-\kappa_{1}^{2}$.
For the charged scalars, we have
\begin{eqnarray}
 H^{+} & = & \frac{1}{\sqrt{1+\xi^{2}}}\left[\frac{\xi}{\kappa_{+}}\left(\kappa_{1}\phi_{1}^{+}+\kappa_{2}\phi_{2}^{+}\right)+\delta_{R}^{+}\right],
\end{eqnarray}
and the fields $\delta_{L}^{+}$, $\delta_{L}^{++}$ and $\delta_{R}^{++}$
with masses
\begin{eqnarray}
m_{H^{+}}^{2} & \simeq & \frac{1}{2}\frac{\kappa_{+}^{2}}{\kappa_{-}^{2}}\alpha_{3}v_{R}^{2}+\frac{1}{4}\alpha_{3}\kappa_{-}^{2},\\
m_{\delta_{L}^{+}}^{2} & \simeq & \frac{1}{2}\left(\rho_{3}-2\rho_{1}\right)v_{R}^{2}+\frac{1}{4}\alpha_{3}\kappa_{-}^{2},\\
m_{\delta_{L}^{++}}^{2} & \simeq & \frac{1}{2}\left(\rho_{3}-2\rho_{1}\right)v_{R}^{2}+\frac{1}{2}\alpha_{3}\kappa_{-}^{2},\\
m_{\delta_{R}^{++}}^{2} & \simeq & 2\rho_{2}v_{R}^{2}+\frac{1}{2}\alpha_{3}\kappa_{-}^{2}.
\end{eqnarray}

The dependence on $\beta$ in our study appears in the $W-W'$ mixing parameter $\xi$. Even for the largest $\xi$ that we consider in case (d), we still keep $\sin\beta \ll 1$, $\epsilon\ll 1$ and hence $\kappa_{-}^{2}\simeq\kappa_{+}^{2}$. See ref. \cite{Maiezza:2016ybz} for the effect of larger $\beta$ on the running of the SM-like Higgs quartic coupling.

\begin{table}[htb]
\setlength{\tabcolsep}{12pt}
\renewcommand{\arraystretch}{1.3}
  \begin{center}
    \begin{tabular}{|c|c||c|c|c|}
      \hline
      Fields & Content & $SU(2)_L$ & $SU(2)_R$ & $U(1)_{B-L}$\\
\hline
$L_{eL},L_{\mu L},L_{\tau L}$ & $\left( \begin{array}{c} \nu_e \\ e\end{array}\right)_L$,\, $\left( \begin{array}{c} \nu_\mu \\ \mu\end{array}\right)_L$,\, $\left( \begin{array}{c} \nu_\tau \\ \tau\end{array}\right)_L$ & 2 & 1 & -1 \\
$L_{eR},L_{\mu R},L_{\tau R}$ & $\left( \begin{array}{c} \nu_e \\ e\end{array}\right)_R$,\, $\left( \begin{array}{c} \nu_\mu \\ \mu\end{array}\right)_R$,\, $\left( \begin{array}{c} \nu_\tau \\ \tau\end{array}\right)_R$ & 1 & 2 & -1 \\

$Q_{uL},Q_{c L},Q_{t L}$ & $\left( \begin{array}{c} u \\ d\end{array}\right)_L$,\, $\left( \begin{array}{c} c \\ s\end{array}\right)_L$,\, $\left( \begin{array}{c} t \\ b \end{array}\right)_L$ & 2 & 1 & 1/3 \\

$Q_{uR},Q_{c R},Q_{t R}$ & $\left( \begin{array}{c} u \\ d\end{array}\right)_R$,\, $\left( \begin{array}{c} c \\ s\end{array}\right)_R$,\, $\left( \begin{array}{c} t \\ b \end{array}\right)_R$ & 2 & 1 & 1/3 \\
\hline
$\mathbf W_L$ & $W^{\pm}_L,W^3_L$ & 3 & 1 & 0 \\
$\mathbf W_R$ & $W^{\pm}_R,W^3_R$ & 1 & 3 & 0 \\
$B$ & $B$ & 1 & 1 & 0\\
\hline
$\mathbf \Phi$ & $\left( \begin{array}{cc} \phi_1^0 & \phi_1^{+} \\ \phi_2^- & \phi_2^0 \end{array}\right)$ & 2 & 2 & 0\\
$\mathbf \Delta_{L}$ & $\left( \begin{array}{cc} \delta_L^+/\sqrt{2} & \delta_L^{++} \\ \delta_L^0 & -\delta_L^+/\sqrt{2}\end{array}\right)$ & 3 & 1 & 2 \\
$\mathbf \Delta_{R}$  & $\left( \begin{array}{cc} \delta_R^+/\sqrt{2} & \delta_R^{++} \\ \delta_R^0 & -\delta_R^+/\sqrt{2}\end{array}\right)$ & 1 & 3 & 2 \\
\hline
\end{tabular}
    \end{center}
  \caption{Field content and quantum numbers for the MLRSM.}
    \label{tab:model}
  \end{table}

\section{Form Factors}
\label{app:formfac}

\begin{figure}
  \centering
  \begin{tikzpicture}
    \begin{feynman}
      \vertex[blob] (a2) at (0, 0) {};
      \vertex (a1) at (-2,-1) {$l$};
      \vertex (a3) at (2,-1) {$l'$};
      \vertex (b1) at (1, 2) {$\gamma$};
      ;
      \diagram* {
        (a1) -- [fermion] (a2) -- [fermion] (a3),
        (a2) -- [photon,momentum'=\(q\)] (b1),
      };
    \end{feynman}
  \end{tikzpicture}
  \caption{Generic vertex  for $l \rightarrow l'\, \gamma$ interaction.}
  \label{fig:PhotonVertex}
\end{figure}
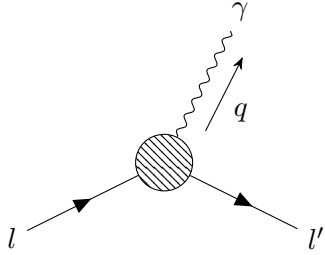

The most general amplitude for a vertex like the one shown in Fig. \ref{fig:PhotonVertex} can be written as $\mathcal{M}^\mu = -e\: \overline{u}(p_2)\Gamma^\mu(q^2) u(p_1)$, with $q\equiv p_1 - p_2$, $p_1$ and $p_2$ being the initial and final fermion momentum, respectively, and where $\Gamma^\mu$ is given by
\be 
\Gamma^\mu(q^2) = \left(F_1(q^2) + G_1(q^2) \gamma^5\right)\left(\gamma^\mu - \frac{\slashed{q}q^\mu}{q^2}\right) + \frac{i\sigma^{\mu \nu} q_\nu}{m_l + m_{l'}}\left(F_2(q^2) + G_2(q^2) \gamma^5\right).
\label{eq:GeneralAmplitude}
\ee

From gauge invariance $F_1(0) = G_1(0) = 0$, so in $l\rightarrow l' \gamma$ decay these form factors do not contribute. During numerical computations we checked explicitly that these conditions were satisfied.

In order to compute $\mu \rightarrow 3e$ contributions through photon exchange, the same vertex appears so we can write the result again as a function of the presented form factors. Differently from $\mu \to e \gamma$, in this case we do not have $q^2 = 0$ but since the momentum exchange is low, the form factors are well described by the $q^2 \rightarrow 0$ limit. As $F_1(q^2),G_1(q^2) \xrightarrow[]{q^2\rightarrow 0} 0$, we can approximate them as 
\be 
F_1(q^2) \approx q^2 \frac{dF_1(0)}{d q^2}\, ,\quad\quad\quad G_1(q^2) \approx q^2 \frac{dG_1(0)}{d q^2}.
\label{eq:F1_G1_small_q}
\ee
Finally, the amplitude of the process $\mu \rightarrow e + \gamma \rightarrow 3e$ can be written in general as
\be
\mathcal{M} = -e\,\left\{\overline{u}(p_2) \Gamma^\mu(q^2)u(p_1)\left(\frac{-i}{q^2}\right) \overline{u}(p_3) \left( -i e \gamma_\mu \right)v(p_4) - (p_2 \leftrightarrow p_3) \right\}.
\ee

The form factors $F_2(0)$ and $G_2(0)$ can
be written as
\be
 F^S_2(0) = \frac{1}{(4\pi)^2} \frac{m^2_\ell}{m^2_S}\,  f^{\ell \ell'}_{S}(a_{S1}^2,a_{S2}^2)\, , 
\label{eq:F2}
\ee
and
\be
 G^S_2(0) = \frac{1}{(4\pi)^2} \frac{m^2_\ell}{m^2_S} \, g^{\ell \ell'}_{S}(a_{S1},a_{S2})\, , 
\label{eq:G2}
\ee
where $f_S^{\ell \ell'}$ and $g_S^{\ell'}$ are functions that depend on the mass of the particles in the corresponding loop diagrams as well as $a_{S1}$ and $a_{S2}$ their two types of couplings to fermions.

While in the numerical evaluations, we use the full expressions for $F_2$ and $G_2$ as given in eqs.~\eqref{eq:F2} and \eqref{eq:G2}, 
here we give the approximate expressions for $f_{S}^{\ell \ell'}$ and $g_{S}^{\ell \ell'}$ for the contributions involving heavy scalars in the loop up to to leading terms in the expansion $m_{\ell}/m_S \ll 1$ and for both regimes $m_{i}/m_S \ll 1$ and $m_{i}/m_S \gg 1$ where $m_{\ell(i)}$ is the charged lepton (neutrino) mass and $m_S$ is the corresponding scalar mass in the loop (the latter case can only happen for $i=4,5,6$). For the charged gauge boson contribution $W'$ ($W$), 
we also show both regimes $m_{i}/m_{W,W'} \ll 1$ and $m_{i}/m_{W,W'} \gg 1$.
The scalars (vector bosons) couplings to fermions are, respectively, the scalar $a_{Ss}$ (vector $a_{Sv}$) and pseudo-scalar $a_{Sp}$ (axial-vector $a_{Sa}$).
We also show the expressions for $F_1$, $G_1$ and $d_\ell$
in the same limits.

\begin{itemize}
    \item Neutral scalars 
    \begin{equation}
        \mathcal{L} \supset \overline{l_\ell}(a_{Ss}^{\ell \ell'}+i a_{Sp}^{\ell \ell'}\gamma_5) l_{\ell'} S
    \end{equation}
    
    \begin{align}
    f_{S}^{\ell \ell'} &\approx \sum_{\ell'' = e, \mu, \tau} a_{Ss}^{\ell \ell''}a_{Ss}^{\ell' \ell''} \left[ \frac{1}{6}-\frac{m_{\ell''}}{m_\ell}\left(\frac{3}{2}+ \ln{\frac{m_{\ell''}^2}{m_S^2}}\right)\right] \nonumber\\
    &+ \sum_{\ell'' = e, \mu, \tau} a_{Sp}^{\ell \ell''}a_{Sp}^{\ell' \ell''} \left[ \frac{1}{6}+\frac{m_{\ell''}}{m_\ell}\left(\frac{3}{2}+ \ln{\frac{m_{\ell''}^2}{m_S^2}}\right)\right], \\
    g_{S}^{\ell \ell'} &\approx \sum_{\ell'' = e, \mu, \tau} i a_{Ss}^{\ell \ell''}a_{Sp}^{\ell' \ell''} \left[ \frac{1}{6}-\frac{m_{\ell''}}{m_\ell}\left(\frac{3}{2}+ \ln{\frac{m_{\ell''}^2}{m_S^2}}\right)\right] \nonumber\\
    &- \sum_{\ell'' = e, \mu, \tau} i a_{Sp}^{\ell \ell''}a_{Ss}^{\ell' \ell''} \left[ \frac{1}{6}+\frac{m_{\ell''}}{m_\ell}\left(\frac{3}{2}+ \ln{\frac{m_{\ell''}^2}{m_S^2}}\right)\right],
    \end{align}
    
    \begin{align}
    F_{1}^{S}(0) &\approx - \dfrac{q^2}{72 \pi^2 m_S^2} \sum_{\ell'' = e, \mu, \tau} (a_{Ss}^{\ell \ell''}a_{Ss}^{\ell' \ell''} + a_{Sp}^{\ell \ell''}a_{Sp}^{\ell' \ell''}) \left[ \frac{3}{2}\ln{\frac{m_{\ell''}^2}{m_S^2}} +2\right], \\
    G_{1}^{S}(0) &\approx - i\dfrac{q^2}{72 \pi^2 m_S^2} \sum_{\ell'' = e, \mu, \tau} (a_{Sp}^{\ell \ell''}a_{Ss}^{\ell' \ell''} - a_{Ss}^{\ell \ell''}a_{Sp}^{\ell' \ell''}) \left[ \frac{3}{2}\ln{\frac{m_{\ell''}^2}{m_S^2}} +2\right],
    \end{align}
    
    \begin{equation}
        d_\ell^S = - \frac{1}{16 \pi^2 m_S^2} \sum_{\ell'' = e, \mu, \tau} {\rm Re}\left[a_{Ss}^{\ell \ell''}a_{Sp}^{\ell \ell''}\right] m_{\ell''} \left[3+2 \ln\frac{m_{\ell''}^2}{m_S^2}\right]\, ,
    \end{equation}
    with the matrices given by
    \begin{eqnarray}
    a_{H s} &=& -\frac{\kappa_+}{2 \kappa^2_{-}} \left( 
    {\cal U}_L\, \hat{M}_\nu \, {\cal U}_R^\dagger +  {\cal U}_R\,  \hat{M}_\nu \, {\cal U}_L^\dagger \right)+ \frac{2 \kappa_1 \kappa_2}{\kappa^2_{-}\kappa_+}\, \hat{M}_{\ell} \, ,\\
     a_{H p} &=& i \frac{\kappa_+}{2 \kappa^2_{-}} \left( 
    {\cal U}_L\, \hat{M}_\nu \, {\cal U}_R^\dagger -  {\cal U}_R\,  \hat{M}_\nu \, {\cal U}_L^\dagger \right)\, ,\\
     a_{A^0 s} &=&  a_{H p}\, , \\
     a_{A^0 p} &=& -  a_{H s}\, , \\
      a_{h s} &=& -\frac{\hat{M}_\ell}{\kappa_+}\, , \\
     a_{h p} &=& 0\, . \\
    \end{eqnarray}

    \item Single-charged scalars 
    \begin{equation}
        \mathcal{L} \supset \overline{\hat\nu_i}(a_{Ss}^{i\ell}+ a_{Sp}^{i\ell}\gamma_5) l_\ell S^+
    \end{equation}
    
   \subitem(i) $m_S \gg m_i$ 
   \begin{eqnarray}   
    f_{S}^{\ell \ell'} &\approx & \sum_{i=1,\dots,6}- \frac{1}{12}\left[
    a_{Ss}^{i\ell}a_{Ss}^{i\ell' *} \left(1 + 6  \, \frac{m_i}{m_\ell}\right)\right.
    + \left . a_{Sp}^{i\ell }a_{Sp}^{i\ell' *} \left(1 - 6 \, \frac{m_i}{m_\ell}\right)
    \right] \, ,\\
   g_{S}^{\ell \ell'} &\approx & \sum_{i=1,\dots,6} \frac{1}{12} 
   \left[
    a_{Sp}^{i\ell }a_{Ss}^{i\ell'*} \left(1-6 \,   \frac{m_i}{m_\ell}\right)\right . 
    + \left . a_{Ss}^{i\ell }a_{Sp}^{i\ell'*} \left(1 + 6 \,  \frac{m_i}{m_\ell}\right)
    \right]\, ,
    \end{eqnarray}
    \begin{align}
    F_{1}^{S}(0) &\approx \dfrac{q^2}{288 \pi^2 m_S^2} \sum_{i=1,\dots,6} (a_{Ss}^{i \ell}a_{Ss}^{i \ell'*} + a_{Sp}^{i \ell }a_{Sp}^{i \ell'*}) , \\
    G_{1}^{S}(0) &\approx \dfrac{q^2}{288 \pi^2 m_S^2} \sum_{i=1,\dots,6} (a_{Sp}^{i \ell}a_{Ss}^{i\ell'*} + a_{Ss}^{i\ell}a_{Sp}^{i\ell'*}),
    \end{align}
    \begin{equation}
        d_\ell^S = \frac{1}{16 \pi^2 m_S^2} \sum_{i=1,\dots,6} {\rm Im}\left[a_{Ss}^{i\ell}a_{Sp}^{i\ell*}\right] m_i\, ,
    \end{equation}
    
    \subitem(ii) $m_i \gg m_S$ for $i=4,5,6$ 
   \begin{eqnarray}   
    f_{S}^{\ell \ell'} &\approx & \sum_{i=1,2,3}- \frac{1}{12}\left[
    a_{Ss}^{i\ell}a_{Ss}^{i\ell' *} \left(1 + 6  \, \frac{m_i}{m_\ell}\right)\right.
    + \left . a_{Sp}^{i\ell }a_{Sp}^{i\ell' *} \left(1 - 6 \, \frac{m_i}{m_\ell}\right)
    \right] \, \nonumber \\
    &-& \sum_{i=4,5,6} \frac{m_S^2}{6 \: m_i^2} \left[a_{Ss}^{i\ell }a_{Ss}^{i\ell' *}\left(1 + 3 \, \frac{m_i}{m_\ell}\right) + a_{Sp}^{i\ell }a_{Sp}^{i\ell' *} \left(1 - 3 \, \frac{m_i}{m_\ell}\right) \right]  ,\\
   g_{S}^{\ell \ell'} &\approx & \sum_{i=1,2,3} \frac{1}{12} 
   \left[
    a_{Sp}^{i\ell }a_{Ss}^{i\ell'*} \left(1-6 \,  
    \frac{m_i}{m_\ell}\right)\right . 
    + \left . a_{Ss}^{i\ell }a_{Sp}^{i\ell'*} \left(1 + 6 \,  \frac{m_i}{m_\ell}\right)
    \right]\, \nonumber \\
    &+& \sum_{i=4,5,6} \frac{m_S^2}{6 \: m_i^2} \left[a_{Sp}^{i\ell }a_{Ss}^{i\ell' *}\left(1 - 3 \, \frac{m_i}{m_\ell}\right) + a_{Ss}^{i\ell }a_{Sp}^{i\ell' *} \left(1 + 3 \, \frac{m_i}{m_\ell}\right) \right]
    \end{eqnarray}
    \begin{align}
    F_{1}^{S}(0) &\approx \dfrac{q^2}{288 \pi^2 m_S^2} \sum_{i=1,2,3} (a_{Ss}^{i \ell}a_{Ss}^{i \ell'*} + a_{Sp}^{i \ell }a_{Sp}^{i \ell'*})  \nonumber \\
    &+ \sum_{i=4,5,6} \dfrac{q^2}{576 \pi^2 m_i^2} (a_{Ss}^{i \ell}a_{Ss}^{i \ell'*} + a_{Sp}^{i \ell }a_{Sp}^{i \ell'*}) \left(6 \ln \frac{m_i^2}{m_S^2} - 11\right) , \\
    G_{1}^{S}(0) &\approx \dfrac{q^2}{288 \pi^2 m_S^2} \sum_{i=1,2,3} (a_{Sp}^{i \ell}a_{Ss}^{i\ell'*} + a_{Ss}^{i\ell}a_{Sp}^{i\ell'*})\nonumber \\
    &+ \sum_{i=4,5,6} \dfrac{q^2}{576 \pi^2 m_i^2} (a_{Sp}^{i \ell}a_{Ss}^{i\ell'*} + a_{Ss}^{i\ell}a_{Sp}^{i\ell'*}) \left(6 \ln \frac{m_i^2}{m_S^2} - 11\right) ,
    \end{align}
    \begin{equation}
        d_\ell^S = \frac{1}{16 \pi^2} \left( \sum_{i=1,2,3} {\rm Im}\left[a_{Ss}^{i\ell}a_{Sp}^{i\ell*}\right] \frac{m_i}{m_S^2}\  + \sum_{i=4,5,6} {\rm Im}\left[a_{Ss}^{i\ell}a_{Sp}^{i\ell*}\right] \frac{1}{m_i} \right),
    \end{equation}
    with the matrices given by
    \begin{eqnarray}
    a_{\delta_L^+s} &=& \frac{1}{2 v_R} \left( {\cal U}_L^TU_L^*U_R^T{\cal U}_R^*\hat{M}_\nu {\cal U}_R^\dagger U_R U_L^\dagger\right)\, ,\\
     a_{\delta_L^+p} &=& - a_{\delta_L^+s} \, ,\\
     a_{H^+ s} &=&- \frac{\zeta^2}{2 v_R\sqrt{1+\zeta^2}}\left( 
     {\cal U}_L^\dagger  {\cal U}_L \hat{M}_\nu {\cal U}_R^\dagger - \frac{2 \kappa_1 \kappa_2}{\kappa^2_+} {\cal U}_L^\dagger \hat M_\ell - {\cal U}_R^\dagger \hat M_\ell \right .\nonumber \\
     &+& \left .\frac{2 \kappa_1 \kappa_2}{\kappa^2_+} 
     {\cal U}_R^\dagger {\cal U}_R\hat{M}_\nu {\cal U}_L^\dagger - \frac{1}{\zeta^2}{\cal U}_R^T{\cal U}_R^* \hat{M}_\nu {\cal U}_R^\dagger\right)\, ,\label{eq:coupHpS}\\
        a_{H^+ p} &=& -\frac{\zeta^2}{2 v_R\sqrt{1+\zeta^2}}\left( 
     {\cal U}_L^\dagger  {\cal U}_L \hat{M}_\nu {\cal U}_R^\dagger - \frac{2 \kappa_1 \kappa_2}{\kappa^2_+} {\cal U}_L^\dagger \hat M_\ell + {\cal U}_R^\dagger \hat M_\ell \right .\nonumber \\
     &-& \left .\frac{2 \kappa_1 \kappa_2}{\kappa^2_+} 
     {\cal U}_R^\dagger {\cal U}_R\hat{M}_\nu {\cal U}_L^\dagger - \frac{1}{\zeta^2}{\cal U}_R^T{\cal U}_R^* \hat{M}_\nu {\cal U}_R^\dagger\right)\, ,\label{eq:coupHpP}
    \end{eqnarray}
     where  $\zeta\equiv \sqrt{2}\, \displaystyle \frac{\kappa_{+}v_R}{\kappa_{-}^2}$.
     
     \item Double-charged scalars 
     \begin{equation}
        \mathcal{L} \supset \overline{l_\ell^c}(a_{Ss}^{\ell \ell'}+ a_{Sp}^{\ell \ell'}\gamma_5) l_{\ell'} S^{++}
    \end{equation}

    \begin{eqnarray}   
    f_{S}^{\ell \ell'} &\approx &\sum_{\ell'' = e, \mu, \tau} -\frac{2}{3} \left[
    a_{Ss}^{\ell \ell''}a_{Ss}^{\ell' \ell''*} \left(
    2-3 \frac{m_{\ell''}}{m_\ell} -6 \, \frac{m_{\ell''}}{m_\ell} \ln{\frac{m_{\ell''}^2}{m_S^2}} \right)\right . \nonumber \\
    &+& \left . a_{Sp}^{\ell \ell''}a_{Sp}^{\ell' \ell''*} 
     \left(
    2+3 \frac{m_{\ell''}}{m_\ell} +6 \, \frac{m_{\ell''}}{m_\ell} \ln{\frac{m_{\ell''}^2}{m_S^2}}\right)
    \right] \, ,\\
   g_{S}^{\ell \ell'} &\approx & \sum_{\ell'' = e, \mu, \tau}\frac{2}{3} \left[
    a_{Ss}^{\ell {\ell''}}a_{Sp}^{\ell' \ell''*} \left(
    2-3 \frac{m_{\ell''}}{m_\ell} -6 \, \frac{m_{\ell''}}{m_\ell} \ln{\frac{m_{\ell''}^2}{m_S^2}} \right)\right . \nonumber \\
    &+& \left . a_{Sp}^{\ell \ell''}a_{Ss}^{\ell' \ell''*} 
     \left(
    2+3 \frac{m_{\ell''}}{m_\ell} +6 \, \frac{m_{\ell''}}{m_\ell} \ln{\frac{m_{\ell''}^2}{m_S^2}}) \right)
    \right] \, ,
    \end{eqnarray}
    \begin{align}
    F_{1}^{S}(0) &\approx \dfrac{q^2}{144 \pi^2 m_S^2} \sum_{\ell'' = e, \mu, \tau} (a_{Ss}^{\ell \ell''}a_{Ss}^{\ell' \ell''*} + a_{Sp}^{\ell \ell''}a_{Sp}^{\ell' \ell''*}) \left[ 3\ln{\frac{m_{\ell''}^2}{m_S^2}} +5\right], \\
    G_{1}^{S}(0) &\approx \dfrac{q^2}{144 \pi^2 m_S^2} \sum_{\ell'' = e, \mu, \tau} (a_{Sp}^{\ell \ell''}a_{Ss}^{\ell' \ell''*} + a_{Ss}^{\ell \ell''}a_{Sp}^{\ell' \ell''*}) \left[ 3\ln{\frac{m_{\ell''}^2}{m_S^2}} +5\right],
    \end{align}
    
    \begin{equation}
        d_\ell^S = \frac{1}{16 \pi^2 m_S^2} \sum_{\ell'' = e, \mu, \tau} {\rm Im}\left[a_{Sp}^{\ell \ell''}a_{Ss}^{\ell \ell''*}\right] m_{\ell''} \left[1+2 \ln\frac{m_{\ell''}^2}{m_S^2}\right] \, ,
    \end{equation}
    with the matrices given by
    \begin{eqnarray}\label{eq:dcs1}
    a_{\delta_{R}^{++}s} &=& \frac{1}{2\sqrt{2}\, v_R} \, {\cal U}_R^* \hat M_\nu {\cal U}_R^\dagger\, ,\\
     a_{\delta_{R}^{++}p} &=& a_{\delta_{R}^{++}s}\, , \\
  a_{\delta_{L}^{++}s} &=& \frac{1}{2\sqrt{2}\, v_R}\,  
U_L^*U_R^T {\cal U}_R^* \hat M_\nu {\cal U}_R^\dagger U_{R}U_L^\dagger  \,,\\   
 a_{\delta_{L}^{++}p} &=& -  a_{\delta_{L}^{++}s} \, .
 \label{eq:dcs2}
    \end{eqnarray}
    
\item Charged gauge bosons

\begin{equation}
        \mathcal{L} \supset \overline{\hat\nu_i}\,\gamma_\mu (a_{Sv}^{i\ell}+ a_{Sa}^{i\ell}\gamma_5) l_\ell \, S_\mu^+
\end{equation}

\subitem(i) $m_S \gg m_i$
\begin{eqnarray}
f_S^{\ell \ell'}& \approx& \sum_{i = 1, \dots, 6}\frac{1}{6} \left[ a_{Sa}^{i\ell}a_{Sa}^{i\ell'*} \left( 5 + 12\frac{m_i}{m_\ell} \right) \right .
+  \left . a_{Sv}^{i\ell }a_{Sv}^{i\ell'*} \left( 5 - 12\frac{m_i}{m_\ell} \right)
\right] \, , \\
g_S^{\ell \ell'}& \approx&\sum_{i=1,\dots, 6} -\frac{1}{6}
\left[ a_{Sa}^{i\ell }a_{Sv}^{i\ell'*} \left( 5 + 12\frac{m_i}{m_\ell} \right) \right .
+  \left . a_{Sv}^{i\ell }a_{Sa}^{i\ell'*} \left( 5 - 12\frac{m_i}{m_\ell} \right)
\right] \, , 
\end{eqnarray}

\begin{align}
    F_{1}^{S}(0) &\approx -\dfrac{5\, q^2}{144 \pi^2 m_S^2} \sum_{i=1,\dots, 6} (a_{Sa}^{i \ell}a_{Sa}^{i \ell'*} + a_{Sv}^{i \ell}a_{Sv}^{i \ell'*}), \\
    G_{1}^{S}(0) &\approx -\dfrac{5\, q^2}{144 \pi^2 m_S^2} \sum_{i=1,\dots, 6} (a_{Sv}^{i \ell}a_{Sa}^{i \ell'*} + a_{Sa}^{i \ell}a_{Sv}^{i \ell'*}) ,
\end{align}
    
\begin{equation}
        d_\ell^S = \frac{1}{4 \pi^2 m_S^2} \sum_{i=1,\dots,6} {\rm Im}\left[a_{Sv}^{i\ell}a_{Sa}^{i\ell*}\right] m_i\, ,
\end{equation}

\subitem(ii) $m_i \gg m_S$ for $i=4,5,6$
\begin{eqnarray}
f_S^{\ell \ell'}& \approx& \sum_{i = 1,2,3}\frac{1}{6} \left[ a_{Sa}^{i\ell}a_{Sa}^{i\ell'*} \left( 5 + 12\frac{m_i}{m_\ell} \right) \right .
+  \left . a_{Sv}^{i\ell }a_{Sv}^{i\ell'*} \left( 5 - 12\frac{m_i}{m_\ell} \right)
\right]  \nonumber \\
& +& \sum_{i = 4,5,6}\frac{1}{6} \left[ a_{Sa}^{i\ell}a_{Sa}^{i\ell'*} \left( 2 + 3\frac{m_i}{m_\ell} \right) \right .
+  \left . a_{Sv}^{i\ell }a_{Sv}^{i\ell'*} \left( 2 - 3\frac{m_i}{m_\ell} \right)
\right] \, , \\
g_S^{\ell \ell'}& \approx&-\sum_{i=1,2,3} \frac{1}{6}
\left[ a_{Sa}^{i\ell }a_{Sv}^{i\ell'*} \left( 5 + 12\frac{m_i}{m_\ell} \right) \right .
+  \left . a_{Sv}^{i\ell }a_{Sa}^{i\ell'*} \left( 5 - 12\frac{m_i}{m_\ell} \right)
\right] \, \nonumber \\
& -& \sum_{i = 4,5,6}\frac{1}{6} \left[ a_{Sa}^{i\ell}a_{Sv}^{i\ell'*} \left( 2 + 3\frac{m_i}{m_\ell} \right) \right .
+  \left . a_{Sv}^{i\ell }a_{Sa}^{i\ell'*} \left( 2 - 3\frac{m_i}{m_\ell} \right)
\right] \, ,
\end{eqnarray}
\begin{align}
    F_{1}^{S}(0) &\approx -\dfrac{5\, q^2}{144 \pi^2 m_S^2} \sum_{i=1,2,3} (a_{Sa}^{i \ell}a_{Sa}^{i \ell'*} + a_{Sv}^{i \ell}a_{Sv}^{i \ell'*}) \nonumber \\
    &+\dfrac{q^2}{576 \pi^2 m_i^2} \sum_{i=4,5,6} (a_{Sa}^{i \ell}a_{Sa}^{i \ell'*} + a_{Sv}^{i \ell}a_{Sv}^{i \ell'*}) \left[6 \ln \frac{m_i^2}{m_S^2} + 1 \right] \, ,\\
    G_{1}^{S}(0) &\approx -\dfrac{5\, q^2}{144 \pi^2 m_S^2} \sum_{i=1,2,3} (a_{Sv}^{i \ell}a_{Sa}^{i \ell'*} + a_{Sa}^{i \ell}a_{Sv}^{i \ell'*}) \nonumber \\
    &+\dfrac{q^2}{576 \pi^2 m_S^2} \sum_{i=4,5,6} (a_{Sv}^{i \ell}a_{Sa}^{i \ell'*} + a_{Sa}^{i \ell}a_{Sv}^{i \ell'*}) \left[6 \ln \frac{m_i^2}{m_S^2} + 1 \right] \, ,
\end{align}
\begin{equation}
    d_\ell^S = \frac{1}{4 \pi^2 m_S^2}\left( \sum_{i=1,2,3} {\rm Im}\left[a_{Sv}^{i\ell}a_{Sa}^{i\ell*}\right] m_i\, + \frac{1}{4} \sum_{i=4,5,6} {\rm Im}\left[a_{Sv}^{i\ell}a_{Sa}^{i\ell*}\right] \left(m_i+6\frac{m_S^2}{m_i} \ln \frac{m_i^2}{m_S^2}\right)\, \right)\nonumber ,
\end{equation}
with the matrices given by
\begin{eqnarray}
a_{Wv} &\approx & \frac{g}{2\sqrt{2}}( {\cal U}_L^\dagger + \frac{\kappa_1 \kappa_2}{v_R^2} \, {\cal U}_R^\dagger)\, , \\
a_{Wa} &\approx& -\frac{g}{2\sqrt{2}}( {\cal U}_L^\dagger - \frac{\kappa_1 \kappa_2}{v_R^2} \, {\cal U}_R^\dagger)\, , \\
a_{W'v} &\approx & \frac{g}{2\sqrt{2}}({\cal U}_R^\dagger - \frac{\kappa_1 \kappa_2}{v_R^2} \, {\cal U}_L^\dagger )\, , \\
a_{W'a} &\approx & \frac{g}{2\sqrt{2}}( {\cal U}_R^\dagger +\frac{\kappa_1 \kappa_2}{v_R^2} \, {\cal U}_L^\dagger )\, .
\end{eqnarray}

\end{itemize}

\newpage

\section{Plots $v_L/v_R$ vs $m_4$ for different scenarios}
\label{app:Plots}

\subsection{Varying the hierarchy of heavy neutrinos}

\begin{figure}[!htb]
     \centering
         \includegraphics[width=0.9\textwidth]{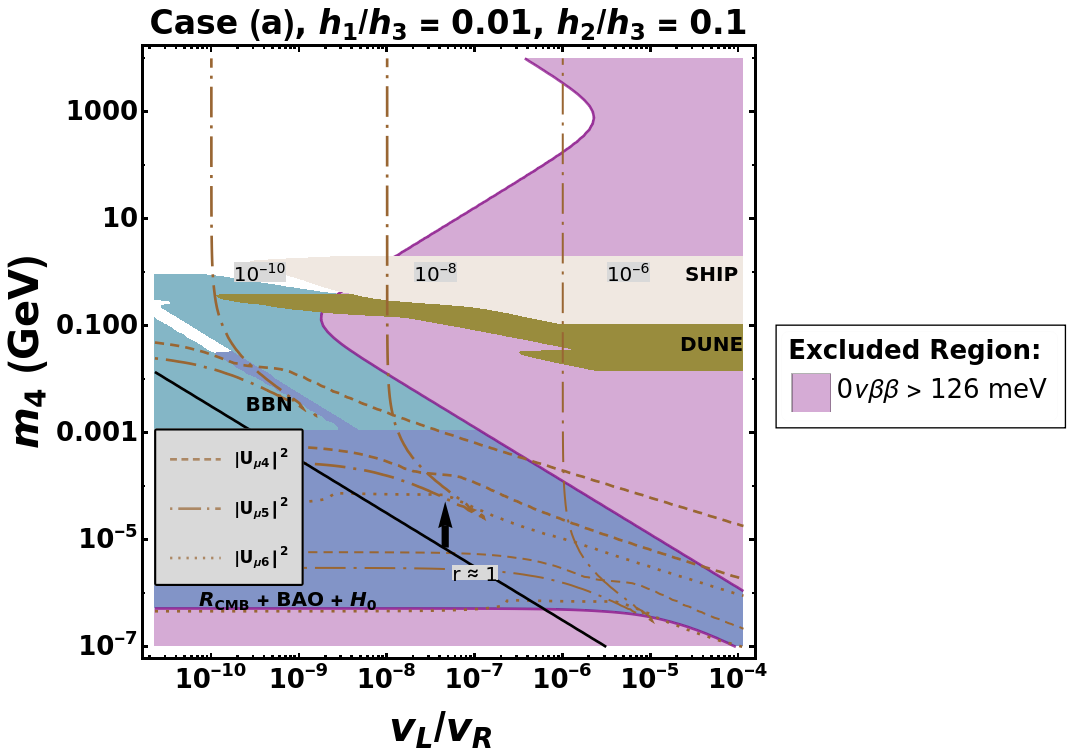}
        \caption{Case (a) for a different heavy neutrino mass hierarchy.}
        \label{fig:finalplotdifHierarchies}
\end{figure}

\pagebreak

\subsection{Varying $m_0$ and light neutrino ordering}
\label{app:lightneutrinosplots}
\begin{figure}[!htb]
         \centering
         \includegraphics[width=0.9\textwidth]{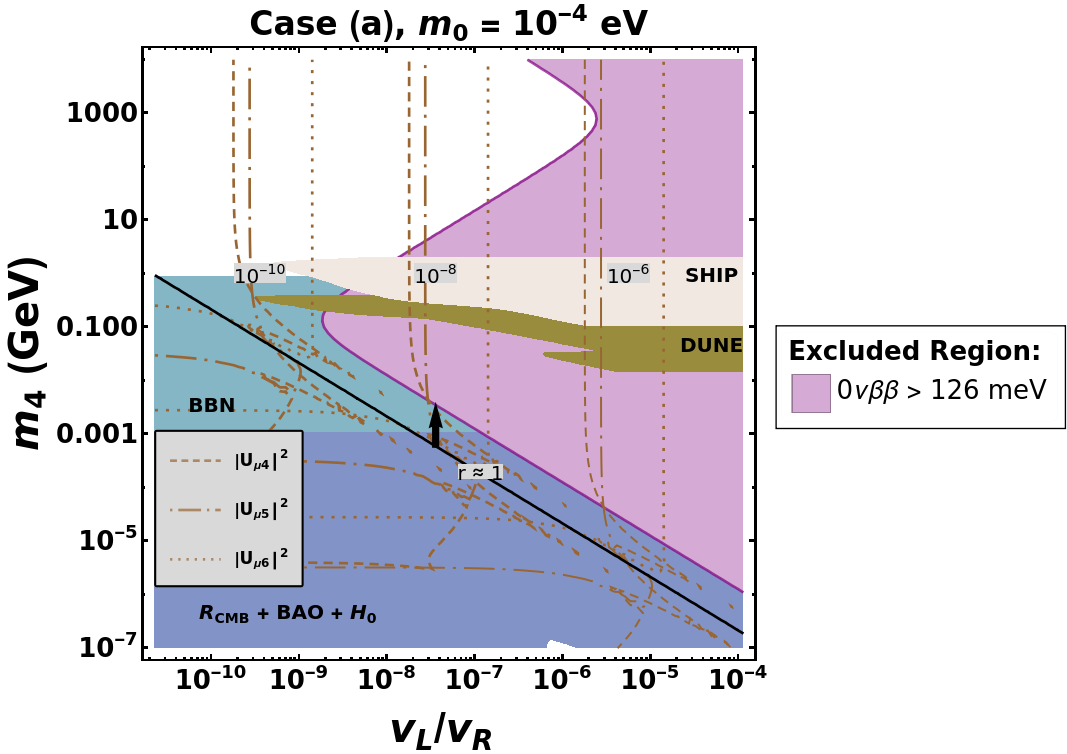}
        \caption{Case (a) with $m_0=10^{-4}$ eV.}
        \label{fig:smallm0}
\end{figure}

\begin{figure}[!hbt]
         \centering
         \includegraphics[width=0.9\textwidth]{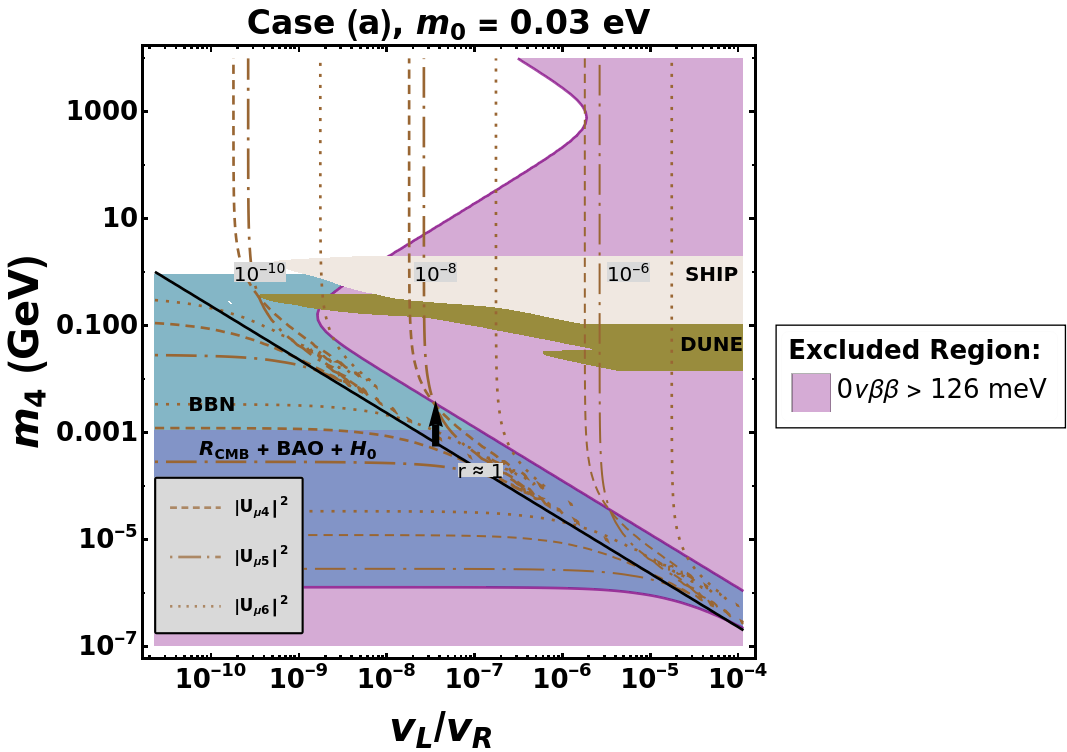}
        \caption{Case (a) with
        $m_0=0.03$ eV.}
        \label{fig:degeneratem0}
\end{figure}

\begin{figure}[!htb]
         \centering
         \includegraphics[width=0.9\textwidth]{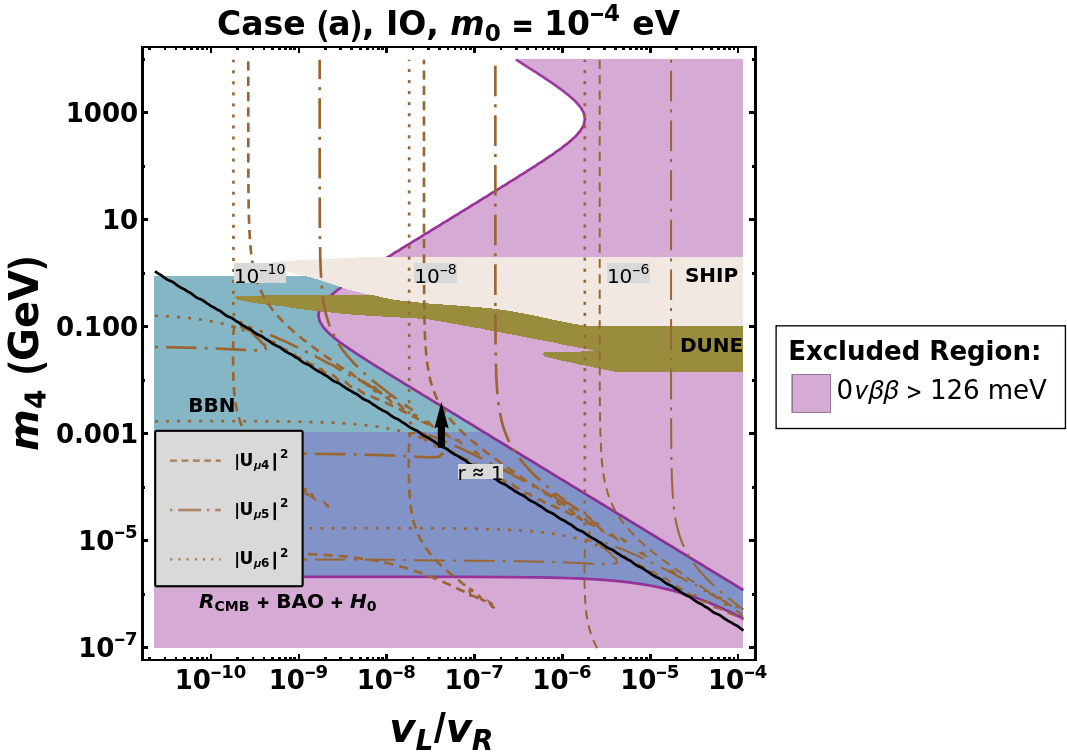}
        \caption{Case (a) but for IO and 
        $m_0= 10^{-4}$ eV.}
        \label{fig:caseAIO}
\end{figure}

\pagebreak

\subsection{$U_L$ and $\phi_i$}

\begin{figure}[!h]
    \centering
    \includegraphics[width = 0.9\textwidth]{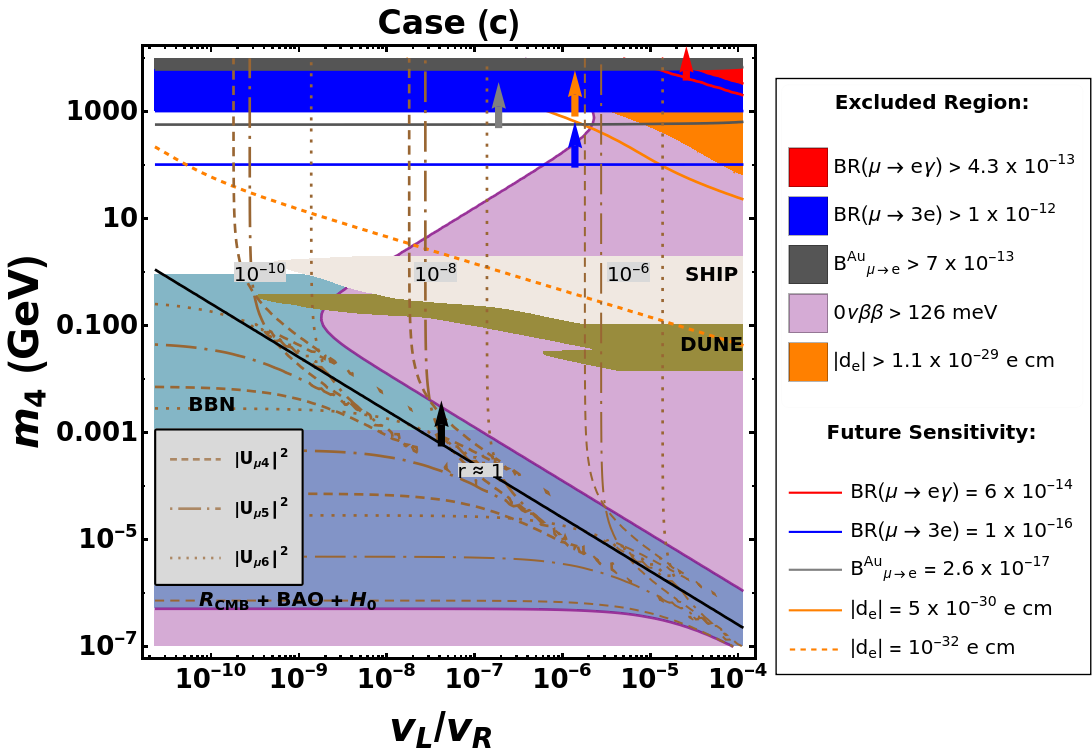}
    \caption{Case (c) where $\phi_i=0$.}
    \label{fig:casecwithoutphases}
\end{figure}

\begin{figure}[!h]
    \centering
    \includegraphics[width =0.9
    \textwidth]{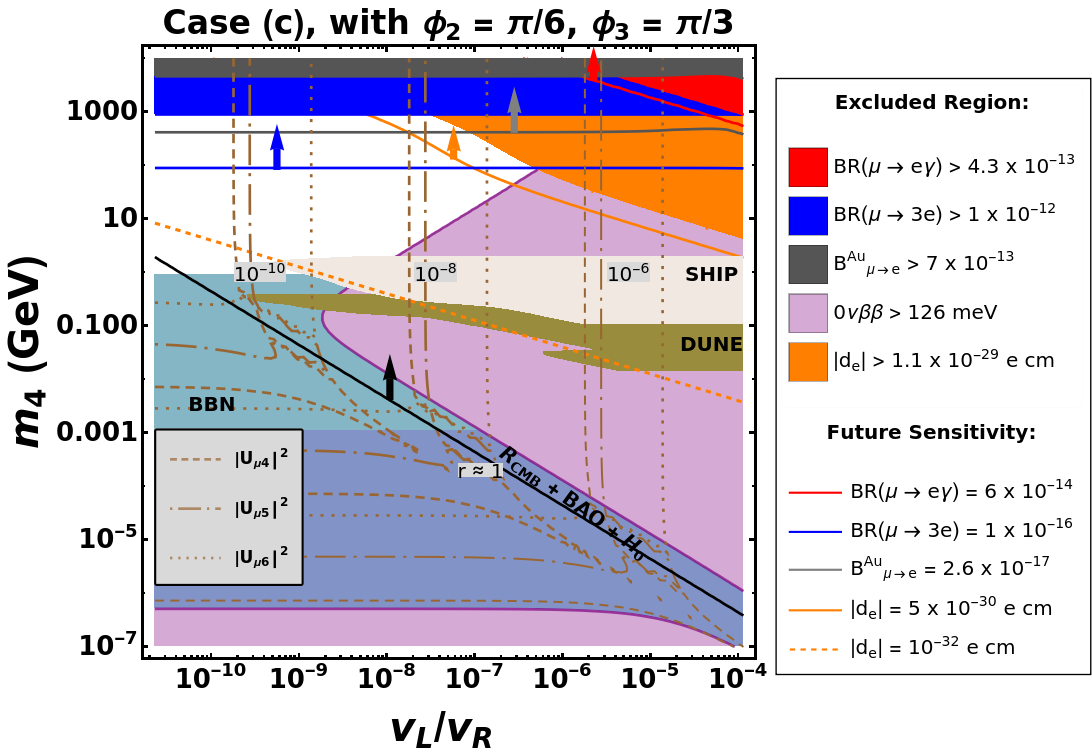}
    \caption{Case (c) with $\phi_2=\pi/6$ and $\phi_3=\pi/3$.}
    \label{fig:casecwithphases}
\end{figure}
\pagebreak

\pagebreak
\bibliographystyle{JHEP}
\bibliography{ENSLR}{}
\end{document}